\documentclass[%
 aip,
 amsmath,amssymb,
preprint,%
]{revtex4-2}

\usepackage{graphicx}% Include figure files
\usepackage{dcolumn}% Align table columns on decimal point
\usepackage{bm}% bold math
\usepackage{amsmath}
\usepackage[utf8]{inputenc}
\usepackage[T1]{fontenc}
\usepackage{mathptmx}
\usepackage[colorlinks=true,linkcolor=blue]{hyperref}%
\usepackage{color}
\usepackage{CJK}
\usepackage{float}
\UseRawInputEncoding

\usepackage{etoolbox}

\makeatletter
\def\@email#1#2{%
 \endgroup
 \patchcmd{\titleblock@produce}
  {\frontmatter@RRAPformat}
  {\frontmatter@RRAPformat{\produce@RRAP{*#1\href{mailto:#2}{#2}}}\frontmatter@RRAPformat}
  {}{}
}%
\makeatother

\begin{document}

\preprint{AIP/123-QED}

\begin{CJK*}{UTF8}{gbsn}

\title{Ensemble data assimilation-based mixed subgrid-scale model for large-eddy simulations}% Force line breaks with \\
\author{Yunpeng Wang (王云朋)$^{1,2,3}$}
\author{Zelong Yuan (袁泽龙)$^{1,2,3}$}
\author{Jianchun Wang (王建春)$^{1,2,3,*}$}
\email[Author to whom correspondence should be addressed:\ ]{wangjc@sustech.edu.cn}
\affiliation{\small 1. National Center for Applied Mathematics Shenzhen (NCAMS), Southern University of Science and Technology, Shenzhen, 518055 China.\\
2. Department of Mechanics and Aerospace Engineering, Southern University of Science and Technology, Shenzhen, 518055 China.\\
3. Guangdong-Hong Kong-Macao Joint Laboratory for Data-Driven Fluid Mechanics and Engineering Applications, Southern University of Science and Technology, Shenzhen, 518055 China.}
% 4. Department of Ocean Science, The Hong Kong University of Science and Technology, Hong Kong, 999077 China.}
\date{\today}% It is always \today, today,

\begin{abstract}

An ensemble Kalman filter (EnKF)-based mixed model (EnKF-MM) is proposed for the subgrid-scale (SGS) closure in the large-eddy simulation (LES) of turbulence. The model coefficients are determined through the EnKF-based data assimilation technique. The direct numerical simulation (DNS) results are filtered to obtain the benchmark data for LES. Reconstructing the correct kinetic energy spectrum of the filtered DNS (fDNS) data has been adopted as the target for the EnKF to optimize the coefficient of the functional part in the mixed model. The proposed EnKF-MM framework is subsequently tested in the LES of both the incompressible homogeneous isotropic turbulence (HIT) and turbulent mixing layer (TML). The performance of LES is comprehensively examined through the predictions of the flow statistics including the velocity spectrum, the probability density functions (PDFs) of the SGS stress, the PDF of the strain rate and the PDF of the SGS energy flux. The structure functions, the evolution of turbulent kinetic energy, the mean flow and the Reynolds stress profile, and the iso-surface of the Q-criterion are also examined to evaluate the spatial-temporal predictions by different SGS models. The results of the EnKF-MM framework are consistently more satisfying compared to the traditional SGS models, including the dynamic Smagorinsky model (DSM), the dynamic mixed model (DMM) and the velocity gradient model (VGM), demonstrating its great potential in the optimization of SGS models for LES of turbulence.

%\begin{description}
%\item[PACS number(s)]
%47.20.Ky, 47.20.Lz, 47.20.Ma
%\end{description}
\end{abstract}

\pacs{Valid PACS appear here}% PACS, the Physics and Astronomy

%\tableofcontents
\maketitle
\end{CJK*}

\section{\label{sec:level1}Introduction}

In recent years, large-eddy simulation (LES) has become an increasingly popular tool in both the research and engineering community thanks to the fast development of modern computers, even though the direct numerical simulation (DNS) is still impractical at high Reynolds number due to the large range of motion scales involved.\cite{Pope2000,Meneveau2000,Yang2021,Moser2021} Compared to the Reynolds-averaged Navier-Stokes (RANS) method which calculates the ensemble average of the flow and models the Reynolds stress,\cite{Pope2000,Durbin2018,Pope1975,Duraisamy2019,Jiang2021} LES can directly solve the major energy-containing large-scale motions above the grid scale while leaving the subgird-scale (SGS) motions handled by the SGS models.\cite{Garnier2009,Moin1991,Germano1992,Smagorinsky1963,Lilly1967,Deardorff1970,Chen2012,Rahman2019,Mons2021,Qi2022,Vadrot2023}

Various SGS models can be found in the literature, including the Smagorinsky model (SM),\cite{Smagorinsky1963,Lilly1967,Hasslberger2021} dynamic Smagorinsky model (DSM),\cite{Moin1991,Germano1991,Germano1992,Lilly1992} scale-similarity model,\cite{Bardina1980,Liu1994} dynamic mixed model (DMM),\cite{Speziale1988,Erlebacher1992,Zang1992,Shi2008,Yu2017} velocity gradient model (VGM),\cite{Clark1979} implicit-LES (ILES),\cite{Boris1992,Garnier1999,Visbal2003, Adams2004,Grinstein2007} non-local eddy viscosity approach,\cite{ClarkDiLeoni2021} SGS helicity equation model,\cite{Qi2021} data assimilation-based methods,\cite{Bauweraerts2021,Buzzicotti2020,Li2022} and machine learning-based models.\cite{Durbin2018,Duraisamy2019,Jiang2021,Wang2018,Maulik2018,Zhou2019,Beck2019,Yang2019,Xie2019,Xie2019b,Xie2019c,Xie2018,Gamahara2017,Vollant2017,Maulik2019,Xie2020d,Park2021,Subel2021,Li2021,Yuan2022} Most of the SGS models can be labeled as either functional or structural models.\cite{Meneveau2000,Sagaut2006} Functional models usually assume an eddy viscosity form for the SGS stress, and mimic the molecular dissipation to remove kinetic energy from the resolved scales. However, the prediction of the SGS term itself is often not satisfying,\cite{Vollant2017} so that the local structure of the flow field can not be well recovered. Meanwhile, the coefficient of functional models is approximately determined through theoretical arguments on simple turbulent flows, which often leads to over estimation of the SGS dissipation. On the other hand, structural modeling is known to be accurate in terms of predicting the local SGS term. However, highly accurate SGS model can be unstable \cite{Nadiga2007}, potentially due to insufficient SGS dissipation or the backscatter of kinetic energy.\cite{Vreman1995}

Among many methods that have been proposed to resolve the stability issue with structural models,\cite{Liu1994,Vreman1997,Beck2019,Zhou2020,Vollant2016,Xie2020b,Xie2020d} a very convenient way is to combine it with a functional (dissipative) part,\cite{Vreman1995} giving rise to the mixed SGS models. For such models, the coefficient of the dissipative part must be properly determined, and this constitutes the major purpose of the current work. 

A typical example of structural modeling is the velocity gradient model (VGM),\cite{Clark1979} which has a much higher SGS accuracy than the eddy viscosity-based models. The VGM model can be rigorously derived through Taylor expansion of the unfiltered flow variables using the filtered ones.\cite{Clark1979,Pope2000,Garnier2009} Unfortunately, it has been found that such a high-accuracy SGS model can perform poorly in practical LES or even diverges if used alone, presumably due to the insufficient SGS dissipation.\cite{Vreman1997} As discussed, this issue is present not only in the VGM model, but also in many other high-accuracy SGS models.\cite{Yuan2020,Wang2021,Wang2021a,Wang2022} Consequently, artificial dissipation must be adopted and adjusted to stablize the numerical simulation, and the mixed model strategy is a typical example of such attempts. 

Traditionally, the coefficients of the mixed turbulent models are determined through dynamic procedures, which is free from benchmark data and can be fully determined using the LES variables.\cite{Speziale1988,Erlebacher1992,Zang1992,Shi2008,Yu2017,Vreman1995,Vreman1997,Wang2021a} Recently, machine learning techniques have also been adopted to calibrate the mixed turbulent models, including the artificial-neural-network (ANN)-based algebraic RANS models,\cite{Ling2016,Zhang2022} the ANN-based mixed SGS model,\cite{Xie2019c} the algebraic SGS model,\cite{Xie2020d,Xu2023}, the spatial gradient SGS model,\cite{Wang2021}, and the gene-expression programming-based SGS models.\cite{Li2021,Wu2022,Schoepplein2018} In closer relevance to present study, data assimilation techniques have been increasingly used in optimizing turbulent models due to fast accumulation of high fidelity turbulent data.\cite{Meldi2017,Kato2015,Deng2021,Strofer2021,Fang2021,Wang2021b,Zhang2021,Foures2014,He2018,He2019,Franceschini2020,He2021,Pawar2020,Li2023} For example, Buzzicotti \emph{et al}. has adopted a nudging method to evaluate the performance of the SGS models by incorporating the DNS data to the LES solver.\cite{Buzzicotti2020} More recently, Yuan \emph{et al}. proposed a adjoint-based variational method to calibrate the coefficients of SGS models.\cite{Yuan2023} An additional term has also been introduced to the adjoint equations for numerical stability. These works have shown that with a proper choice of the model coefficients, the accuracy of LES can be significantly augmented.

In the current work, following the mixed model strategy, we combine the Smagorinsky-type functional model with the VGM model to enhance its performance. Meanwhile, we adopt the EnKF method to calibrate the coefficient of the functional part. As a data assimilation technique, EnKF is more convenient and efficient to implement compared to adjoint-based variational methods. In RANS community, EnKF has already been successfully applied in the optimization of the model coefficients,\cite{Kato2013,Deng2018,Chen2019,Deng2021,Fang2021,Wang2021b} for integrating disparate observations in data assimilation,\cite{Zhang2021} in the optimization of sensor locations for data assimilation,\cite{Deng2021} and in updating the weights of a neural network.\cite{Zhang2022} In contrast, the EnKF method is investigated to a much lesser extent in LES,\cite{Labahn2020} and has rarely been used to optimize the SGS model.

In the LES, the filtered DNS (fDNS) data are used as the benchmark data. Further, as the spatial-temporal point-to-point convergence is not feasible for turbulence due to its chaotic nature, we take the important turbulent statistics, namely, the turbulent kinetic energy spectrum as the target to optimize the model coefficient. The filter width ($\overline{\Delta}$) to grid spacing ($h_{LES}$) ratio (FGR) in the LES is taken to be FGR $=\overline{\Delta}/h_{LES}=2$ to ensure that the influence of the numerical errors is much less compared to that of the SGS model.\cite{Pope2004,Radhakrishnan2012,Piomelli2015} The proposed framework is tested in both the incompressible homogeneous isotropic turbulence (HIT) and incompressible turbulent mixing layer (TML).

The rest of the paper is structured as follows. In Section II, the governing equations for LES and some important SGS models are introduced, followed by a description of EnKF framework. Sections III and IV present the results of the EnKF-MM-based LES in the cases of incompressible HIT and TML, respectively. Section V offers a brief discussion on the merits and characteristics of different data-driven and optimization methods for SGS modeling. Finally, Section VI summarizes the paper and gives the major conclusions.

\section{Governing equations of large-eddy simulation and the ensemble Kalman filter-based mixed model}

In the current section, the governing equations of large-eddy simulation are introduced. The unclosed SGS stress is discussed with emphasis on some important SGS models in the literature. Finally, a detailed description of the EnKF-based mixed model (EnKF-MM) is presented.

\subsection{Governing equations of large-eddy simulation and the subgrid closure problem}

The mass and momentum for incompressible turbulence are governed by the Navier-Stokes (NS) equations, namely\cite{Pope2000,Ishihara2009,Buzzicotti2018}
 \begin{equation}
  \frac{\partial u_{i}}{\partial x_{i}}=0,
  \label{mass}
\end{equation}
 \begin{equation}
  \frac{\partial u_{i}}{\partial t}+\frac{\partial u_{i} u_{j}}{\partial x_{j}}=-\frac{\partial p}{\partial x_{i}}+\nu\frac{\partial^{2} u_{i}}{\partial x_{j}\partial x_{j}}+\mathcal{F}_{i},
  \label{momentum}
\end{equation}
where $u_{i}$ is the $i$-th velocity component, $p$ is the pressure divided by the constant density, $\nu$ is the kinematic viscosity, and $\mathcal{F}$ is the large-scale forcing. The summation convention is assumed throughout the paper unless otherwise specified.

The Taylor-scale Reynolds number $Re_{\lambda}$ is defined as\cite{Ishihara2009}
\begin{equation}
  Re_{\lambda}=\frac{u^{rms}\lambda}{\sqrt{3}\nu},
  \label{ReMt}
\end{equation}
where $u^{rms}$ is the root-mean-square (rms) velocity calculated as $u^{rms}=\sqrt{\langle u_{i}u_{i}\rangle}$, with $\langle \cdot \rangle$ denoting a spatial average over all the homogeneous directions (i.e. the entire physical domain for the case of HIT). The Taylor length scale $\lambda$ is defined as
\begin{equation}
  \lambda=\sqrt{\frac{5\nu}{\epsilon}}u^{rms},
  \label{lambda}
\end{equation}
where $\epsilon$ is the dissipation rate of kinetic energy, given by $\epsilon=2\nu\langle S_{ij}S_{ij}\rangle$. Here $S_{ij}=\frac{1}{2}(\partial{u_{i}}/\partial{x_{j}}+\partial{u_{j}}/\partial{x_{i}})$ is the strain rate. In addition, the Kolmogorov length scale $\eta$ and the integral length scale $L_{I}$ are respectively defined as\cite{Pope2000,Ishihara2009}
\begin{equation}
  \eta=(\frac{\nu^{3}}{\epsilon})^{1/4},
  \label{kolmo}
\end{equation}
\begin{equation}
  L_{I}=\frac{3\pi}{2(u^{rms})^{2}}\int_{0}^{\infty}\frac{E(k)}{k}dk,
  \label{inte}
\end{equation}
with the energy spectrum $E(k)$ satisfying $\int_{0}^{\infty}E(k)dk=(u^{rms})^{2}/2$. Additionally, the large-eddy turnover time is defined as $\tau=L_I/u^{rms}$.

The governing equations for large-eddy simulation (LES) are obtained through a filtering operation as follows
\begin{equation}
  \overline{f}(\mathbf{x})=\int_{D}f(\mathbf{x}-\mathbf{r})G(\mathbf{r},\mathbf{x};\overline{\Delta})d\mathbf{r},
  \label{filtering}
\end{equation}
where $f$ can be any physical quantity of interest, G is the filter kernel, $\overline{\Delta}$ is the filter width and D is the physical domain. Applying Eq.~(\ref{filtering}) to Eqs.~(\ref{mass}) and (\ref{momentum}) gives rise to the governing equations of LES, namely\cite{Vreman1995,Meneveau2000,Sagaut2006,Garnier2009}
\begin{equation}
  \frac{\partial \overline{u}_{i}}{\partial x_{i}}=0,
  \label{fiteredmass}
\end{equation}
 \begin{equation}
  \frac{\partial \overline{u}_{i}}{\partial t}+\frac{\partial \overline{u}_{i} \overline{u}_{j}}{\partial x_{j}}=-\frac{\partial \overline{p}}{\partial x_{i}}-\frac{\partial\tau_{ij}}{\partial x_{j}}+\nu\frac{\partial^{2} \overline{u}_{i}}{\partial x_{j}\partial x_{j}}+\overline{\mathcal{F}}_{i}.
  \label{filteredmomentum}
\end{equation}
Clearly, Eq.~(\ref{filteredmomentum}) contains an unclosed SGS stress $\tau_{ij}$,
\begin{equation}
  \tau_{ij}=\overline{u_{i}u_{j}}-\overline{u}_{i}\overline{u}_{j},
  \label{tau}
\end{equation}
which appears due to the nonlinear interactions between the resolved and subgrid motions, and whose modeling in terms of the resolved variables constitutes a major issue of LES. Once the SGS model becomes available, the LES equations can be solved.

An economic way to solve the LES equations is to straightforwardly neglect $\tau_{ij}$, which is the merit of implicit-LES (ILES) since no SGS modeling is required.\cite{Boris1992} However, with ILES, the numerical dissipation must be entirely responsible for the SGS energy transfer from the resolved to the subgrid scales. In turn, the solution depends on both the applied grid and the numerical approach,\cite{Pope2000} and extra numerical viscosity is often necessary in case of insufficient numerical dissipation by the numerical scheme.\cite{Xie2020b}

The explicit treatments of the SGS stress as functions of the LES variables can be categorized into functional and structural modelings.\cite{Sagaut2006} A typical example of functional modeling is the Smagorinsky model (SM),\cite{Smagorinsky1963} namely

\begin{equation}
 \tau^{A}_{ij}=\tau_{ij}-\frac{\delta_{ij}}{3}\tau_{kk}=-2C^{2}_{Smag}\overline{\Delta}^{2}|\overline{S}|\overline{S}_{ij},
  \label{tauDSM}
\end{equation}
with $\overline{\Delta}$ being the filter width and $\overline{S}_{ij}$ the filtered strain rate. $|\overline{S}|=\sqrt{2\overline{S}_{ij} \overline{S}_{ij}}$ is the characteristic filtered strain rate. The classical value for the Smagorinsky coefficient is $C_{Smag}=0.16$, which can be determined through theoretical arguments for isotropic turbulence.\cite{Smagorinsky1963,Lilly1967,Pope2000}

The Smagorinsky model is known to be over-dissipative in the transition regime of turbulence.\cite{Pope2000} In this regard, a dynamic version of the model has been proposed using the Germano identity,\cite{Germano1992,Pope2000} giving rise to the dynamic Smagorinsky model (DSM).\cite{Moin1991,Lilly1992,Meneveau2000,Sagaut2006} Through a least-square method, the coefficient $C^{2}_{Smag}$ is determined by
\begin{equation}
  C^{2}_{Smag}=\frac{\langle L_{ij} M_{ij}\rangle}{\langle M_{kl} M_{kl}\rangle},
  \label{C}
\end{equation}
with $L_{ij}=\widetilde{\overline{u}_{i}\overline{u}_{j}}-\widetilde{\overline{u}}_{i}\widetilde{\overline{u}}_{j}, \alpha_{ij}=-2\overline{\Delta}^{2}|\overline{S}|\overline{S}_{ij}, \beta_{ij}=-2\widetilde{\overline{\Delta}}^2|\widetilde{\overline{S}}|\widetilde{\overline{S}}_{ij}$ and $M_{ij}=\beta_{ij}-\widetilde{\alpha}_{ij}$.
The overbar denotes the filtering at scale $\overline{\Delta}$, and a tilde denotes a coarser filtering ($\widetilde{\Delta}=2\overline{\Delta}$). Clearly, the model coefficient is fully determined by the filtered variables.

As discussed in the introduction, with the functional models, even though the overall energy exchange between the resolved and subgrid scales can be represented, the prediction of the SGS stress itself is often inadequate, leading to poor predictions of the local flow structures. In contrast, the structural models are more accurate in terms of predicting the local SGS stresses even though it can be numerically less stable. To benefit from both types of models, the mixed models have also been proposed in the literature, such as the dynamic mixed model (DMM),which is a combination of the Smagorinsky model and a scale-similarity part which plays the role of the structural model.\cite{Bardina1980,Speziale1988} The DMM model can be written as $\tau_{ij}^{A}=C_{1}h^{A}_{1,ij}+C_{2}h^{A}_{2,ij}$, where $h^{A}_{1,ij}=-2\overline{\Delta}^{2}|\overline{S}|\overline{S}_{ij}$, $h^{A}_{2,ij}=h_{2,ij}-\delta_{ij}h_{2,kk}/3$ and $h_{2,ij}=\widetilde{\overline{u}_{i}\overline{u}_{j}}-\widetilde{\overline{u}}_{i}\widetilde{\overline{u}}_{j}$. Again, the model coefficients are determined through the least-square algorithm by exploiting the Germano identity. The details can be well found in the literature and not reproduced.\cite{Germano1992,Xie2020b,Yuan2020}

In the present work, we adopt a high-accuracy structural model, namely, the velocity gradient model (VGM),\cite{Clark1979} given by
\begin{eqnarray}
\tau_{ij}=\frac{\overline{\Delta}^2}{12}\frac{\partial \overline{u}_{i}}{\partial x_{k}}\frac{\partial \overline{u}_{j}}{\partial x_{k}},
\label{tauVGM}
\end{eqnarray}
which can be deduced from the Taylor expansion of the filtered variables using the unfiltered ones.\cite{Clark1979,Pope2000} The VGM model has been shown to be more accurate than the scale-similarity model (SSM) in the \emph{a priori} analysis.\cite{Wang2021} While the VGM model have high correlation coefficients with the true SGS stress, it is numerically unstable for certain flow conditions due to insufficient SGS dissipation.\cite{Vreman1995,Nadiga2007} To resolve this issue, we combine the VGM model with a functional part, yielding a mixed model, namely

\begin{equation}
 \tau_{ij}=\frac{\overline{\Delta}^2}{12}\frac{\partial \overline{u}_{i}}{\partial x_{k}}\frac{\partial \overline{u}_{j}}{\partial x_{k}}-C_{D}\overline{\Delta}^{2}|\overline{S}|\overline{S}_{ij},
  \label{taumix}
\end{equation}
where $C_D$ is the coefficient for dissipative functional part. Certainly, $C_D$ can be determined through the aforementioned dynamic procedure.\cite{Vreman1997} However, to more accurately calibrate the coefficient of functional part, we will use fDNS data to optimize the coefficient through the EnKF-based data assimilation technique.\cite{Lewis2006} We also note that the structural part of the mixed model is not modified so as to retain as much its SGS accuracy as possible. Besides, our previous adjoint-based variational work has also shown that, in the mixed model, the coefficient of the structural part is very close to the original value even if the optimization of the structural coefficient is considered.\cite{Yuan2023} Hence, we shall only optimize the functional coefficient $C_D$ to avoid unnecessary complexity.

\subsection{The ensemble Kalman filter procedure for the coefficient optimization of the mixed SGS model}

In the current work, the ensemble Kalman filter (EnKF) is used to determine the value of the coefficient $C_D$. The Kalman filter (KF) is a sequential data assimilation technique based on the minimum error variance assumption and the least-square approach.\cite{Kato2015} In the EnKF, the covariance matrix is replaced by the sample covariance matrix to avoid the computational cost in updating the covariance matrix. Since the spatial-temporal point-to-point convergence is not feasible for turbulence due to its chaotic nature, the kinetic energy spectrum is chosen as the target for the EnKF. Based on our tests, in the current mixed SGS model, if the correct energy spectrum can be recovered, many important flow statistics can be adequately reconstructed, except for the SGS stress, the prediction of which by the EnKF-MM model is slightly worse compared to that by the pure structural model. The core of EnKF is the ``prediction and analysis'' process.

For the current problem, in the prediction step, each simulation (LES) sample in the ensemble is performed using different values of $C_D$. In the analysis step, we first calculate the ensemble covariance matrix $P$ as follows

\begin{equation}
 \mathbf{P}=\frac{1}{N-1} \sum_{i=1}^{N} (\mathbf{e}_i- \overline{\mathbf{e}})(\mathbf{e}_i- \overline{\mathbf{e}})^T,
  \label{covarianceP}
\end{equation}
where $N$ is the size of the ensemble that need to be tested in EnKF. For the $i$-th LES sample, the state vector $\mathbf{e}_i$ is of size $n\times k_{max}+1$ where $k_{max}$ is maximum resolved wavenumber, and contains the computed energy spectrum at all the $n$ selected time instants spaced at $\Delta t_{Obs}$ and the model coefficient $C_{D,i}$. $\overline{\mathbf{e}}=\frac{1}{N}\sum_{i=1}^{N} \mathbf{e}_i$ is the sample mean. Here, we have directly treated the energy spectrum as state variables to avoid the complexity of configuring a complicated observation matrix that maps the flow variables to the energy spectrum. In effect, the current treatment takes direct advantage of the relation between the model coefficient $C_D$ and the energy spectrum. Consequently, the Navier-Stokes equations only enter the problem when solving for the flow variables, from which the energy spectrum is subsequently obtained. Since the EnKF is already an approximate treatment for nonlinear non-Gaussion processes like turbulent flow in the first place, the current treatment is not unreasonable as long as the prediction error of energy spectrum can be reduced through iteration. To optimize a constant model coefficient, the kinetic energy spectrum for all the saved data are considered. As shall be seen later, this strategy turns out quite satisfying in the tested cases.

The analysis (filtering) is performed according to

\begin{equation}
 \mathbf{e}_i^a=\mathbf{e}_i+\mathbf{K}(\mathbf{q}- \mathbf{H}\mathbf{e}_i +\mathbf{w}_i),
  \label{analysis}
\end{equation}
where the analysis vector $\mathbf{e}_i^a$ contains the updated model coefficient, and the observation vector $\mathbf{q}$ for the energy spectrum is of size $n\times k_{max}$. $\mathbf{H}$ is the observation matrix of size [$n\times k_{max}$, $n\times k_{max}+1$]. $\mathbf{w}_i$ is the observation noise. In the current case, $\mathbf{w}_i$ is on the machine error level since the observations are taken from the high-fidelity DNS field. The Kalman gain $\mathbf{K}$ is calculated as

\begin{equation}
 \mathbf{K}=\mathbf{PH}^T(\mathbf{HPH}^T+\mathbf{R})^{-1},
  \label{gain}
\end{equation}
which can be derived through the minimum error variance theory and a least-square approach.\cite{Lewis2006} Here $\mathbf{R}$ is the covariance matrix of observation noise, defined by

\begin{equation}
 \mathbf{R}=\frac{1}{N-1} \sum_{i=1}^{N} \mathbf{w}_i\mathbf{w}_i^T.
  \label{covarianceR}
\end{equation}
Assuming that each observation is independent, the covariance matrix of observation noise $\mathbf{R}$ can be also approximated by a diagonal matrix. The major iteration loop of the current EnKF framework is described as follows:

(1) Assigning different values of $C_{D,i}$ to each sample LES member according to a uniform distribution in the range between 0 and $C_{D}^{max}$, where $C_{D}^{max}$ is the maximum value for $C_{D}$.

(2) Initializing all the LESs in the ensemble using the same flow field obtained from fDNS at $t=t_0$.

(3) Running the sample LES simulations using the corresponding values for $C_{D,i}$.

(4) Saving the LES data at selected time instants of interval $\Delta t_{Obs}$: $t_0+\Delta t_{Obs}$, $t_0+2\Delta t_{Obs}$, $t_0+3\Delta t_{Obs}$ ... $t_f$, where $t_f$ is the final (end) time of LES. 

(5) Calculating the energy spectrum at all the selected time instants for each LES sample, as well as obtaining the corresponding spectrum for the benchmark fDNS.

(6) Performing the EnKF to update $C_{D,i}$ for each LES sample.

(7) Ending the loop if the number of iteration exceeds the maximum allowed value $N_{iter}$; otherwise, going back to step (2) with the updated values for $C_{D,i}$.

\begin{figure}\centering
\includegraphics[width=1.0\textwidth]{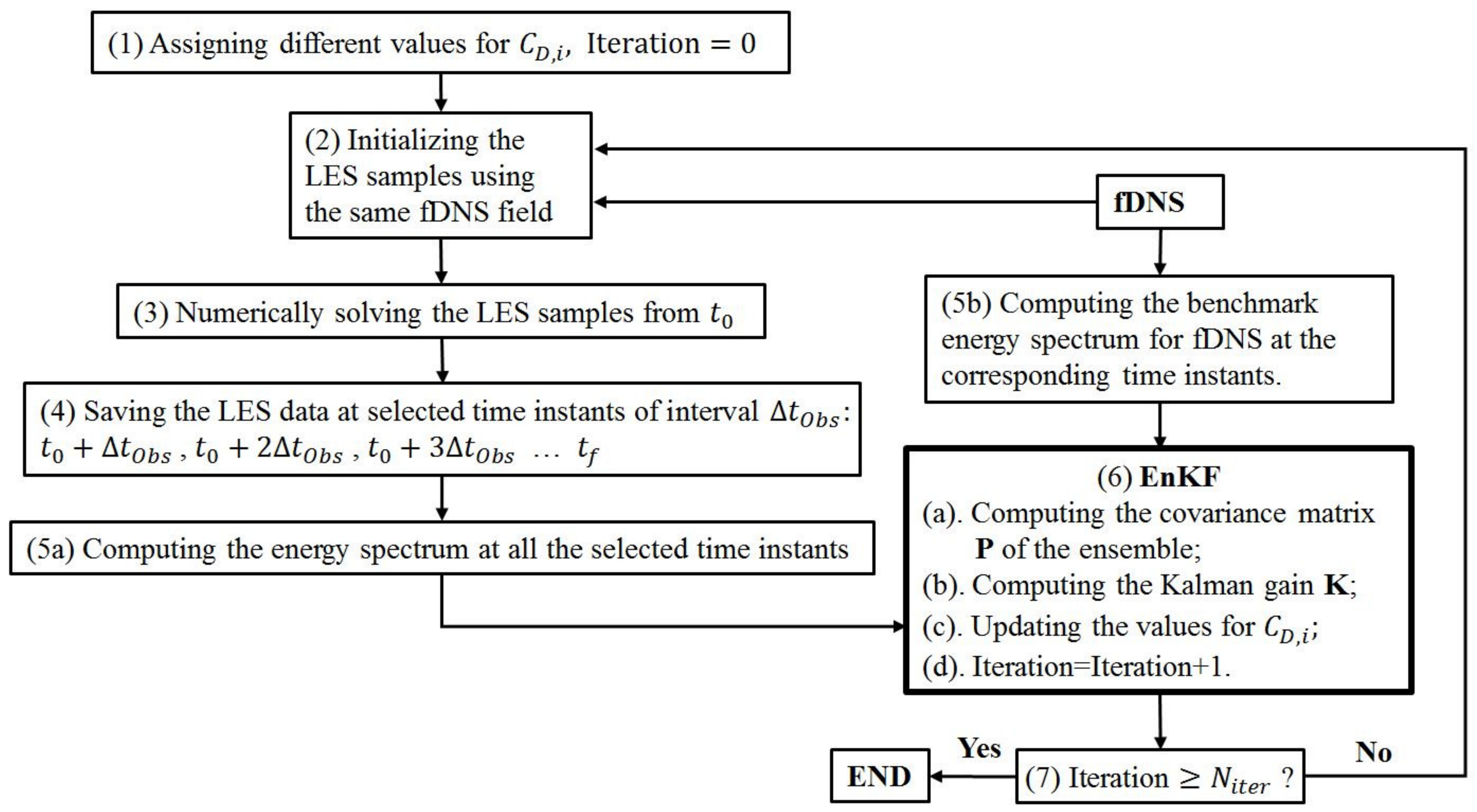}
 \caption{The flowchart of the EnKF-MM framework.}\label{fig_EnKF}
\end{figure}

Fig. \ref{fig_EnKF} displays the flowchart for EnKF-MM framework. In the main body of this work, the Gaussian filter is adopted in the fDNS and LES, and the corresponding filter kernel in one dimension can be written as
\begin{equation}
  G(r)=(\frac{6}{\pi \Delta^2})^{1/2} \exp (-\frac{6r^2}{\Delta^2}).
  \label{gauss}
\end{equation}
For an investigation on the influence of the filter type, the reader is referred to Appendix A. More details on the setup of the parameters for the implementation of EnKF will be provided in following Sections III and IV for the LESs of HIT and TML, respectively. It is important to note that, for significantly different flow configurations, the EnKF-MM model should be calibrated separately for better accuracy. However, for similar flow configurations but with different flow parameters, the EnKF-MM model can still be applied with acceptable accuracy. This is demonstrated in Appendix B where the influence of the Reynolds number, filter width and forcing are investigated.

\section{The numerical results of incompressible homogeneous isotropic turbulence}

The DNS is performed in a $(2\pi)^{3}$ cubic box using $512^{3}$ grid, with periodic boundary conditions and a second-order two-step Adams-Bashforth time scheme. The large-scale forcing is applied to the two lowest wavenumbers such that the kinetic energy in the two wavenumbers satisfies the $-5/3$ scaling law.\cite{Wang2010} More specifically, $E(1)=1.242477$ and $E(2)=0.391356$ are enforced throughout the numerical simulation.\cite{Wang2010} A pseudospectral method is adopted, and dealiasing is achieved through the two-thirds rule.\cite{Patterson1971} The kinematic viscosity is taken at $\nu=0.0025$, yielding a Taylor Reynolds number $Re_{\lambda}\approx160$. The details of the DNS are given in Table~\ref{tab_hit_DNS}.

\begin{table*}
\begin{center}
\small
\begin{tabular*}{0.85\textwidth}{@{\extracolsep{\fill}}cccccc}
\hline
Reso. &$Re_{\lambda}$   &$\nu$ &$\Delta t$ &$u^{rms}$ &$\epsilon$\\ \hline
$512^{3}$ &160    &0.0025 &0.0004 &2.3 &0.72\\ \hline
\end{tabular*}
\normalsize
\caption{Numerical simulation parameters and one-point statistical quantities for incompressible isotropic turbulence at $Re_{\lambda}=160$.}
\label{tab_hit_DNS}
\end{center}
\end{table*}

The LESs for the forced incompressible HIT are performed at the same Taylor Reynolds number $Re_{\lambda}\approx160$ with filter width $\overline{\Delta}=16h_{DNS}$. Here $h_{DNS}$ is the grid spacing in the DNS. Since the LES is governed by the filtered Navier-Stokes equations, the LESs with different SGS models are computed using the same initial field obtained from the filtered DNS data. We note that, the initial conditions of LES are taken from the fully developed state instead of the initial state for DNS. In this way, we can circumvent the difficulty in simulating the initial transition period, which is very sensitive to grid resolution and challenging to predict in a coarse-grid simulation like LES. In fact, as the initial field for DNS is artificially generated with random noises, it may not satisfy the Navier-Stokes equations. This poses another difficulty in LES if its initial condition is taken from the initial field of DNS.

In the LES, the filter width ($\overline{\Delta}$) to grid spacing ($h_{LES}$) ratio $FGR$ ($=\overline{\Delta}/h_{LES}$) is taken at $FGR=2$ to ensure that the influence of the numerical errors is much less compared to that of the SGS model.\cite{Pope2004,Radhakrishnan2012,Piomelli2015} In turn, a $64^3$ grid has been adopted for LES. The ratio of the time step in the LES to that in the DNS is taken at $\Delta t_{LES}/\Delta t_{DNS}=10$. Here the time steps are chosen such that the computational cost of time integration can be reduced while the CFL condition\cite{Wang2010} is still satisfied for numerical stability. The LESs are run for approximately six large-eddy turnover time in total. Most of the results are time-averaged values due to the statistically stationary nature of forced HIT unless the otherwise noted.

The setup of the parameters for the implementation of EnKF in the LES of forced HIT is provided in Table \ref{tab_hit_EnKF}, where $N_{iter}$ is the total number of iteration, $\Delta t_{Obs}$ is the interval between observations, $t_0$ and $t_f$ are the starting and end time of LES, respectively. As discussed, the difference between each LES samples are achieved by setting different values for $C_D$. The maximum value for $C_D$ is based on the original classical Smagorinsky coefficient $C_{Smag} =0.16$ since $C_D^{max} =2C^{2}_{Smag} \approx 0.05$. The original Smagorinsky model is known to be over-dissipative, especially in the transitional flow regime.\cite{Pope2000} In the mixed model, with confidence, we expect the optimum value for $C_D$ should be lower than the classical value since the structural part of the model also contributes to the SGS dissipation.\cite{Garnier2009} The initial values for $C_D$ are uniformly distributed in the range as we do not know \emph{a priori} the ideal value for the coefficient. Meanwhile, because the observations are taken from the DNS, the observation noise $\mathbf{w}_i$ should be on the machine error level which is of order $10^{-7}$ for a single-precision solver. The multiplier $\mathbf{U}(-1,1)$ in $\mathbf{w}_i$ stands for a vector composed of the uniformly distributed random numbers between -1 and 1 since the numerical truncation is mainly responsible for the observation noise.   

\begin{table*}
\begin{center}
\small
\begin{tabular*}{0.95\textwidth}{@{\extracolsep{\fill}}ccccccc}
\hline
Range of $C_D$ &Ensemble size (N) &$N_{iter}$ &$\Delta t_{Obs}$ &$t_0$   &$t_f$ &$\mathbf{w}_i$\\ \hline
$0<C_D<0.05$ &10, 20 &50 &$0.3\tau$ &0 &$6\tau$ &$\mathbf{U}(-1,1)\times 10^{-7}$\\ \hline
\end{tabular*}
\normalsize
\caption{Detailed parameters for the implementation of the ensemble Kalman filter in the LES of forced HIT.}
\label{tab_hit_EnKF}
\end{center}
\end{table*}

\begin{figure}\centering
\includegraphics[height=.35\textwidth]{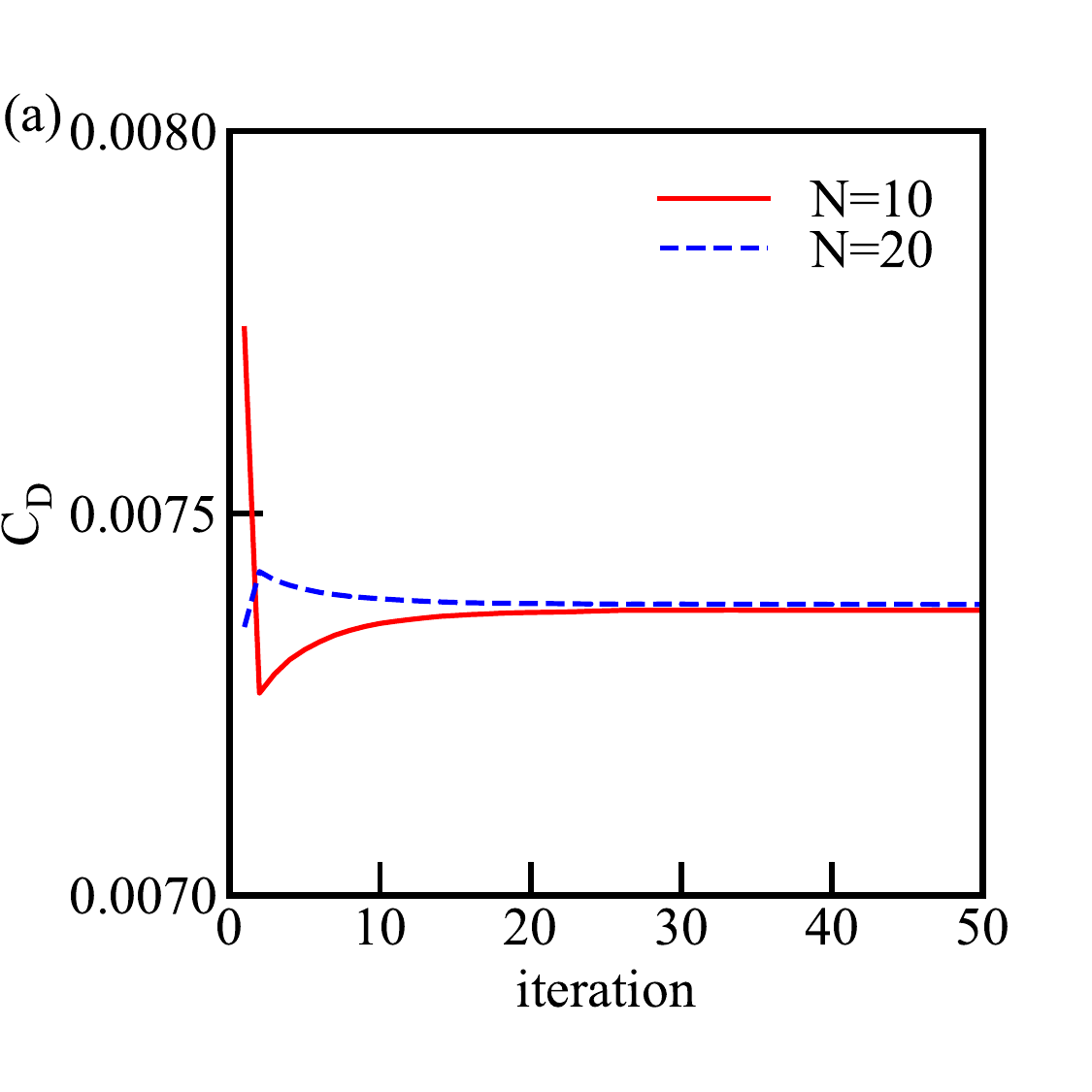}\hspace{-0.2in}
\includegraphics[height=.35\textwidth]{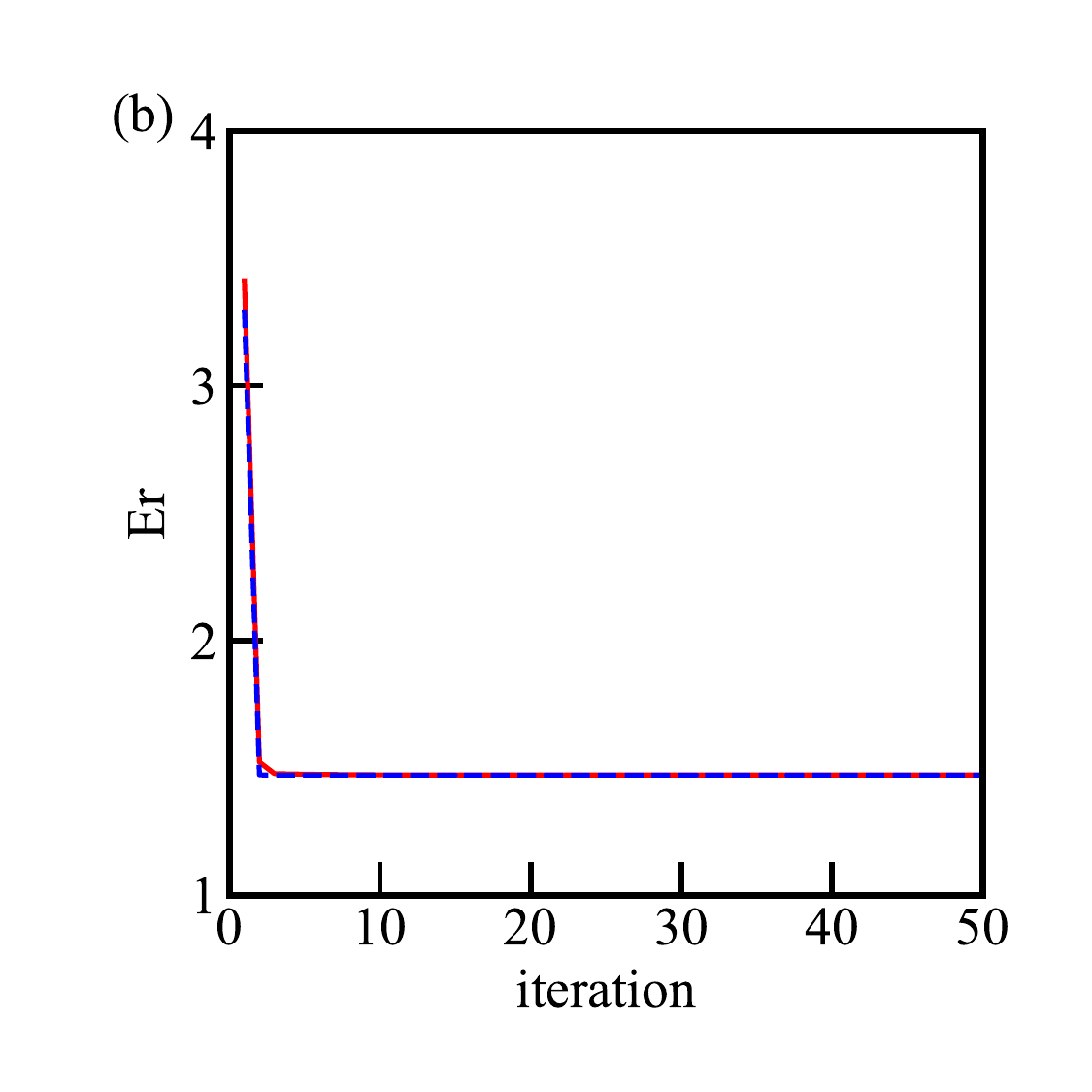}\hspace{-0.2in}
\includegraphics[height=.35\textwidth]{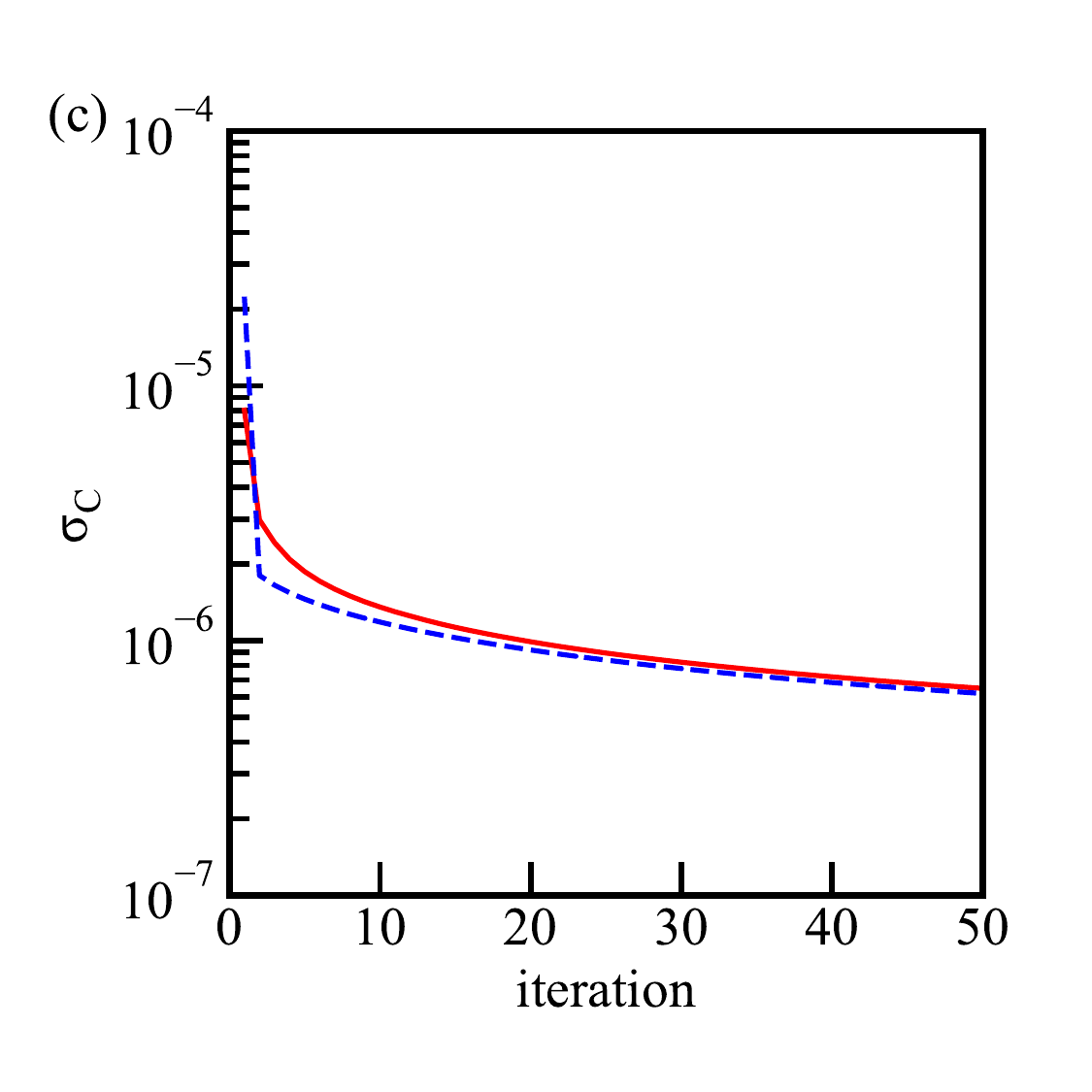}\hspace{-0.2in}
 \caption{The evolution curves of EnKF in the case of forced HIT: (a) the coefficient $C_D$ in the mixed SGS model, (b) the magnitude of the total prediction error in the kinetic energy spectrum, (c) standard deviation of $C_D$.}\label{fig_hit_evocurve}
\end{figure}

The energy spectrum of fDNS at equally spaced time instants (at every $0.3 \tau$) are taken as the state variables for the EnKF process. To get converged results, the EnKF is iterated after each run of the LES. The ensemble of LESs at each iteration are re-initialized using the same fDNS field with the updated coefficient $C_D$. Different sizes of the ensemble $N=10$ and $20$ are tested. Fig.~\ref{fig_hit_evocurve}a shows the evolution curves for the coefficient $C_D$, and Fig.~\ref{fig_hit_evocurve}b displays the corresponding prediction errors for the energy spectrum, defined by

\begin{equation}
 Er=\sum_{t=t_0}^{t_f}\sum_{k=1}^{k_{max}} |E(k)-E^{true}(k)|,
  \label{Er_abs}
\end{equation}
where $E^{true}(k)$ represents the true energy spectrum. Meanwhile, Fig. \ref{fig_hit_evocurve}c displays the evolution of the standard deviation $\sigma_C$ of the model coefficient $C_D$, which gradually decreases and saturates with time, suggesting the shrinking of the ensemble and the convergence of the EnKF. As can be seen, the converged results for $N=10$ and $20$ are very similar with $C_D\approx 0.00738$, which will be used in the EnKF-MM model in the LES. Compared to the value of $C_D$ based on the classical Smagorinsky coefficient 0.16 (i.e. $C_D = 2C^{2}_{Smag}\approx 0.05$),\cite{Smagorinsky1963,Lilly1967,Pope2000} the value of $C_D$ in the EnKF-MM model is much less, suggesting the dominance of the structural part. Nevertheless, we have also tested the case for which the functional model is used alone with a tuned coefficient, and the details can be found in Appendix C.
 
\begin{figure}\centering
\includegraphics[width=.45\textwidth]{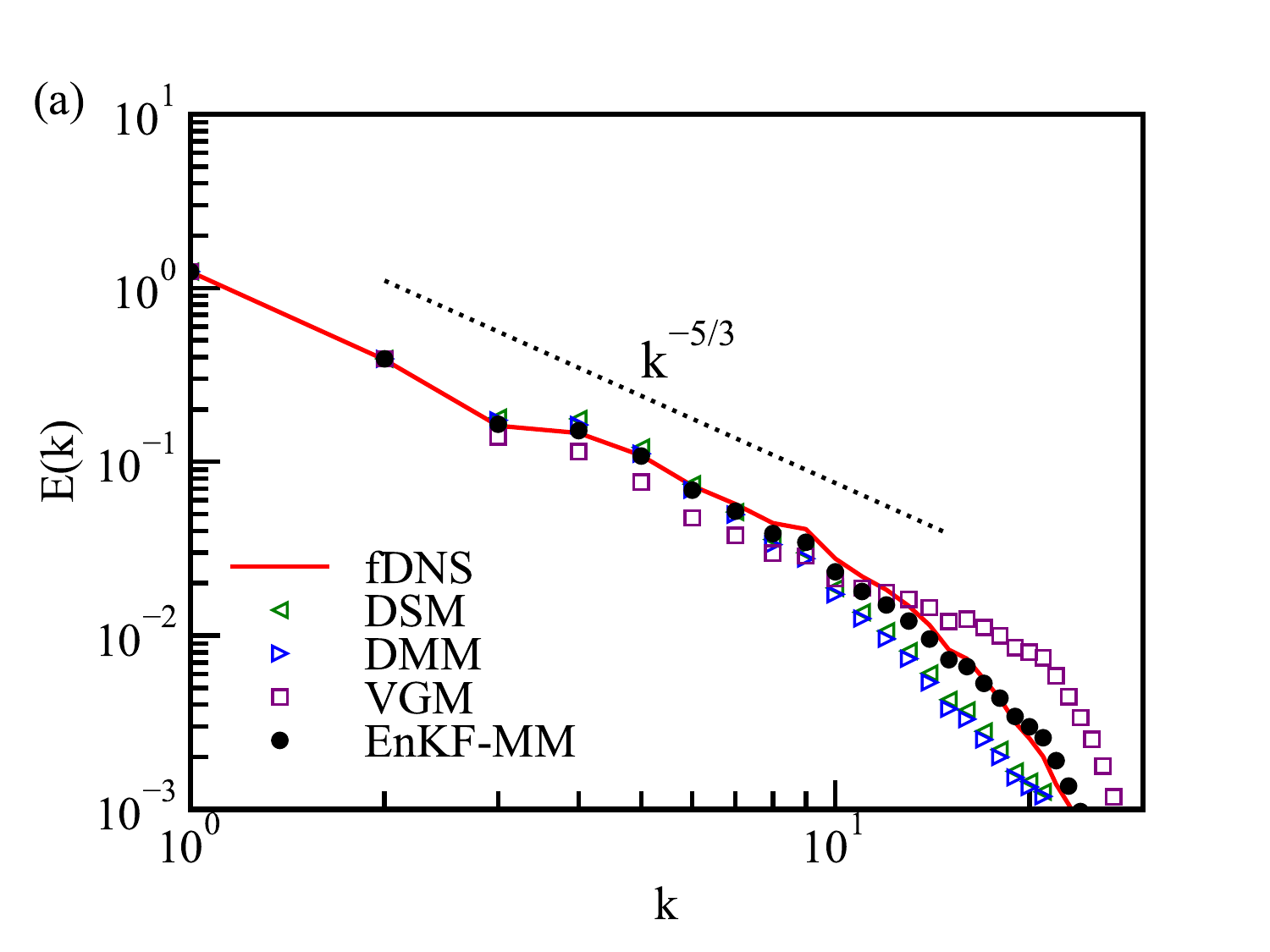}
\includegraphics[width=.45\textwidth]{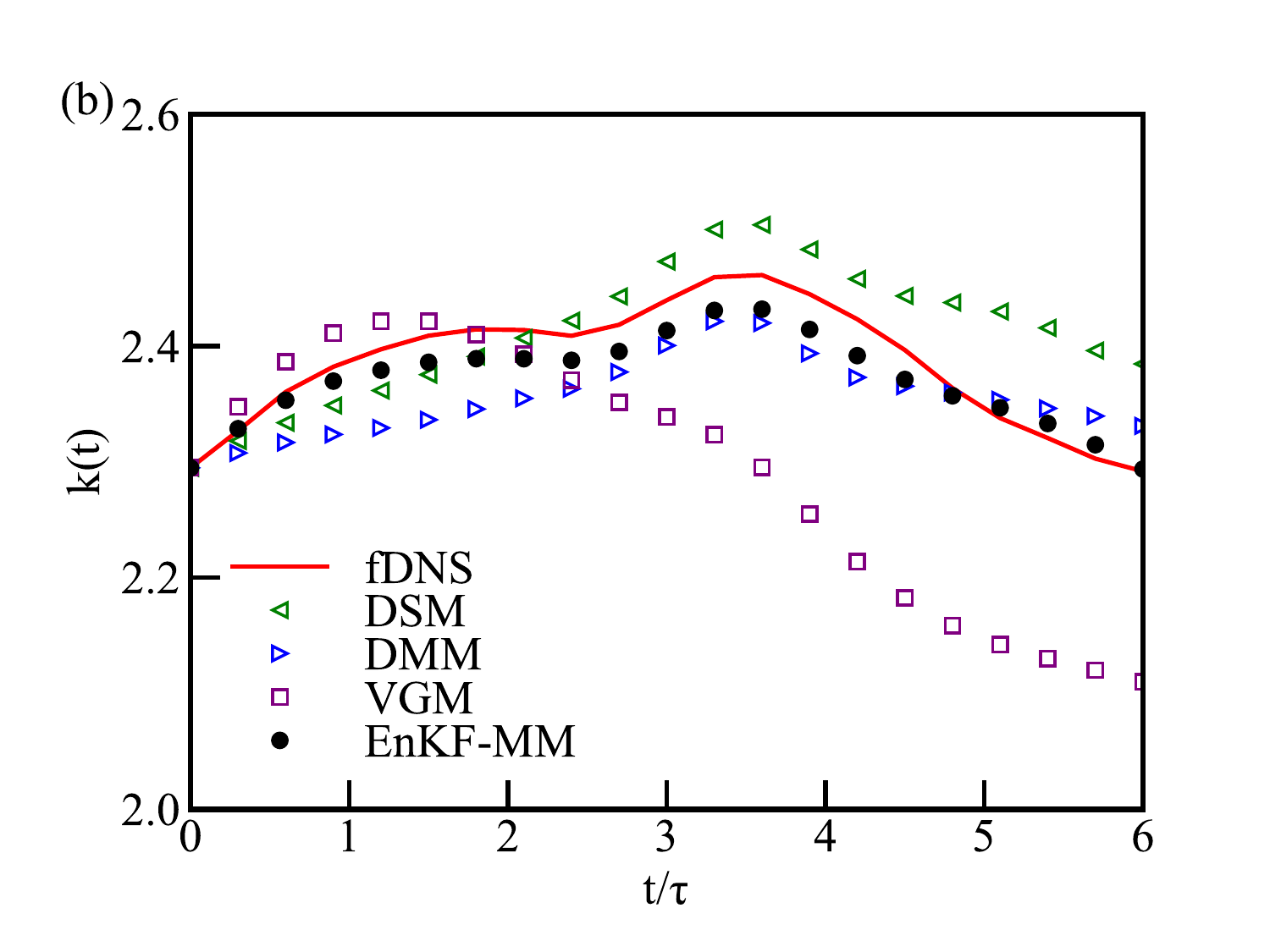}
 \caption{The kinetic energy spectrum and kinetic energy evolution in the LES of forced HIT using different SGS models: (a) kinetic energy spectrum, (b) evolution of kinetic energy.}\label{fig_hit_Ek_kt}
\end{figure}

Once the coefficient has been obtained through the EnKF process, we are in a position to examine the performance of the resulted EnKF-MM model in the LES. Since the LES equations are the filtered Navier-Stokes equations, the fDNS results are commonly adopted as benchmarks for the LES. The turbulent kinetic energy spectra of LES are shown in Fig.~\ref{fig_hit_Ek_kt}a. All the results retain an inertial range satisfying approximately the $k^{-5/3}$ scaling law.\cite{Pope2000}  As Fig.~\ref{fig_hit_Ek_kt}a shows, the DSM and DMM models slightly overestimate the kinetic energy compared to the fDNS result at large scales while underestimate it at small scales due to the strong SGS dissipation. Opposite to the predictions by the DSM and DMM models, the VGM model strongly underestimates the kinetic energy at large scales but overestimates it at small scales due to the insufficient SGS dissipation. In comparison, the EnKF-MM model has an overall closer agreement with the fDNS result. The influence of the SGS models on the evolution of the kinetic energy is shown in Fig. \ref{fig_hit_Ek_kt}b. Again, the EnKF-MM model gives the most satisfying prediction, which is consistent with its best prediction of the energy spectrum.

\begin{figure}\centering
\includegraphics[width=.45\textwidth]{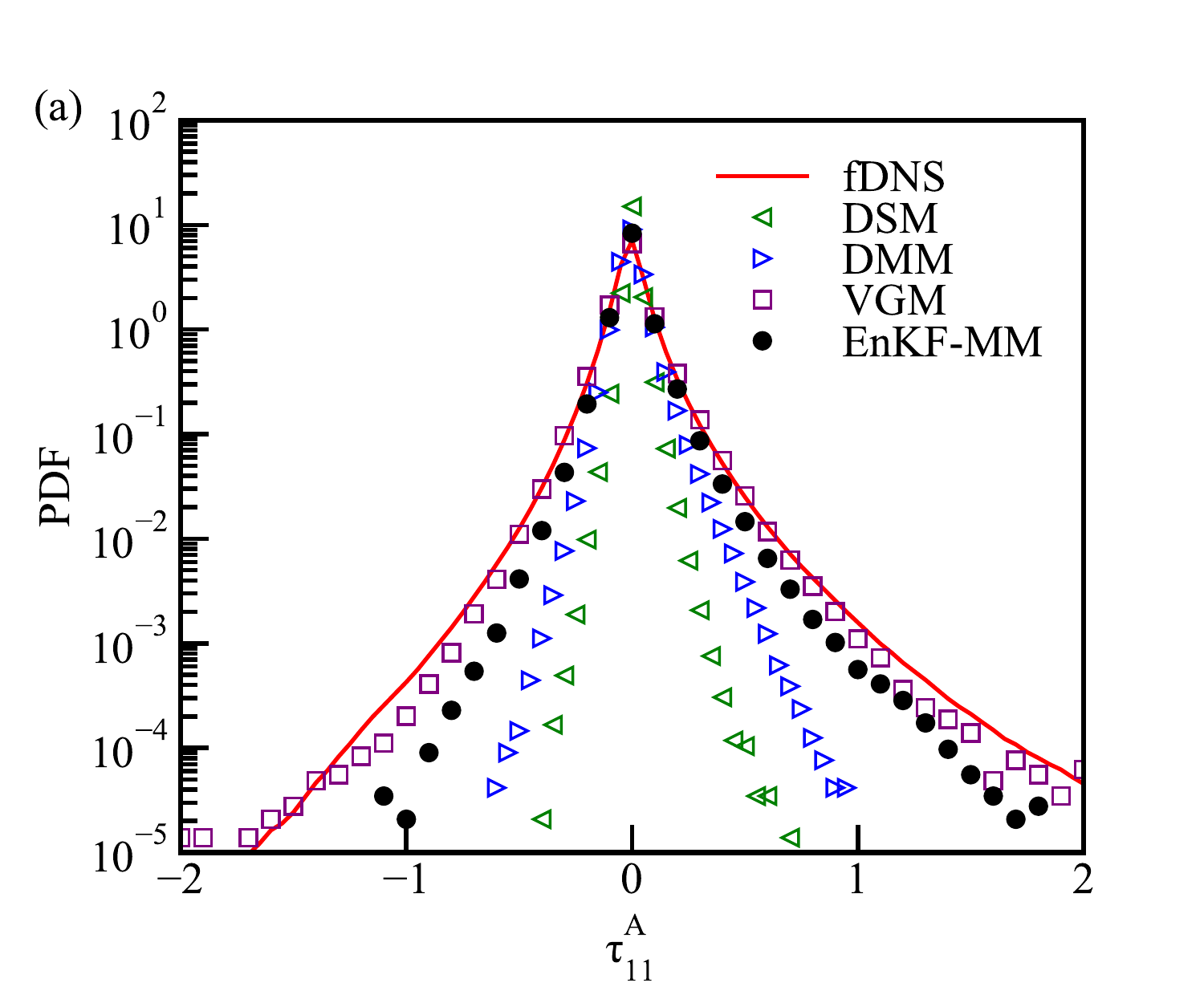}
\includegraphics[width=.45\textwidth]{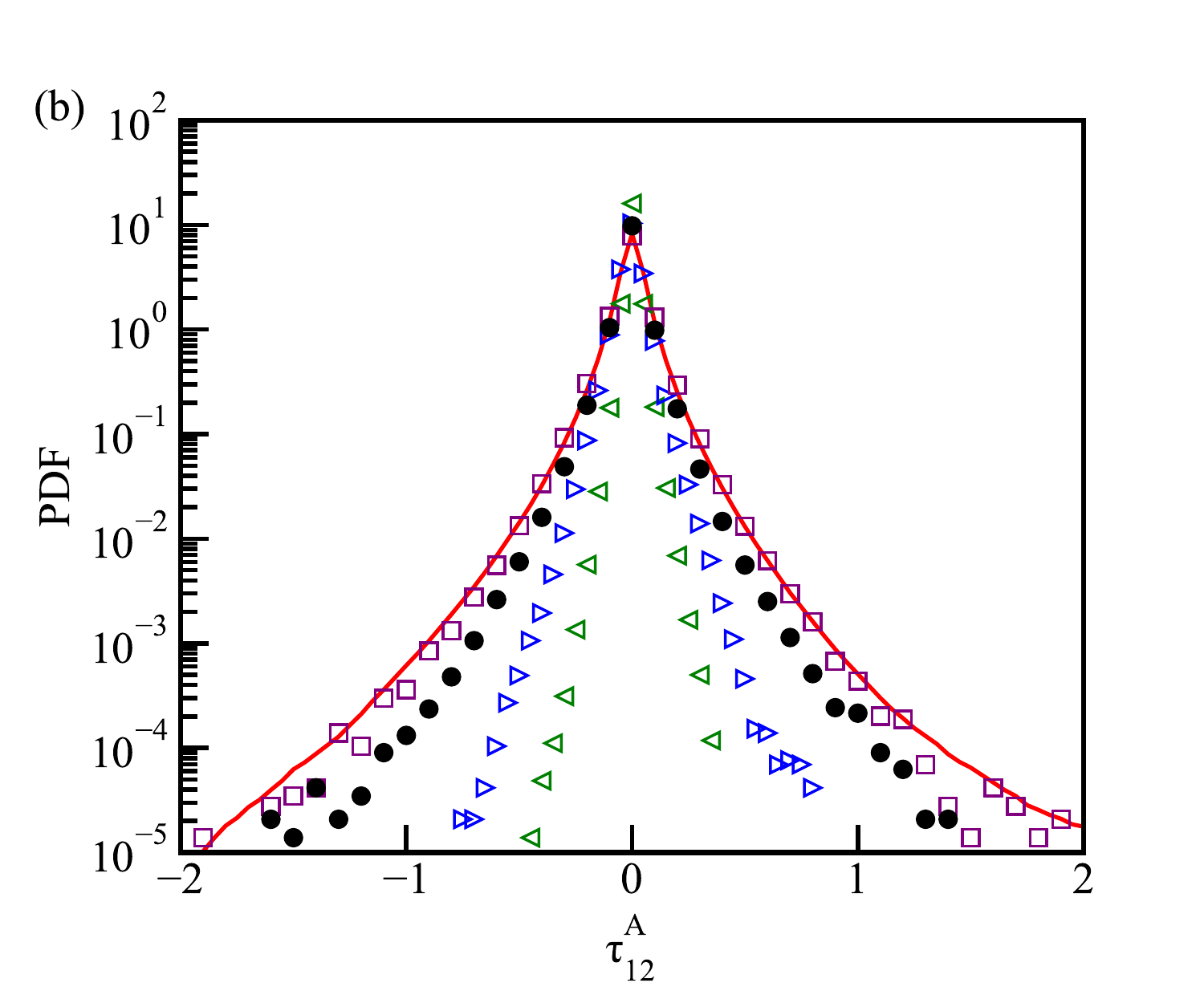}
 \caption{PDFs of normal and shear components of the SGS stresses in the LES of forced HIT: (a) normal component, (b) shear component.}\label{fig_hit_tau}
\end{figure}

Next, in Fig.~\ref{fig_hit_tau}, we evaluate the predictions of the anisotropic SGS stress $\tau^{A}_{ij}$ ($=\tau_{ij}-\frac{\delta_{ij}}{3}\tau_{kk}$) by different SGS models through the probability density function (PDF). The PDFs of normal and shear stresses are shown in Figs.~\ref{fig_hit_tau}a and \ref{fig_hit_tau}b, respectively. As observed, the PDFs of $\tau^{A}_{ij}$ are nearly symmetric for isotropic turbulence. To some extent, adding the functional part to the VGM model indeed deteriorates its accuracy on prediction of the SGS stress, since the reconstructed PDF by the EnKF-MM model is slightly worse compared to that by the VGM model. However, as we recall from Fig. \ref{fig_hit_Ek_kt}, the prediction of the kinetic energy by the VGM model is the worst among all the tested models. Hence, the accuracy of the predicted SGS stress alone cannot fully quantify the practical performance of a SGS model. Further, as Fig. \ref{fig_hit_tau} shows, the predicted SGS stress by the EnKF-MM model is still better than the DSM and DMM models, which both predict narrower PDFs. Meanwhile, we also observe that the DMM model performs better than the DSM model.

Here, it is important to note that accurately reconstructing the SGS stress may not guarantee the adequate prediction of other important flow statistics. Clearly, the VGM model gives the best prediction for the SGS stress. However, as will be seen soon, many important flow statistics are poorly predicted by the VGM model. In comparison, the EnKF-MM model aims at accurately recovering the energy distribution at different scales (i.e. the energy spectrum), and its predictions turn out to be more satisfying compared to other traditional SGS models.

\begin{figure}\centering
\includegraphics[width=.45\textwidth]{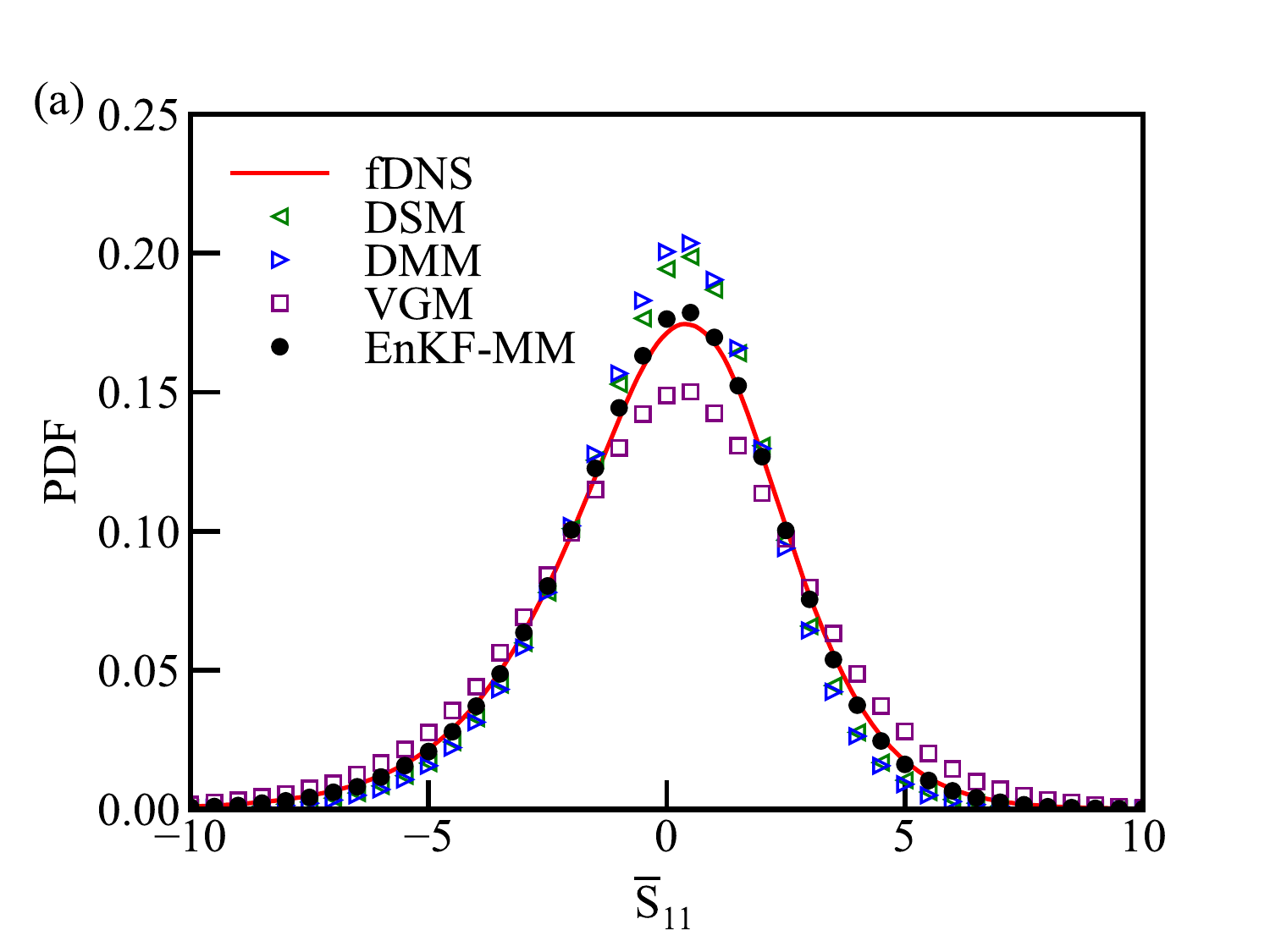}
\includegraphics[width=.45\textwidth]{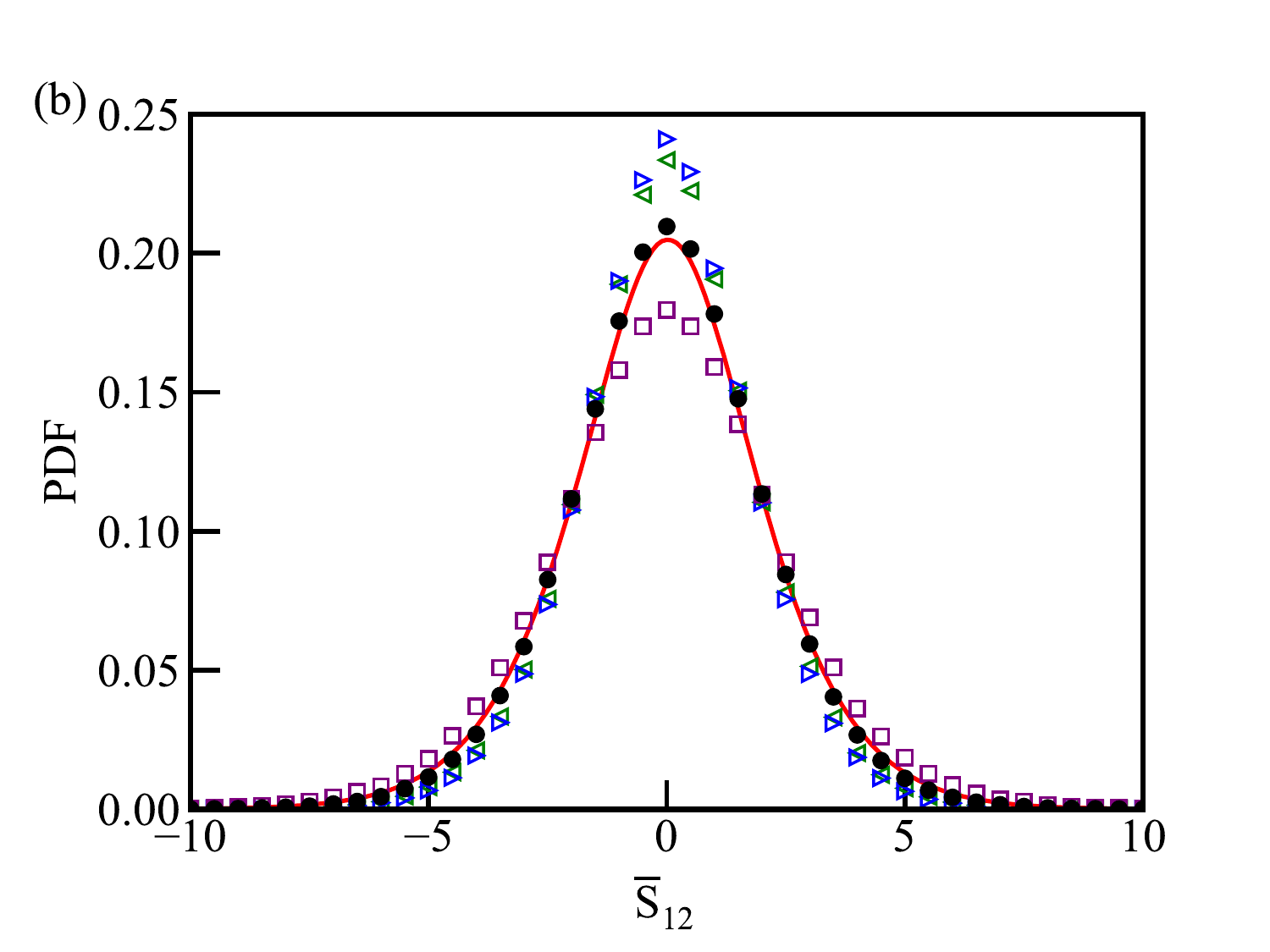}
 \caption{The PDFs of the rate of strain tensor in the LES of forced forced HIT using different SGS models: (a) normal component, (b) shear component.}\label{fig_hit_strain}
\end{figure}

\begin{figure}\centering
\includegraphics[width=.5\textwidth]{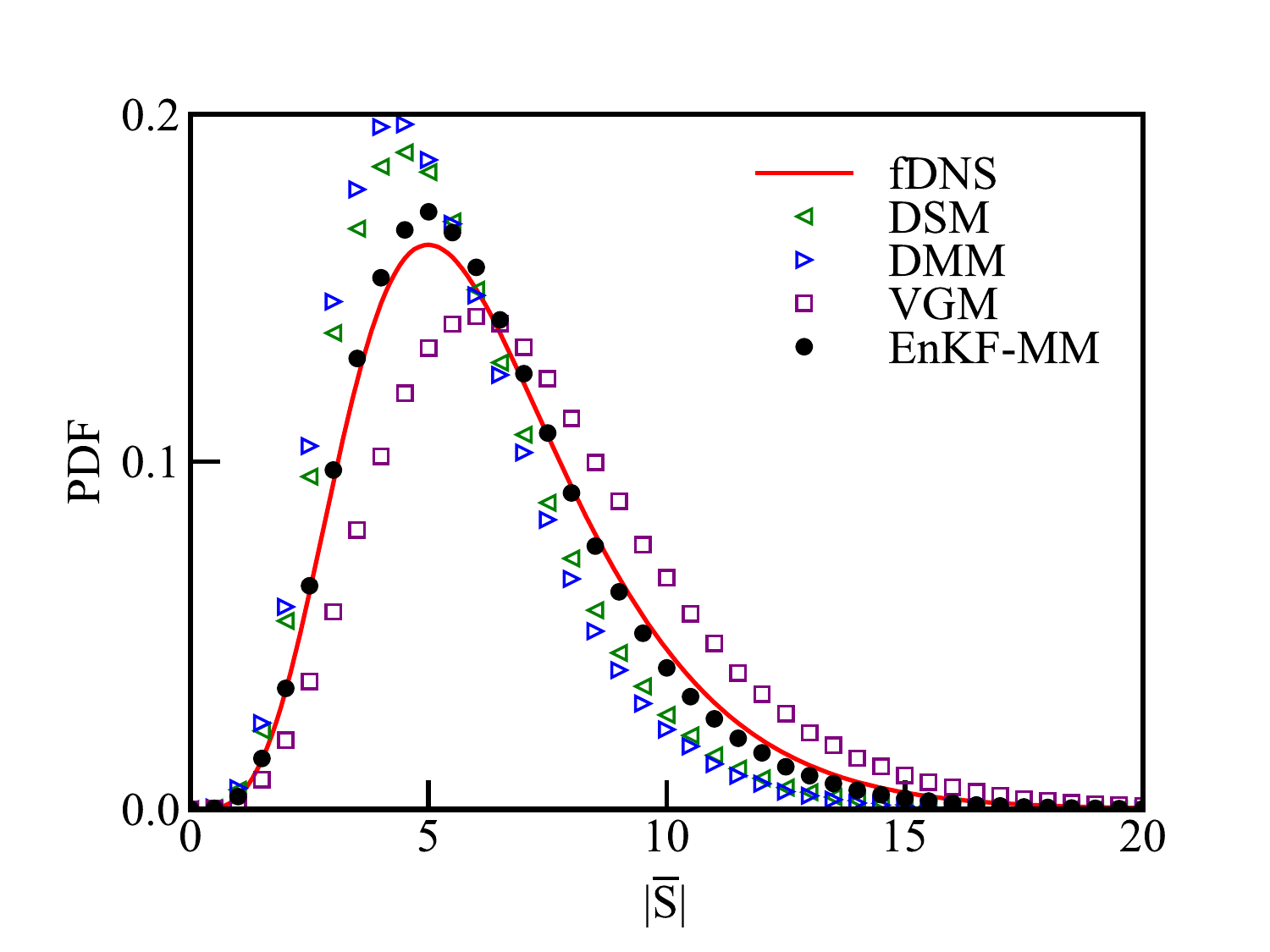}
 \caption{PDFs of the characteristic strain rate in the LES of forced HIT using different SGS models.}\label{fig_hit_sb}
\end{figure}

\begin{figure}\centering
\includegraphics[width=.45\textwidth]{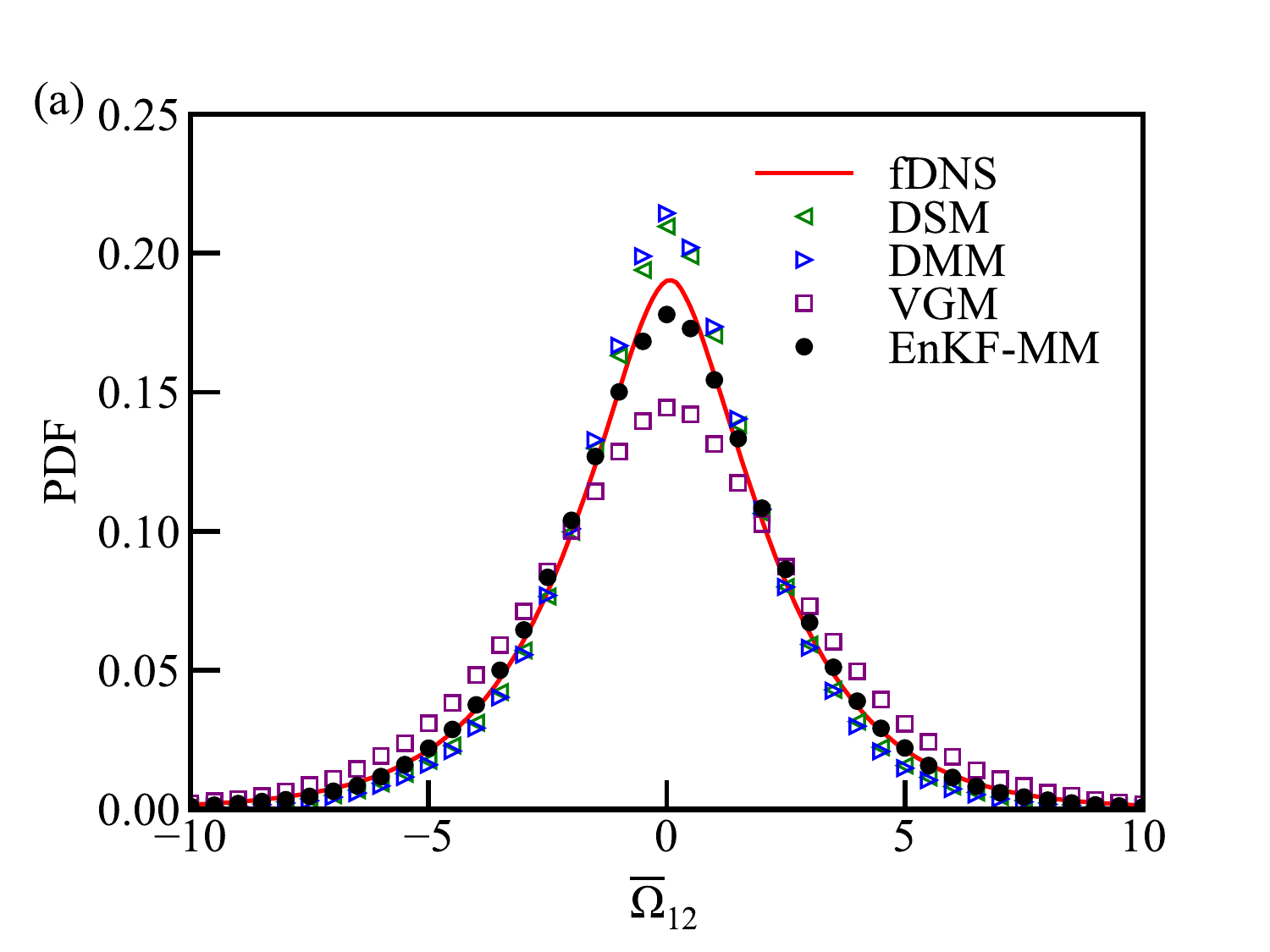}
\includegraphics[width=.45\textwidth]{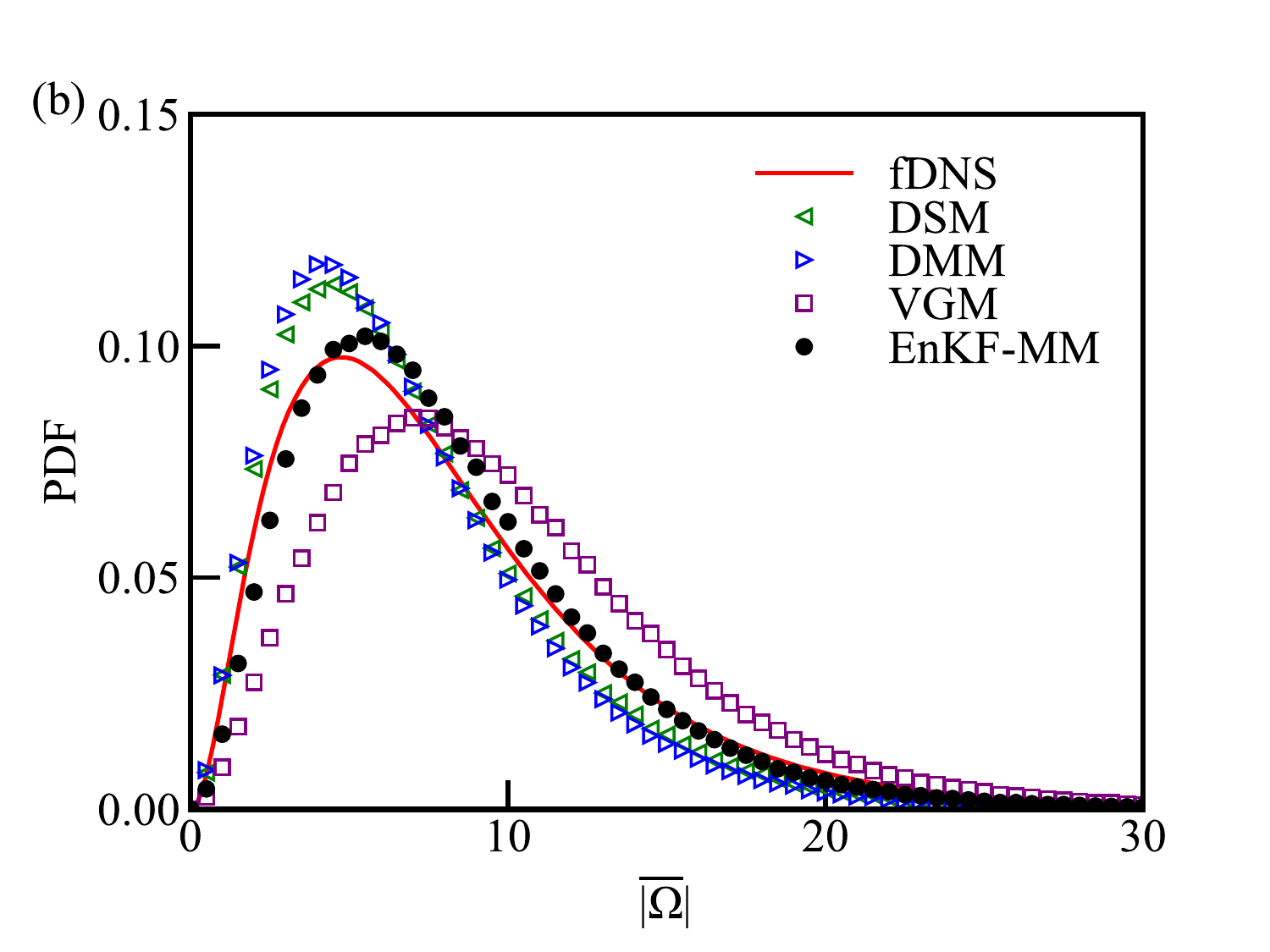}
 \caption{The PDFs of $\overline{\Omega}_{12}$ and $|\overline{\Omega}|$ in the LES of forced HIT using different SGS models: (a) PDFs of $\overline{\Omega}_{12}$, (b) PDFs of $|\overline{\Omega}|$.}\label{fig_hit_rot}
\end{figure}

The filtered strain rate $\overline{S}_{ij}$ is defined as $\overline{S}_{ij}=\frac{1}{2}(\partial \overline{u}_{i}/\partial x_{j}+\partial \overline{u}_{j}/\partial x_{i})$, which directly governs the local straining in LES. The PDFs of the strain rate for the normal and shear components of $\overline{S}_{ij}$ are displayed in Figs.~\ref{fig_hit_strain}a and \ref{fig_hit_strain}b, respectively. The PDFs of DSM and DMM models are slightly narrower than the fDNS result, while the peak values are overestimated. Meanwhile, the VGM model predicts a wider PDF compared to the fDNS result but the peak of PDF is underestimated. In comparison, the prediction by the EnKF-MM model agrees the best with the fDNS result. Further, the PDF of the characteristic filtered strain rate, defined by $|\overline{S}|=\sqrt{2\overline{S}_{ij}\overline{S}_{ij}}$, is shown in Fig.~\ref{fig_hit_sb}. As observed, the DSM and DMM models recover reasonably well the location of the peak for the PDF but overestimate the value of the peak, while the DSM model slightly outperforms the DMM model. For the VGM model, not only the location of the peak is poorly predicted, the value of the peak is also under-predicted.  In contrast, the EnKF-MM model gives an overall better prediction compared to the other models. In addition, the filtered rotation rate is defined as $\overline{\Omega}_{ij}=\frac{1}{2}(\partial{\overline{u}_{i}}/\partial{x_{j}}-\partial{\overline{u}_{j}}/\partial{x_{i}})$. Figs. \ref{fig_hit_rot}a and \ref{fig_hit_rot}b display the PDFs of filtered rotation rate $\overline{\Omega}_{12}$ and the characteristic filtered rotation rate ($|\overline{\Omega}|=\sqrt{2\overline{\Omega}_{ij}\overline{\Omega}_{ij}}$), respectively. Again, the predictions by the EnKF-MM model is the closest to the fDNS results compared to other SGS models.

\begin{figure}\centering
\includegraphics[width=.5\textwidth]{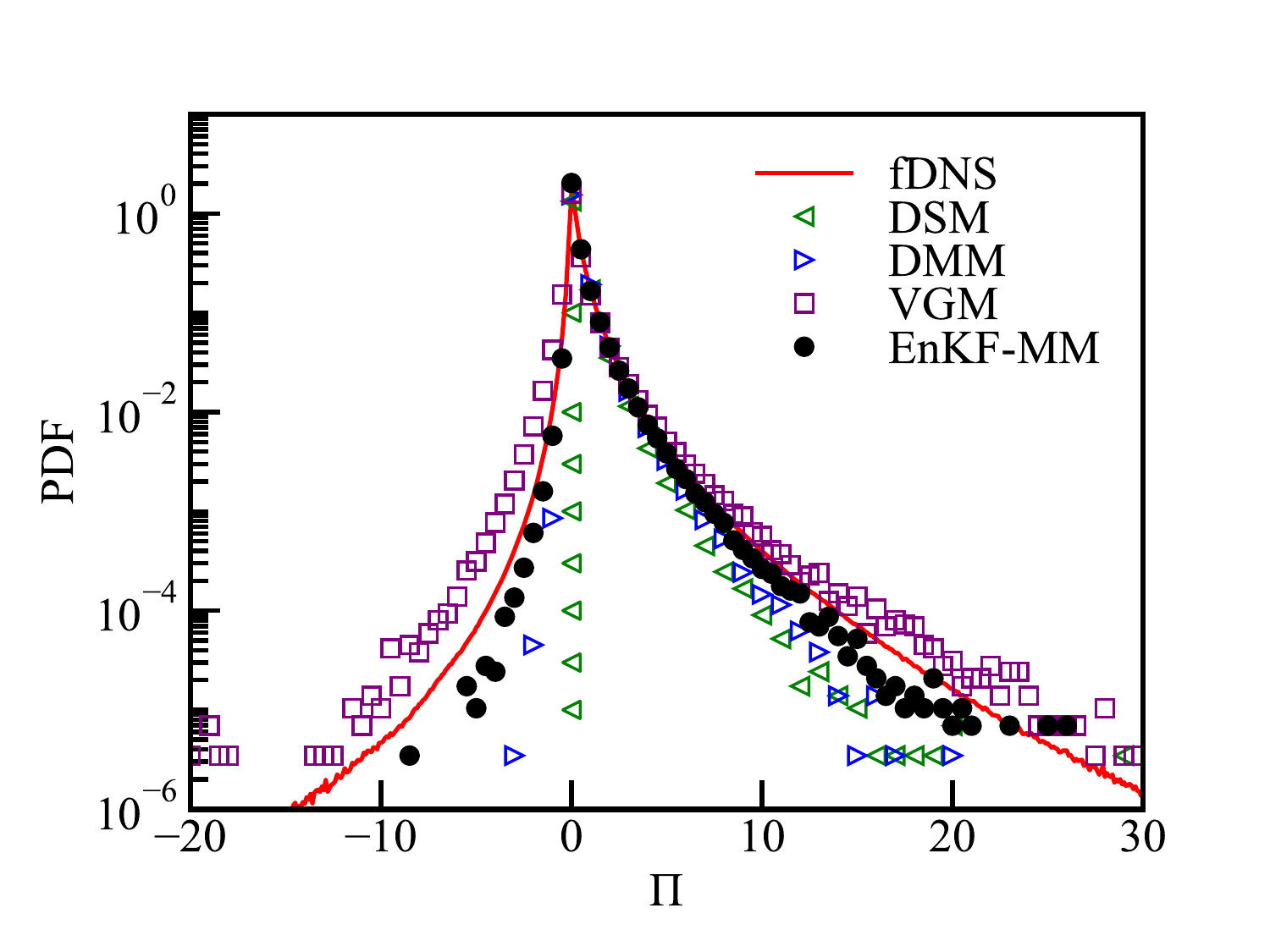}
 \caption{PDF of the SGS energy flux in the LES of forced HIT using different SGS models.}\label{fig_hit_pi}
\end{figure}

To quantify the kinetic energy transfer between the resolved and the subgrid scales, we calculate the SGS energy flux as follows\cite{Yuan2020}
\begin{equation}
  \Pi=-\tau_{ij} \overline S_{ij}.
  \label{SGSflux}
\end{equation}
A positive SGS energy flux indicates the energy transfer from the resolved field to the subgrid field while a negative value indicates the energy transfer in the opposite direction (backscatter).\cite{Wangj2018} The PDFs of the SGS energy flux are shown in Fig.~\ref{fig_hit_pi}. Notably, the SGS energy flux predicted by the DSM model is strictly positive due to the eddy-viscosity formulation of the model. The DMM model performs tangibly better than the DSM model, and some backscatter of kinetic energy is also predicted due to the similarity part of the model. Meanwhile, the VGM model predicts a wider PDF than the fDNS result, and the overestimation of the backscatter may cause numerical-instability problem.\cite{Vreman1995,Vreman1997,Nadiga2007} Recalling that the prediction of the SGS stress by the VGM model is quite satisfying (cf. Fig. \ref{fig_hit_tau}), its relatively poor prediction of the SGS flux is mostly caused by the bad prediction of the filtered strain rate as shown in Fig. \ref{fig_hit_strain}. In comparison, the EnKF-MM model predicts more accurately the SGS energy flux since it has adequate predictions for both the SGS stress and the strain rate.

\begin{figure}\centering
\includegraphics[width=.45\textwidth]{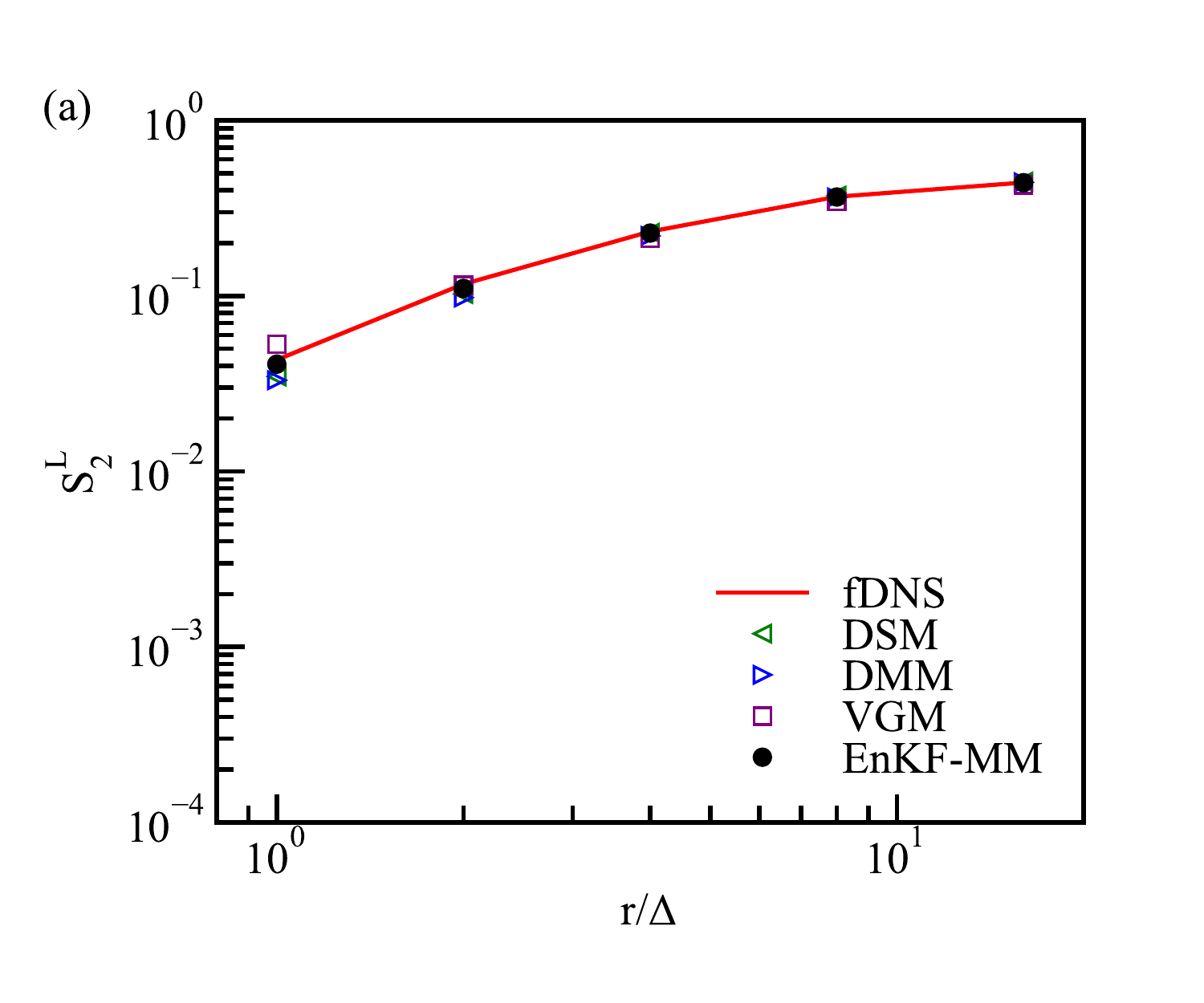}\hspace{-0.2in}
\includegraphics[width=.45\textwidth]{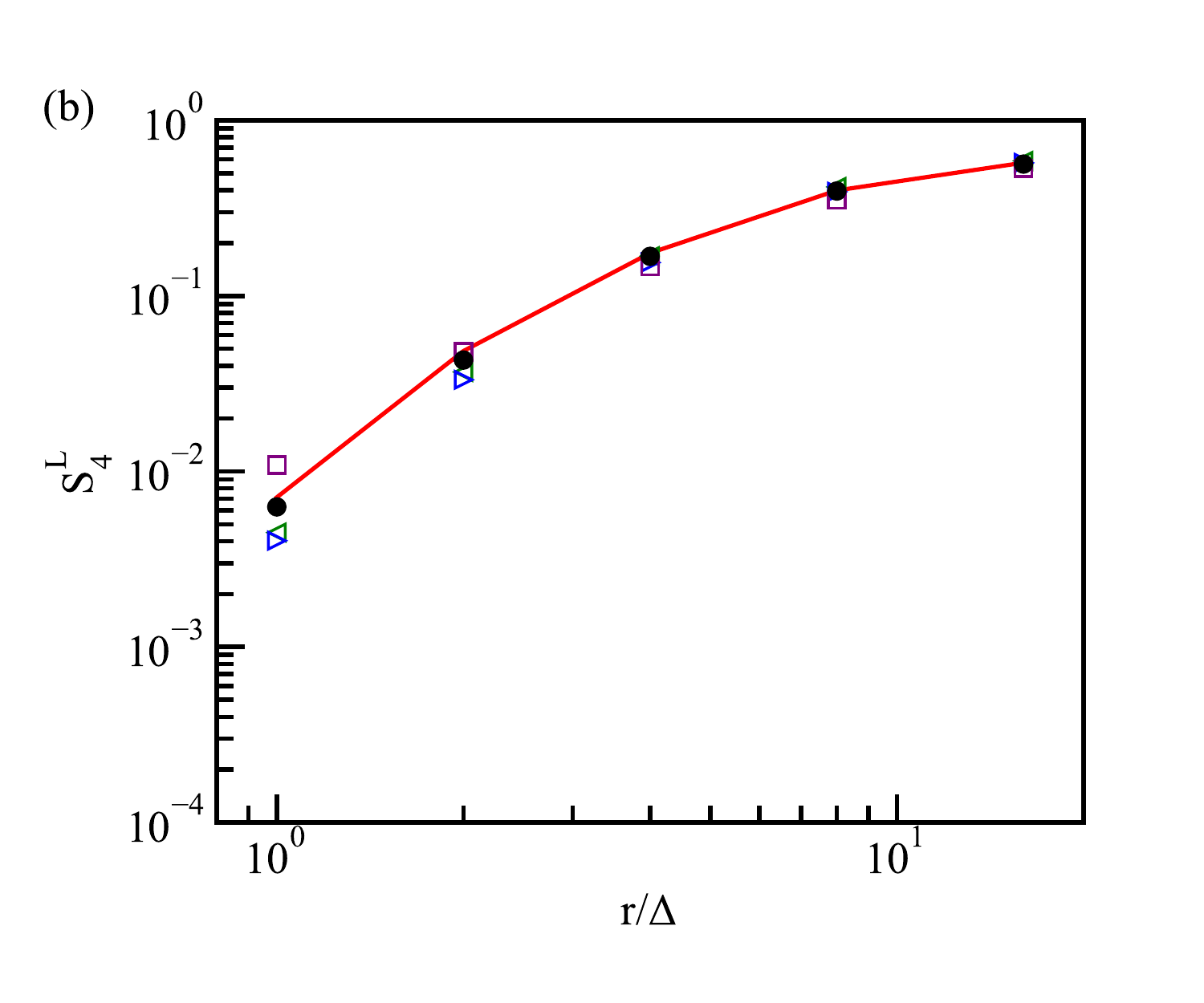}\hspace{-0.2in}
\includegraphics[width=.45\textwidth]{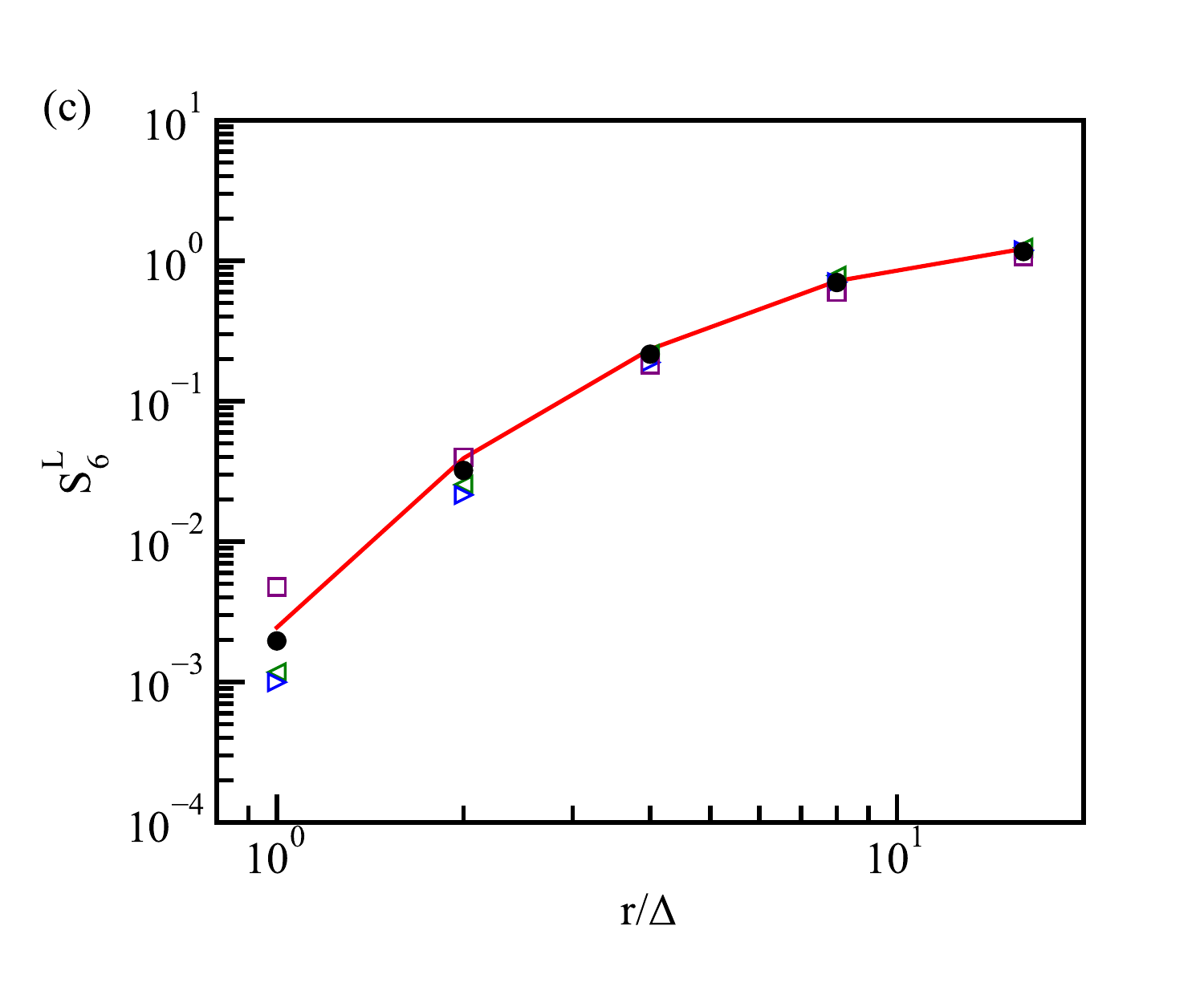}\hspace{-0.2in}
\includegraphics[width=.45\textwidth]{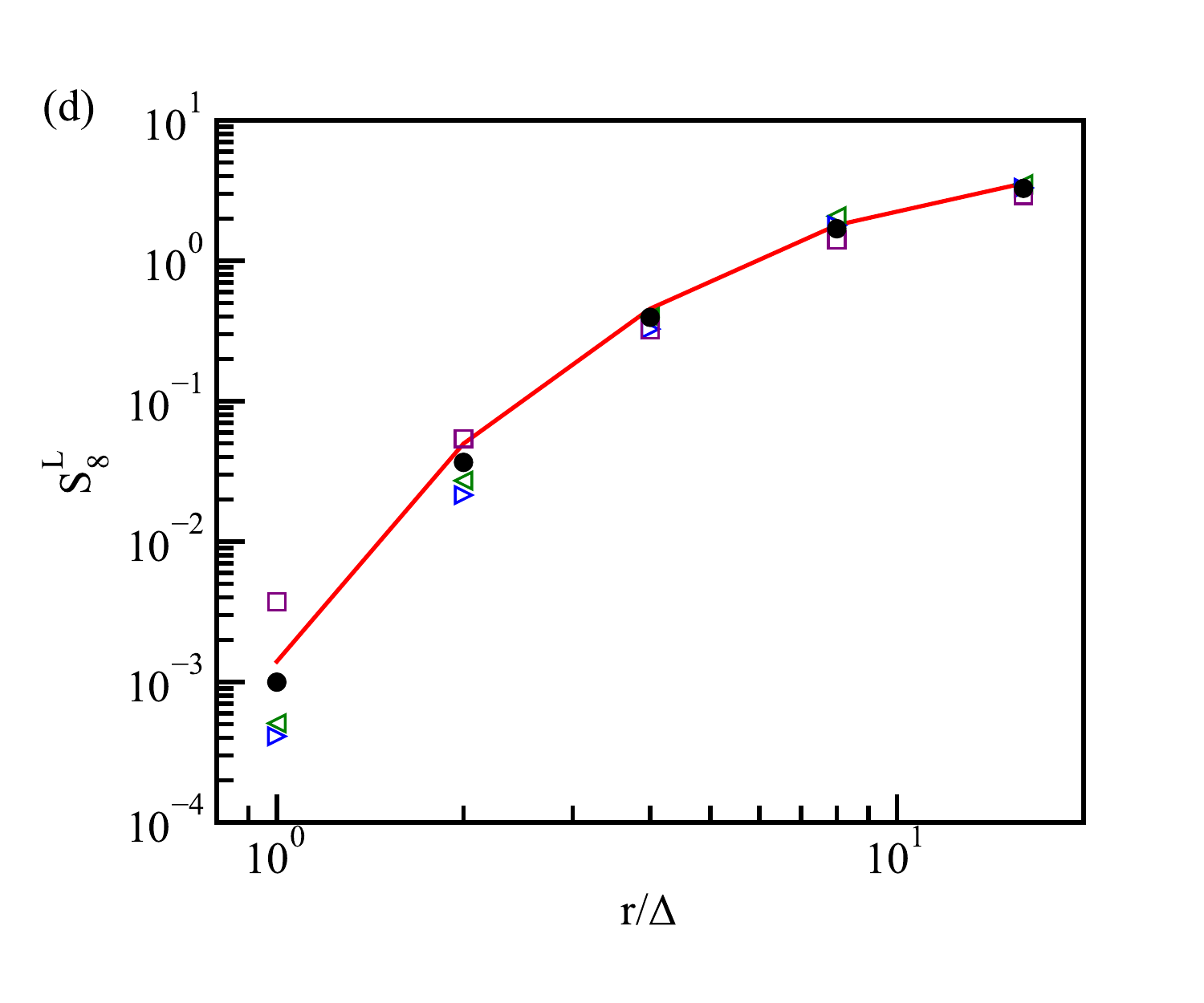}
 \caption{The structure functions in the LES of forced HIT using different SGS models: (a) $S^L_2$, (b) $S^L_4$, (c) $S^L_6$, (d) $S^L_8$.}\label{fig_hit_stru}
\end{figure}

To examine the spatial structures of the flow field in LES, we calculate the longitudinal structure functions of velocity, defined by
\begin{eqnarray}
&& S_{n}^{L}(r)\equiv\langle|\frac{\delta_{r}\overline{u}}{\overline{u}^{rms}}|^n\rangle,
\label{snr}
\end{eqnarray}
where $\delta_{r}\overline{u}=[\mathbf{\overline{u}}(\mathbf{x}+\mathbf{r})-\mathbf{\overline{u}}(\mathbf{x})]\cdot\hat{\mathbf{r}}$ is the longitudinal increment of the velocity at a separation distance $\mathbf{r}$ with $\hat{\mathbf{r}}=\mathbf{r}/|\mathbf{r}|$. Here the velocity increment is normalized by the rms velocity $\overline{u}^{rms}$. The direction of $\mathbf{r}$ can be arbitrarily chosen for isotropic turbulence and is taken to be along the x direction in the current case. The structure functions are displayed in Fig.~\ref{fig_hit_stru}. We observe that the structure functions at large separations can be accurately predicted by all the models. However, as ${r}$ decreases, VGM does a poor job, predicting a value higher than the fDNS result, while the structure functions predicted by the DSM and DMM models are lower than the fDNS result. This trend becomes increasingly more pronounced as the order of structure function $n$ increases. In contrast, the EnKF-MM model can predict reasonably well the structure functions of all orders at the whole range of ${r}$. Finally, it is necessary to point out that the prediction by the EnKF-MM model also becomes worse as the order of the structure function increases even though it is still better compared to other models. This can be a result of targeting the energy spectrum which is a second-order statistical quantity, and the higher-order statistics may not be perfectly predicted using the current method.

\section{The numerical results of incompressible turbulent mixing layer}

\begin{figure}\centering
\includegraphics[width=.7\textwidth]{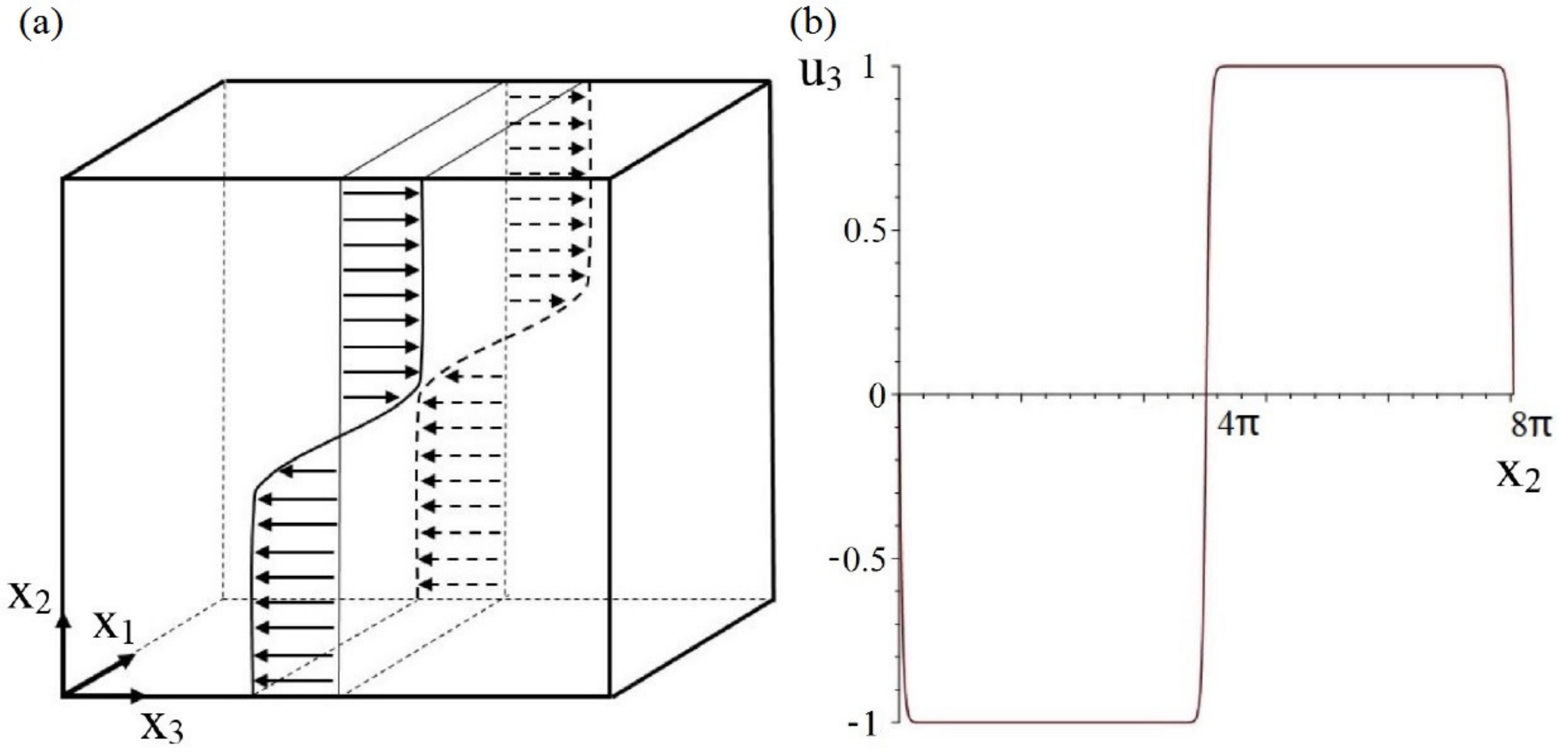}
 \caption{The configuration of TML: (a) schematic configuration; (b) the variation of streamwise velocity in the normal direction.}\label{fig_sl_config}
\end{figure}

In the current section, we test the performance of the EnKF-MM framework at a different flow configuration, namely, the incompressible turbulent mixing layer (TML). Here, the TML is simulated by configuring two streams of equal and opposite velocities as displayed in Fig.~\ref{fig_sl_config}a. To reduce the computational cost, the temporally evolving TML is considered, instead of the spatially evolving TML which requires a much larger computational domain.\cite{Vaghefi2014}

The reference velocity $U_f$ is taken as $U_f=\Delta U/2$ where $\Delta U$ is velocity difference between the upper and lower streams. The turbulent mixing layer develops at the interface between the two streams due to shear layer instability.\cite{Vreman1997} The simulation is performed in a cuboid with the side lengths $L_1$, $L_2$ and $L_3$ equal to $4\pi$, $8\pi$ and $8\pi$ in the $x_1$ (spanwise), $x_2$ (normal) and $x_3$ (streamwise) directions, respectively. The initially imposed velocity profile varies only in the normal direction as follows:

\begin{equation}
  u_3(x_2)=[\text{tanh}(\frac{x_2-L_2/2}{2 C_{\theta} \theta_0})-\text{tanh}(\frac{x_2-L_2}{2 C_{\theta} \theta_0})-\text{tanh}(\frac{x_2}{2 C_{\theta} \theta_0})],~~\text{for}~0<x_2\leq L_2,
  \label{sl_tanh}
\end{equation}
where $\theta_0$ is the specified initial momentum thickness and $C_{\theta}$ is a coefficient to be determined such that the initial momentum thickness equals to $\theta_0$. Based on the initial momentum thickness, the reference time scale is defined by $\tau_{\theta}=\theta_0/\Delta U$. For incompressible TML, the momentum thickness $\theta$ can be calculated as\cite{Vreman1997}

\begin{equation}
\theta=\frac{1}{4}\int_{-L_2/4}^{L_2/4}[\frac{1}{4}-(\frac{\langle\ u_3\rangle}{\langle\Delta U\rangle})^2]dx_2,
  \label{eq_theta}
\end{equation}
where $\langle \cdot \rangle$ represents a spatial average over all the homogeneous directions (i.e. the $x_1$ and $x_3$ directions). In current work, $\theta_0=0.08$. The corresponding velocity is shown in Fig.~\ref{fig_sl_config}b.

With the current velocity profile, periodical boundary conditions can be conveniently adopted in all coordinate directions. To avoid the emergence of a second mixing layer,\cite{Sharan2019} numerical diffusion zones are configured near the top and bottom boundaries.\cite{Reckinger2016,Wang2022a} The initial perturbations are generated using a digital filter method.\cite{Klein2003,Wang2022a,Wang2022,Yuan2023} The algorithm for the configuration of the digital filter is described in previous works in detail\cite{Vaghefi2014,Yuan2023} and not reproduced here. The DNS of incompressible TML is performed using $256$, $512$ and $512$ uniform grid points in the spanwise, normal and streamwise directions, respectively. The detailed parameters for the DNS is presented in Table~\ref{tab_sl_DNS}. 

\begin{table*}
\begin{center}
\small
\begin{tabular*}{0.85\textwidth}{@{\extracolsep{\fill}}cccccc}
\hline
Reso.($N_{x_1}\times N_{x_2}\times N_{x_3}$) &$L_1\times L_2 \times L_3$ &$\nu$ &$\Delta U$ &$\Delta t$ &$\theta_0$ \\ \hline
$256\times512\times512$ &$4\pi \times 8\pi \times 8\pi$ &0.0005 &2 &0.002 &0.08 \\ \hline
\end{tabular*}
\normalsize
\caption{Numerical parameters for the DNS of incompressible TML.}
\label{tab_sl_DNS}
\end{center}
\end{table*}

The LESs are performed at filter width $\overline{\Delta}=8h_{DNS}$, and the filter to grid ratio (FGR) is taken at $FGR=\overline{\Delta}/h_{LES}=2$ to weaken the influence of numerical errors. In turn, $64$, $128$ and $128$ uniform grid points are adopted for the LES in the spanwise, normal and streamwise directions, respectively. The initial fields of LES are obtained by filtering the DNS field at $t/\tau_{\theta}=50$. The ratio of the time steps in the LES to that in the DNS is taken at $\Delta t_{LES}/\Delta t_{DNS}=4$. Again, the time step ratio results from the balance between computational efficiency of LES and numerical stability due to the CFL condition.\cite{Wang2010} The LESs are run until $t/\tau_{\theta}=800$.

\begin{figure}\centering
\includegraphics[height=.35\textwidth]{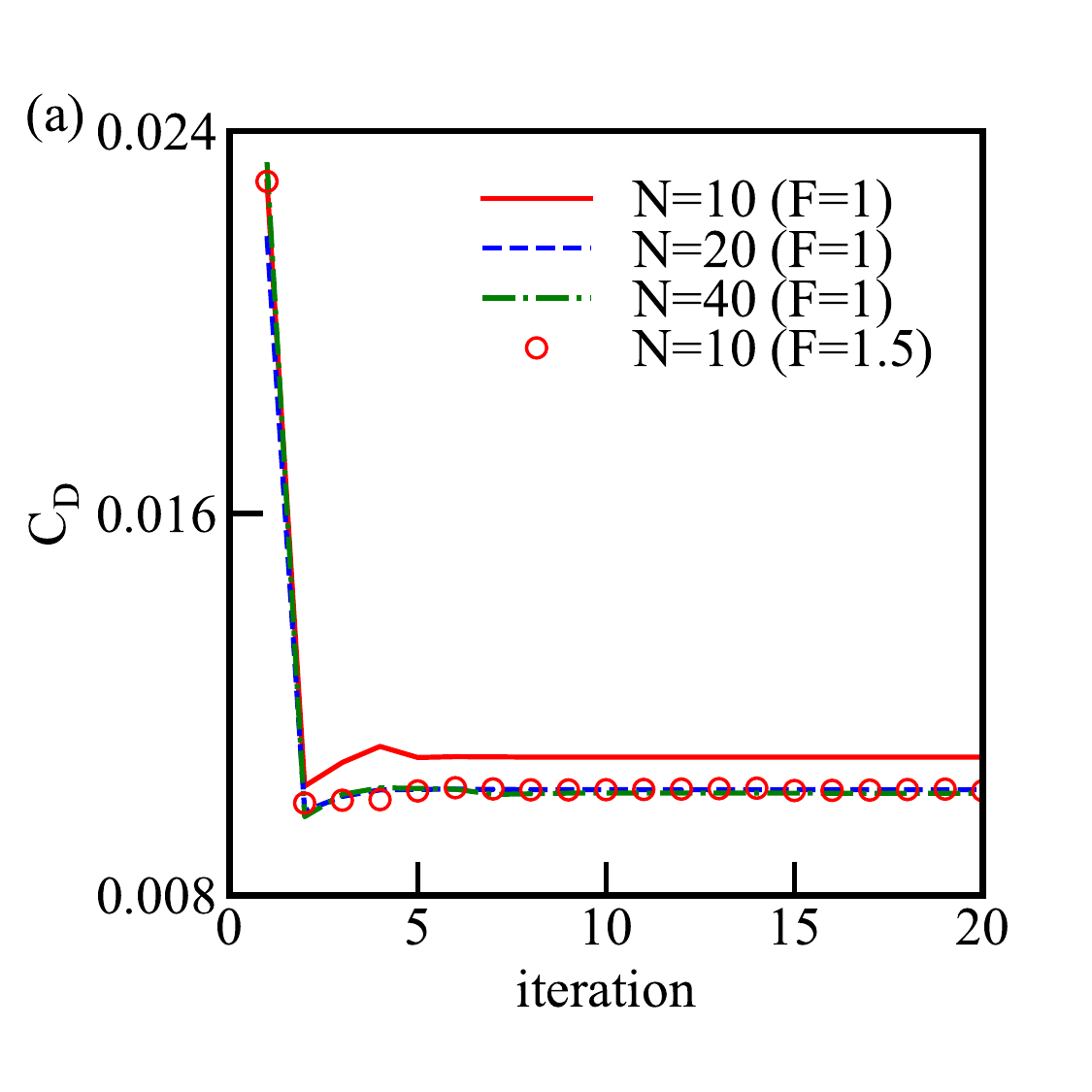}\hspace{-0.2in}
\includegraphics[height=.35\textwidth]{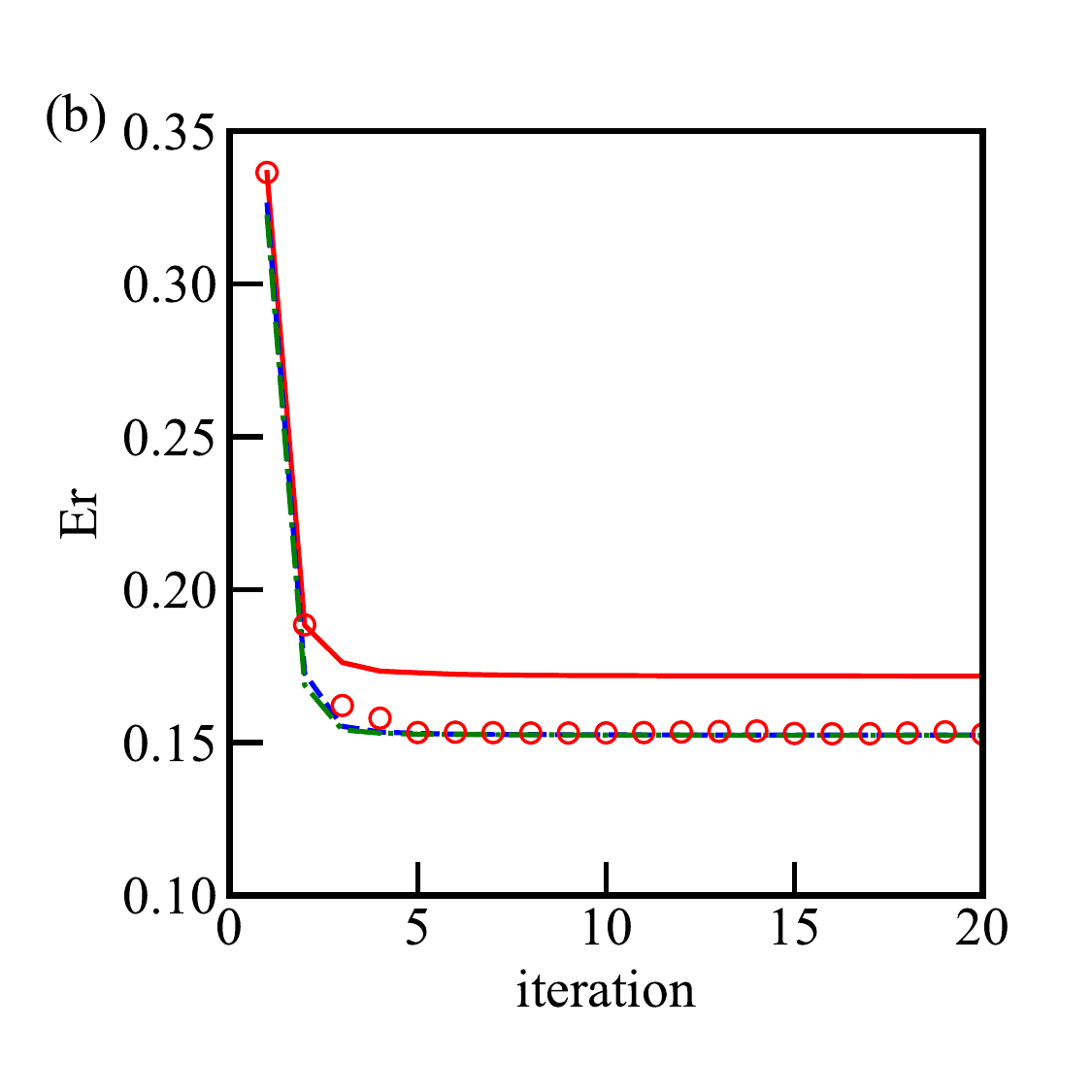}\hspace{-0.2in}
\includegraphics[height=.35\textwidth]{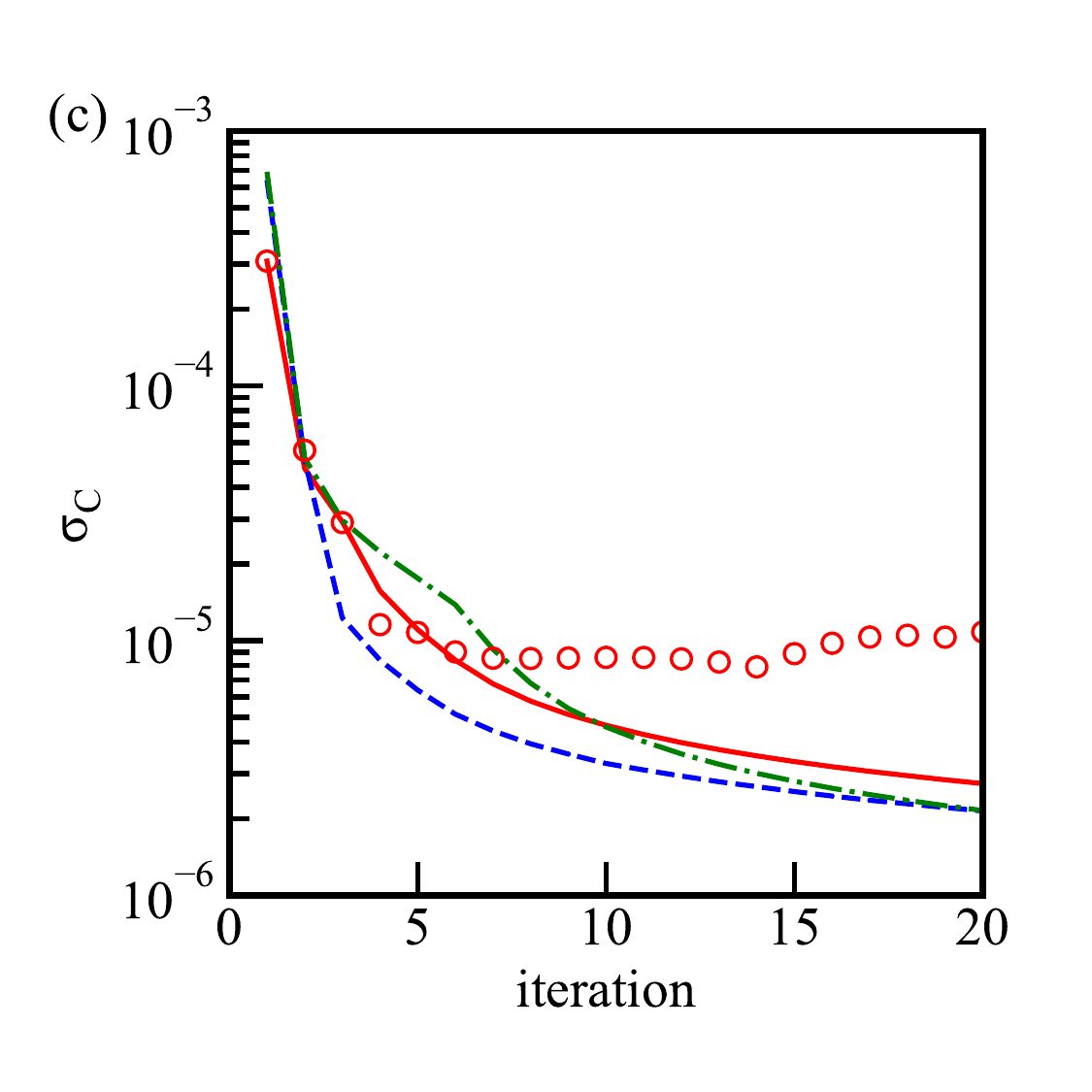}\hspace{-0.2in}
 \caption{The evolution curves of EnKF in the case of TML: (a) the coefficient $C_D$ in the mixed SGS model, (b) the magnitude of the total prediction error in the kinetic energy spectrum, (c) standard deviation of $C_D$.}\label{fig_sl_evocurve}
\end{figure}

The details on the setup of the parameters for the EnKF is provided in Table \ref{tab_tml_EnKF}. Again, the energy spectrum of fDNS at equally spaced time instants (at every $10\tau_{\theta}$) are taken as the state variables for the EnKF process. In the case of TML, to obtain converged results in EnKF, a shifting procedure must be additionally incorporated according to

\begin{equation}
C_{D,i}^{S}=C_{D,i}+(C_{D,min}-C_{D,mean}),
\label{shift}
\end{equation}
where $C_{D,i}^{S}$ is the shifted coefficient, $C_{D,mean}$ is the sample mean of the coefficients and $C_{D,min}$ represents the coefficient of the LES sample that has the minimum prediction error for the energy spectrum in the corresponding iteration. This procedure effectively drag the mean of ensemble of coefficients towards the optimum value of each iteration. Further, for better convergence, we can adopt an ensemble inflation scheme.\cite{Hamill2001} Here we should note that, as a statistical minimization approach, EnKF does not to the same degree suffer from the typical problem of falling into a local minima as in the traditional parameter-estimation methods.\cite{Evensen2009} Meanwhile, being the target of EnKF, the energy spectrum is also not so chaotic compared to the flow field variables. Nevertheless, the estimated model parameter may still diverge from the desired value for a small ensemble size. In such cases, the following ensemble inflation scheme can be adopted, namely\cite{Hamill2001}

\begin{equation}
 C_{D,i}^{I}=C_{D,mean}+F (C_{D,i}-C_{D,mean}),
\label{inflation}
\end{equation}
where $C_{D,i}^{I}$ is the inflated coefficient, $F$ is an inflation factor. Ensemble sizes $N=10$, $20$ and $40$ are tested. Fig. \ref{fig_sl_evocurve}a shows the evolution curves for the coefficient $C_D$, and Fig. \ref{fig_sl_evocurve}b displays the corresponding evolution curves for the total prediction error in the energy spectrum. Clearly, converged results can be obtained for $N=20$ and $40$ in the absence of ensemble inflation ($F=1$), while $N=10$ is insufficient in this case. However, as shown in Fig. \ref{fig_sl_evocurve}, with a proper ensemble inflation ($F=1.5$), the performance of the $N=10$ case can be improved to the same level as the $N=20$ and $40$ cases, with $C_D\approx 0.01$. Further, we show the evolution of the standard deviation $\sigma_C$ of the model coefficient $C_D$ in Fig. \ref{fig_sl_evocurve}c. Apparently, $\sigma_C$ in all cases decreases and saturates with time as the EnKF converges. However, in the case of ensemble inflation, the standard deviation saturates to a tangibly higher level compared to the other cases. This is an expected result due to the effect of the inflation scheme, and it leads to an improved estimation of $C_D$.

\begin{table*}
\begin{center}
\small
\begin{tabular*}{0.95\textwidth}{@{\extracolsep{\fill}}ccccccc}
\hline
Range of $C_D$ &Ensemble size (N) &$N_{iter}$ &$\Delta t_{Obs}$ &$t_0$   &$t_f$ &$\mathbf{w}_i$\\ \hline
$0<C_D<0.05$ &10, 20, 40 &20 &$10\tau_{\theta}$ &$50\tau_{\theta}$ &$800\tau_{\theta}$ &$\mathbf{U}(-1,1)\times 10^{-7}$\\ \hline
\end{tabular*}
\normalsize
\caption{Detailed parameters for the implementation of the ensemble Kalman filter in the LES of incompressible TML.}
\label{tab_tml_EnKF}
\end{center}
\end{table*}

\begin{figure}\centering
\includegraphics[width=.5\textwidth]{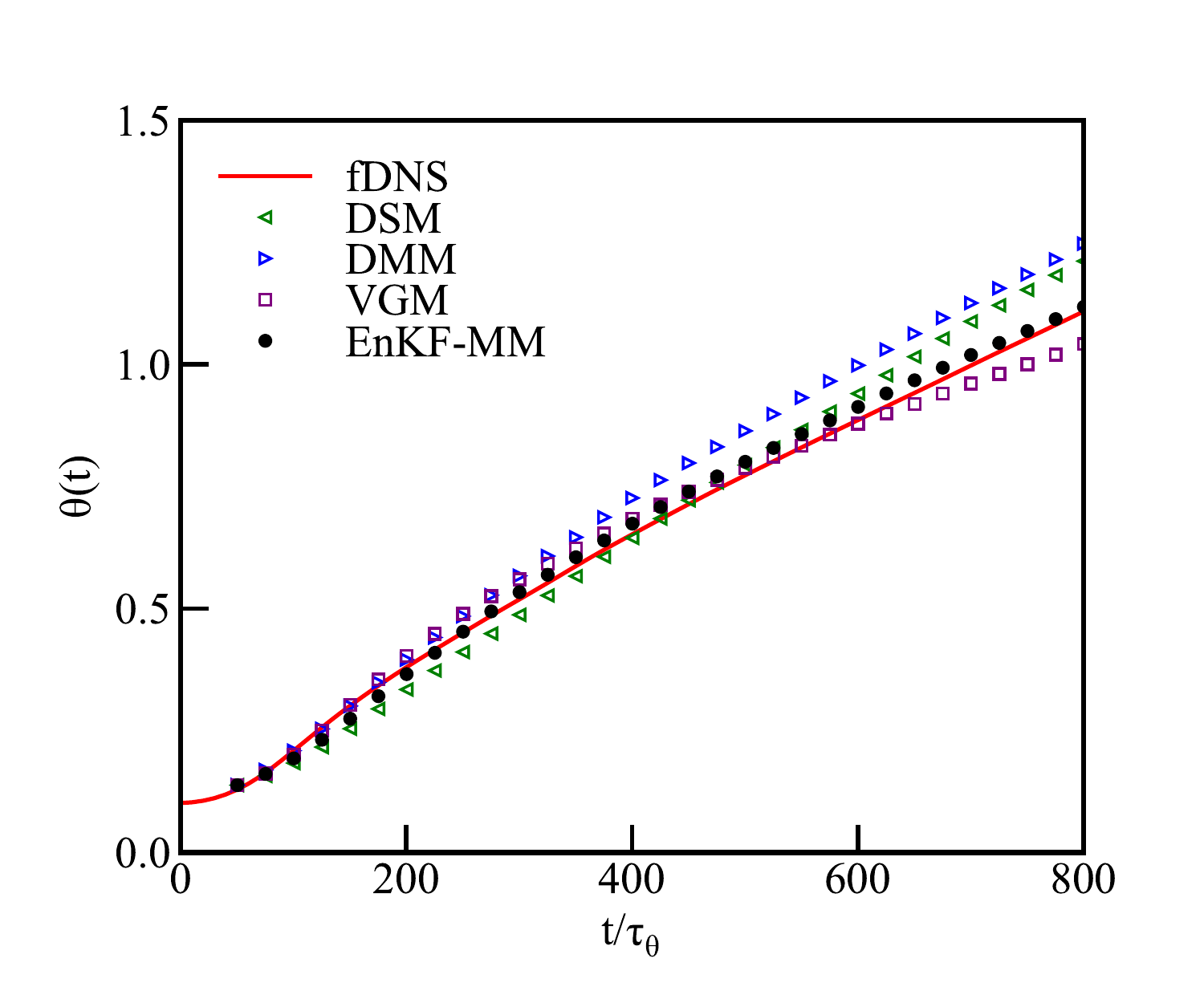}
 \caption{The evolution of momentum thickness in the LES of TML using different SGS models.}\label{fig_sl_theta}
\end{figure}

With the obtained EnKF-MM model, we now examine the results of LES in the case of TML. Fig.~\ref{fig_sl_theta} displays the temporal evolution of the momentum thickness (Eq. (\ref{eq_theta})) in the LES using different SGS models along with the fDNS result. Clearly, a linear growth region is predicted by all SGS models after the transition region. However, the transition region predicted by the DSM model is slightly delayed compared to the fDNS result. On the other hand, the DMM captures the transition region better compared to the DSM model, but it deviates from the fDNS result more strongly with time. Meanwhile, the VGM overestimates the evolution curve in the initial period, but underestimates it later. In comparison, the EnKF-MM model gives the best prediction of the growth of the mixing layer in both the transition and linear growth regions.

\begin{figure}\centering
\includegraphics[width=.45\textwidth]{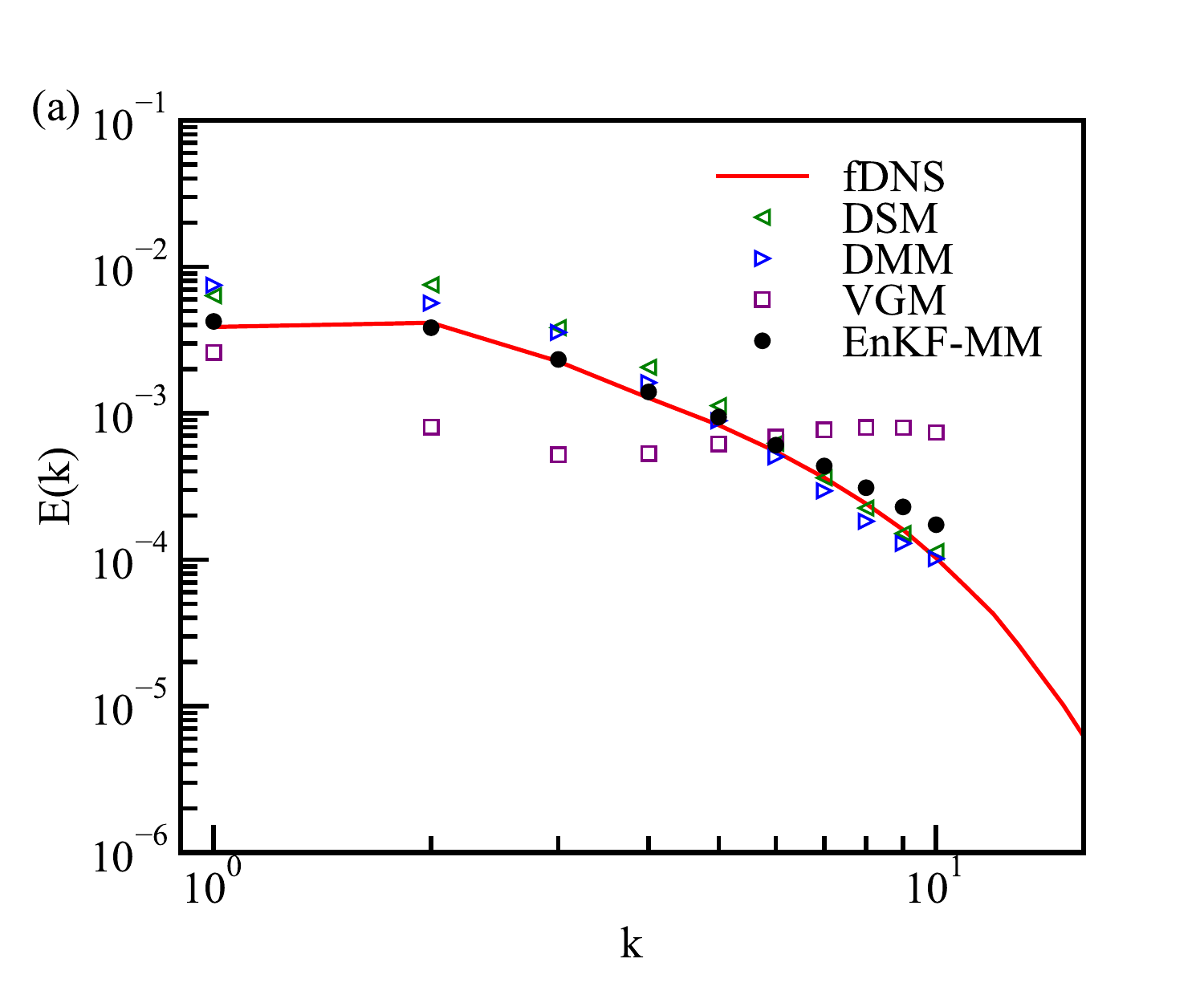}
\includegraphics[width=.45\textwidth]{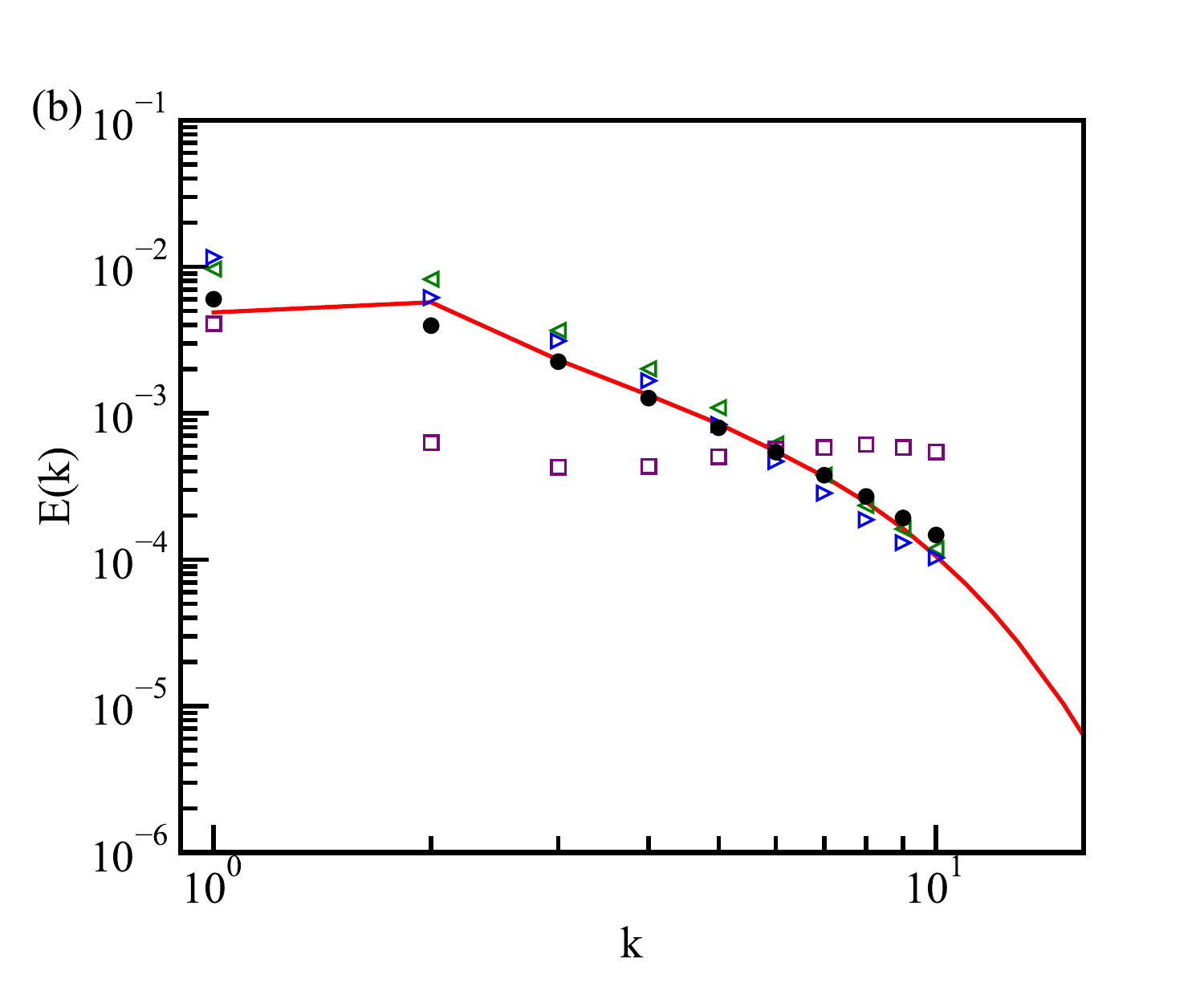}
 \caption{The kinetic energy spectrum in the LES of TML using different SGS models: (a) $t/\tau_{\theta}=600$, (b) $t/\tau_{\theta}=800$.}\label{fig_sl_Ek}
\end{figure}

The kinetic energy spectrum in the LES of TML at two time instants $t/\tau_{\theta}=600$ and $800$ (near the end of the self-similar region) are shown in Figs.~\ref{fig_sl_Ek}a and \ref{fig_sl_Ek}b, respectively. Notably, the VGM model performs much worse compared to other models. Meanwhile, its prediction is also worse in the case of TML compared to that in the LES of forced HIT, presumably due to the absence of control by the large-scale forcing. Both the DSM and DMM models overestimate the kinetic energy at large scales, while the DMM model performs slightly better than the DSM model. In contrast, the EnKF-MM model has a more reasonable prediction of the energy spectrum at both time instants, even though the high-wavenumber kinetic energy is slightly overestimated at $t/\tau_{\theta}=600$.

\begin{figure}\centering
\includegraphics[width=.5\textwidth]{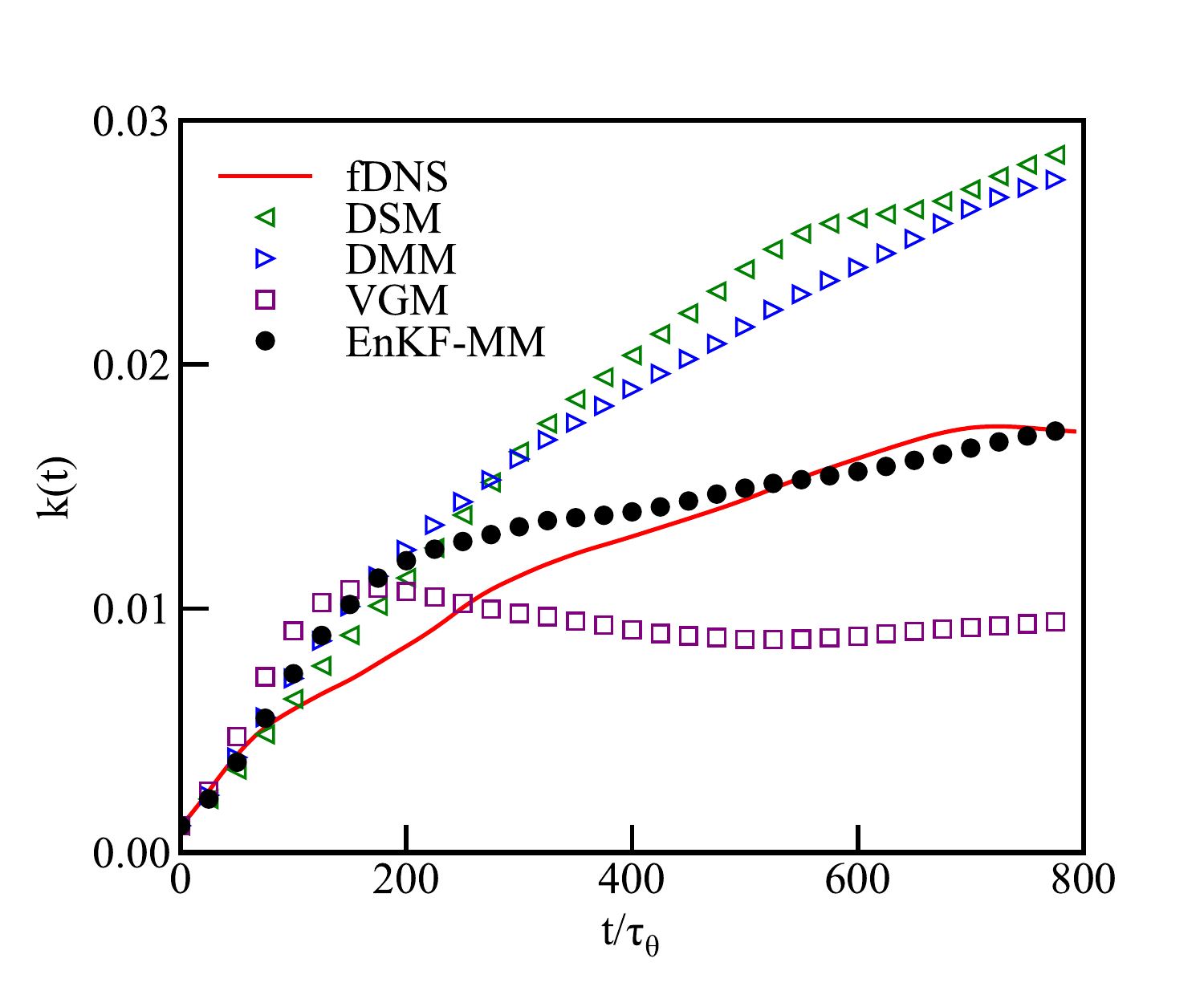}
 \caption{The evolution of kinetic energy in the LES of TML using different SGS models.}\label{fig_sl_kt}
\end{figure}

The evolution of the turbulent kinetic energy is shown in Fig.~\ref{fig_sl_kt}. It is not surprising that the turbulent kinetic energy evolution starts from zero since the turbulent intensity is initially zero. Meanwhile, from the fDNS result, we also observe that the growth rate of the kinetic energy in the linear growth region becomes smaller compared to that in the transition region. As expected, this trend is also reasonably captured by the EnKF-MM model. In contrast, the predictions by the DSM and DMM models are much higher than the fDNS result in the linear growth region. On the other hand, the predicted kinetic energy by the VGM model becomes much lower than the fDNS result. These trends are consistent with the observations in the energy spectrum shown in Fig.~\ref{fig_sl_Ek}.

\begin{figure}\centering
\includegraphics[width=.45\textwidth]{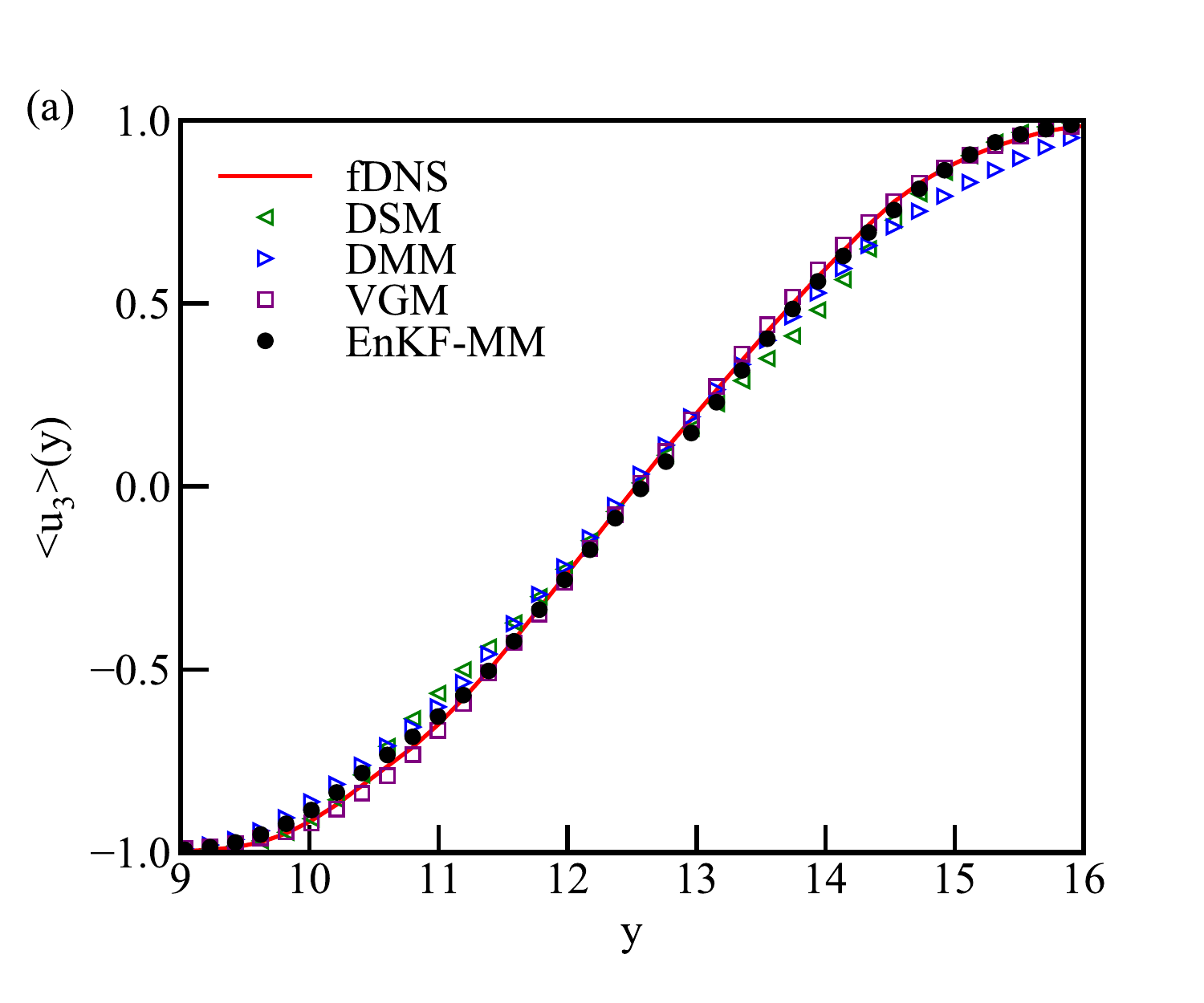}
\includegraphics[width=.45\textwidth]{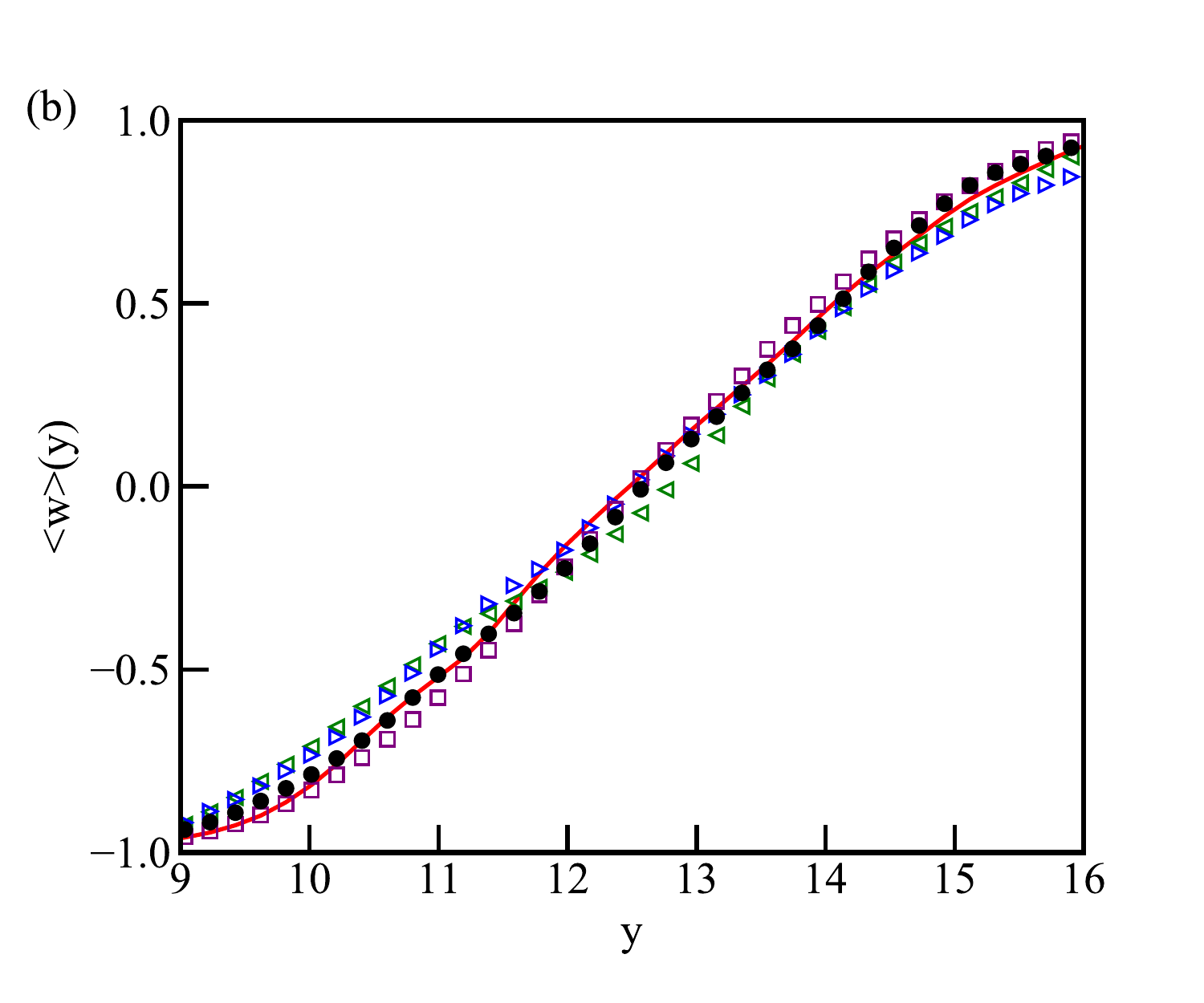}
 \caption{The instantaneous mean streamwise velocity profile across the normal direction: (a) $t/\tau_{\theta}=600$, (b) $t/\tau_{\theta}=800$.}\label{fig_sl_meanW}
\end{figure}

Next, we examine the cross-stream mean streamwise velocity profiles at $t/\tau_{\theta}=600$ and $800$ in Figs. \ref{fig_sl_meanW}a and \ref{fig_sl_meanW}b, respectively. We first observe that the mean velocity profile can be reasonably predicted by all the SGS models, even though the predictions of the DSM and DMM models are slightly worse compared to the EnKF-MM model. Meanwhile, the deviations from the fDNS result for both the DSM and DMM models are more pronounced at $t/\tau_{\theta}=800$ compared to that at $t/\tau_{\theta}=600$. Interestingly, the prediction of the mean velocity profile by the VGM model is quite satisfying compared to the DSM and DMM models, even though its prediction on the kinetic energy is much worse (cf. Figs \ref{fig_sl_Ek} and \ref{fig_sl_kt}). However, this is not so surprising since the lower-order statistics for turbulent flow is relatively easier to predict compared to higher-order statistics such as the energy spectrum. 

\begin{figure}\centering
\includegraphics[width=.45\textwidth]{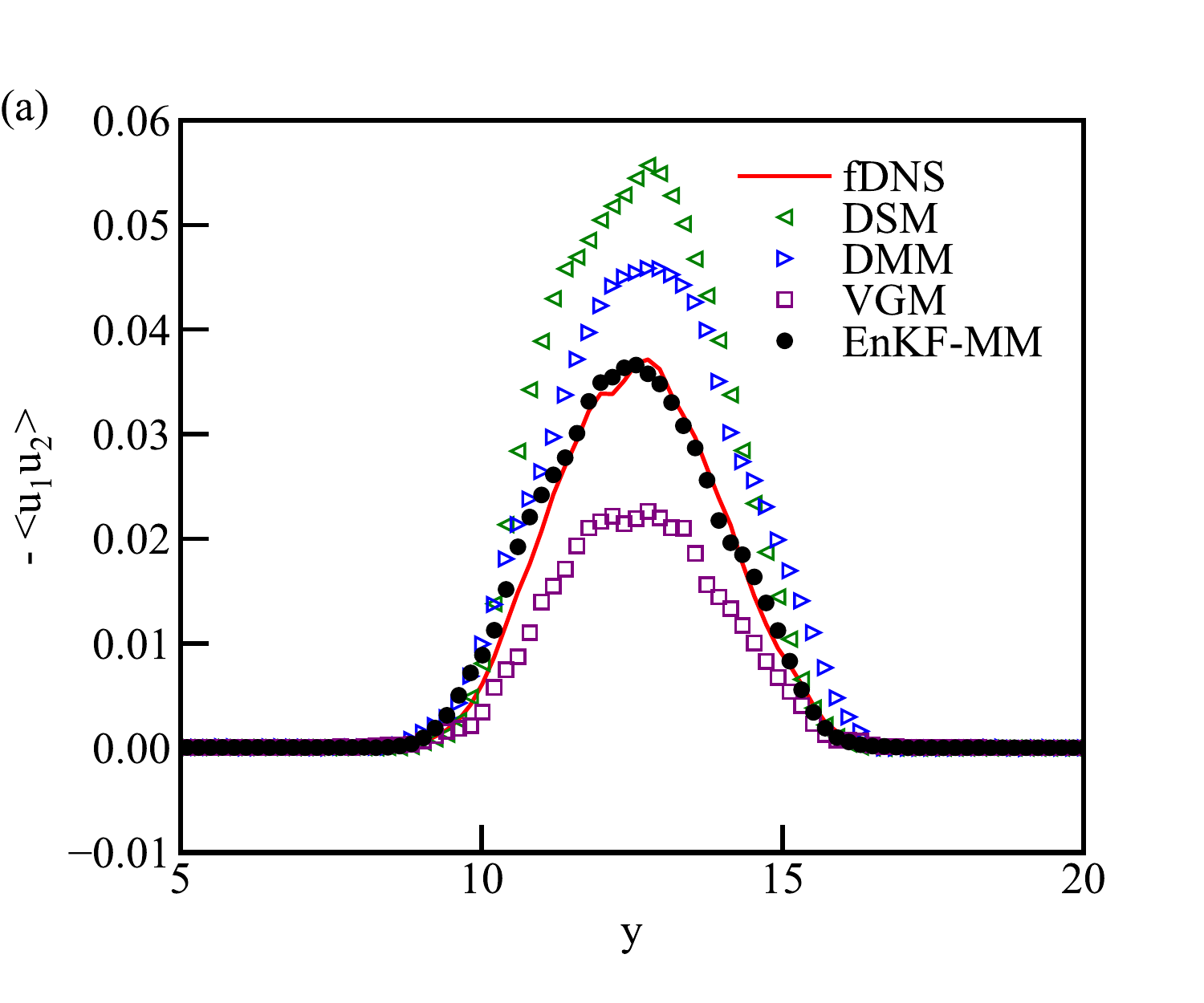}
\includegraphics[width=.45\textwidth]{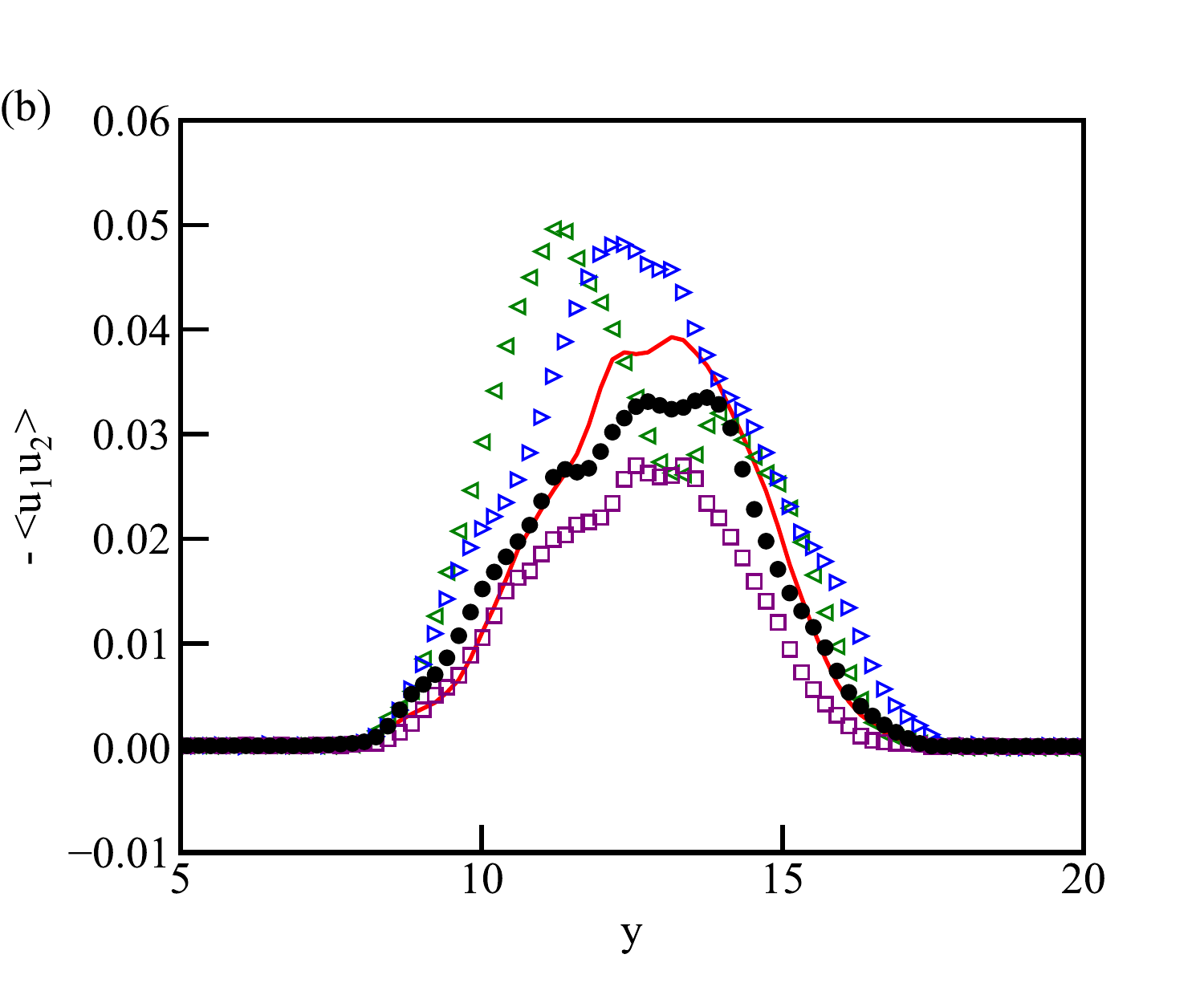}
 \caption{The profile of the Reynolds stress $-\langle u_1 u_2 \rangle$ along the cross-stream direction in the LES of incompressible TML using different SGS models: (a) $t/\tau_{\theta}=600$, (b) $t/\tau_{\theta}=800$.}\label{fig_sl_Reynolds}
\end{figure}

Further, we show the cross-stream variation of the Reynolds stress in Fig.~\ref{fig_sl_Reynolds}. Here the most dominant Reynolds stress term $-\langle u_1 u_2 \rangle$, which is related to the strong shear along the normal direction in the mixing layer, is displayed. As the figure shows, the EnKF-MM model can reasonably recover the profile of the Reynolds stress at both time instants $t/\tau_{\theta}=600$ and $800$ even though the discrepancy is slightly larger at $t/\tau_{\theta}=800$. Both the predictions by the DSM and DMM models are tangibly worse than that by the EnKF-MM model. Meanwhile, the VGM model strongly underestimates the Reynolds stress.

\begin{figure}\centering
\includegraphics[width=.45\textwidth]{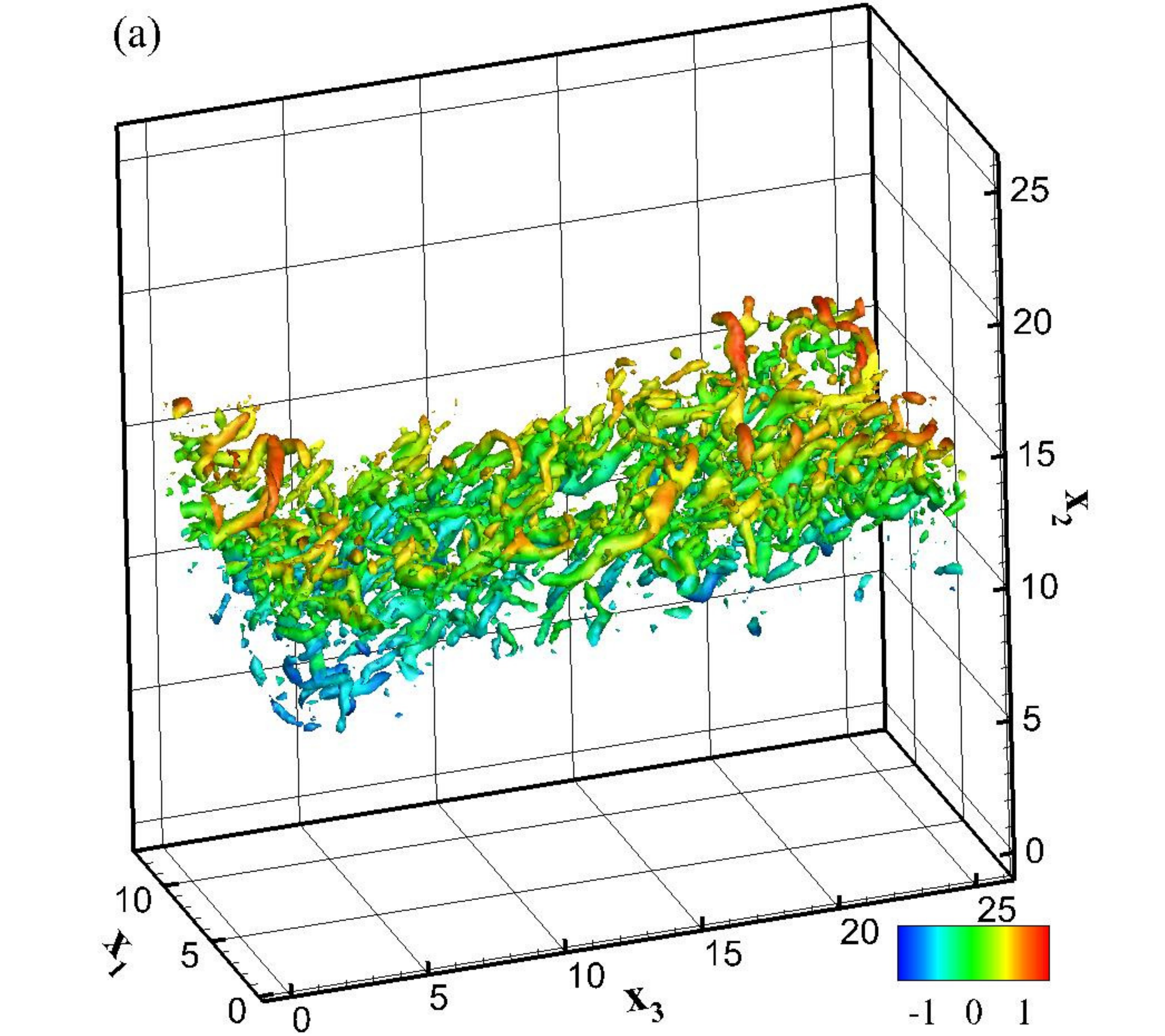}
\includegraphics[width=.45\textwidth]{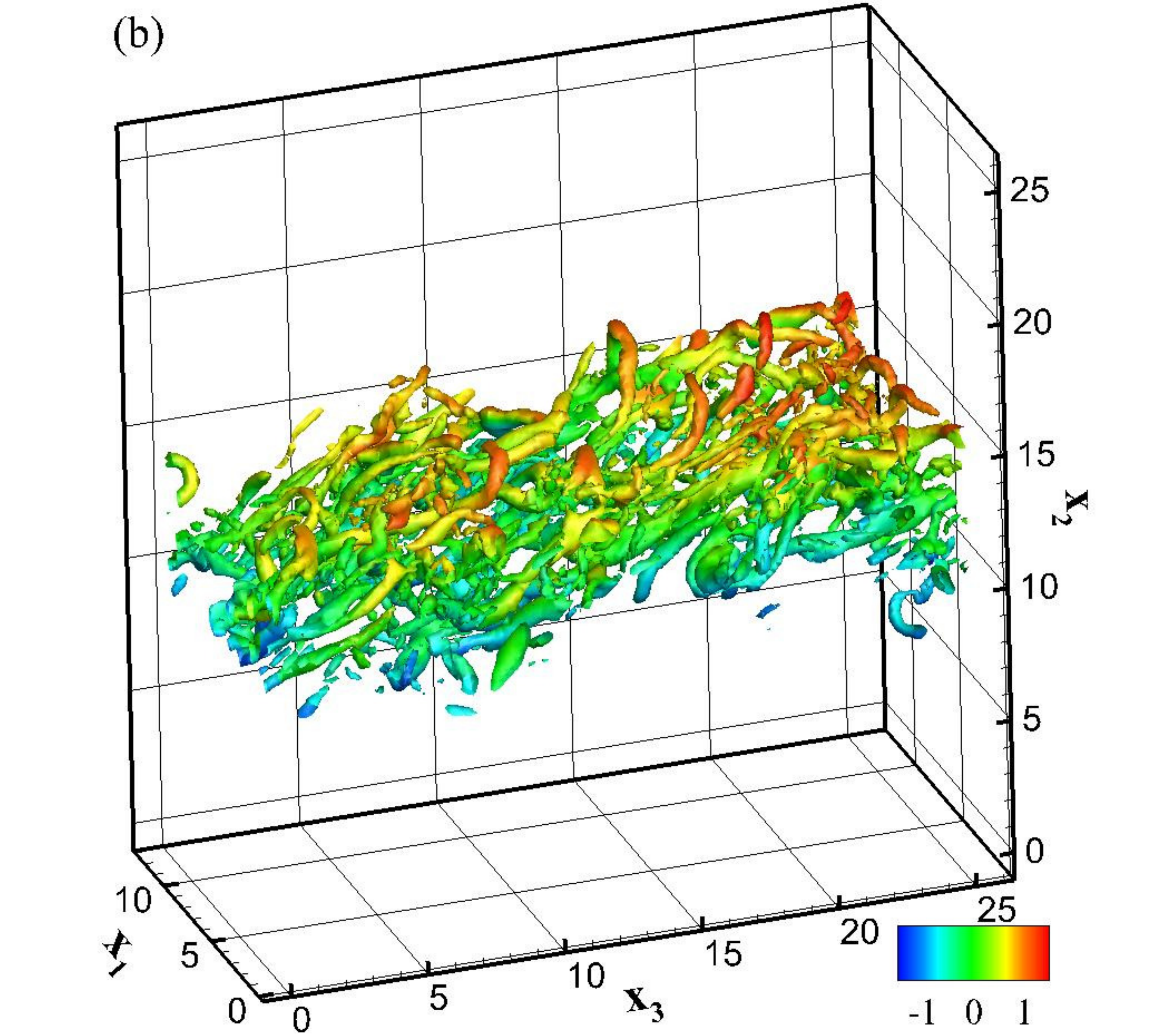}
\includegraphics[width=.45\textwidth]{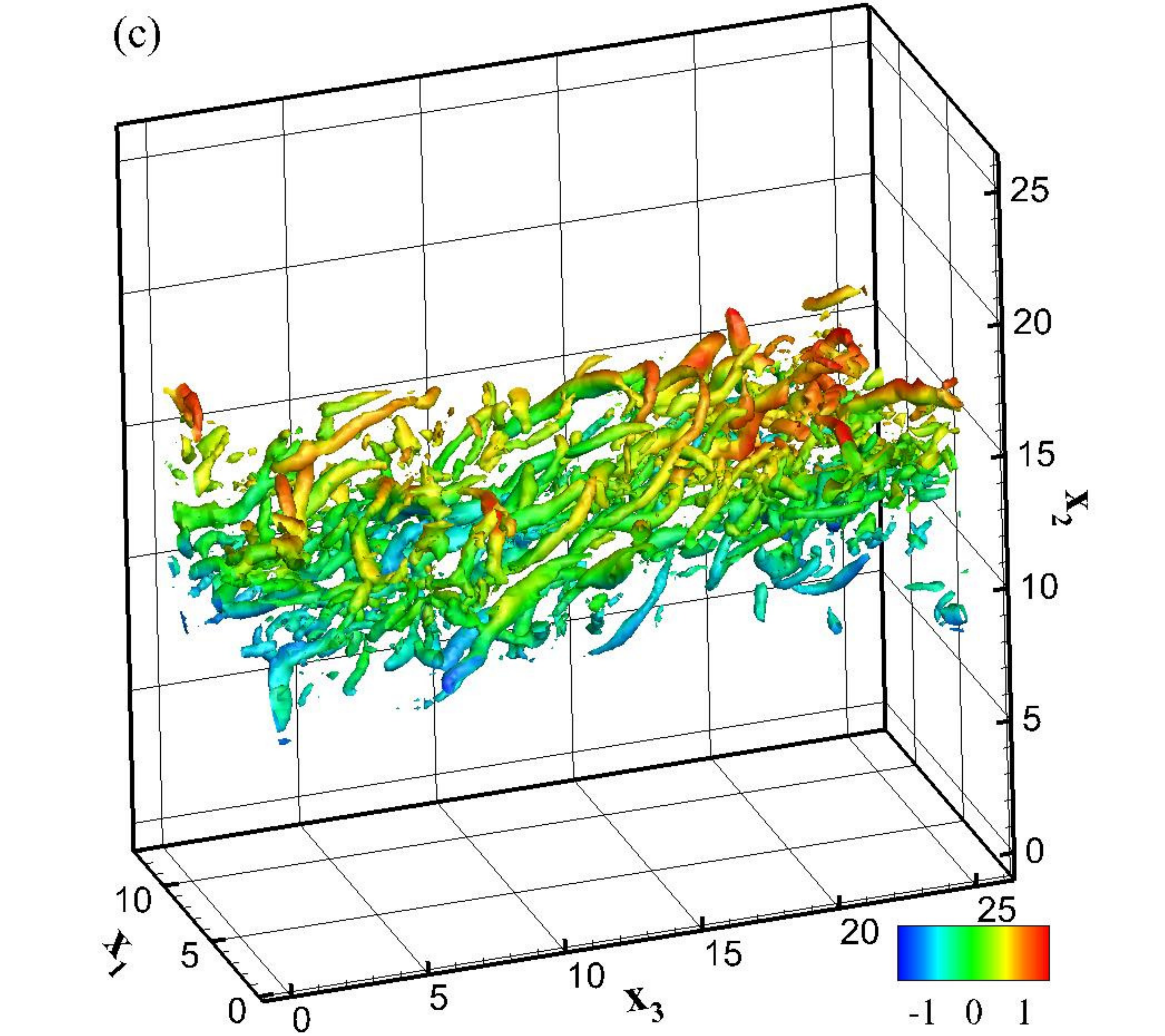}
\includegraphics[width=.45\textwidth]{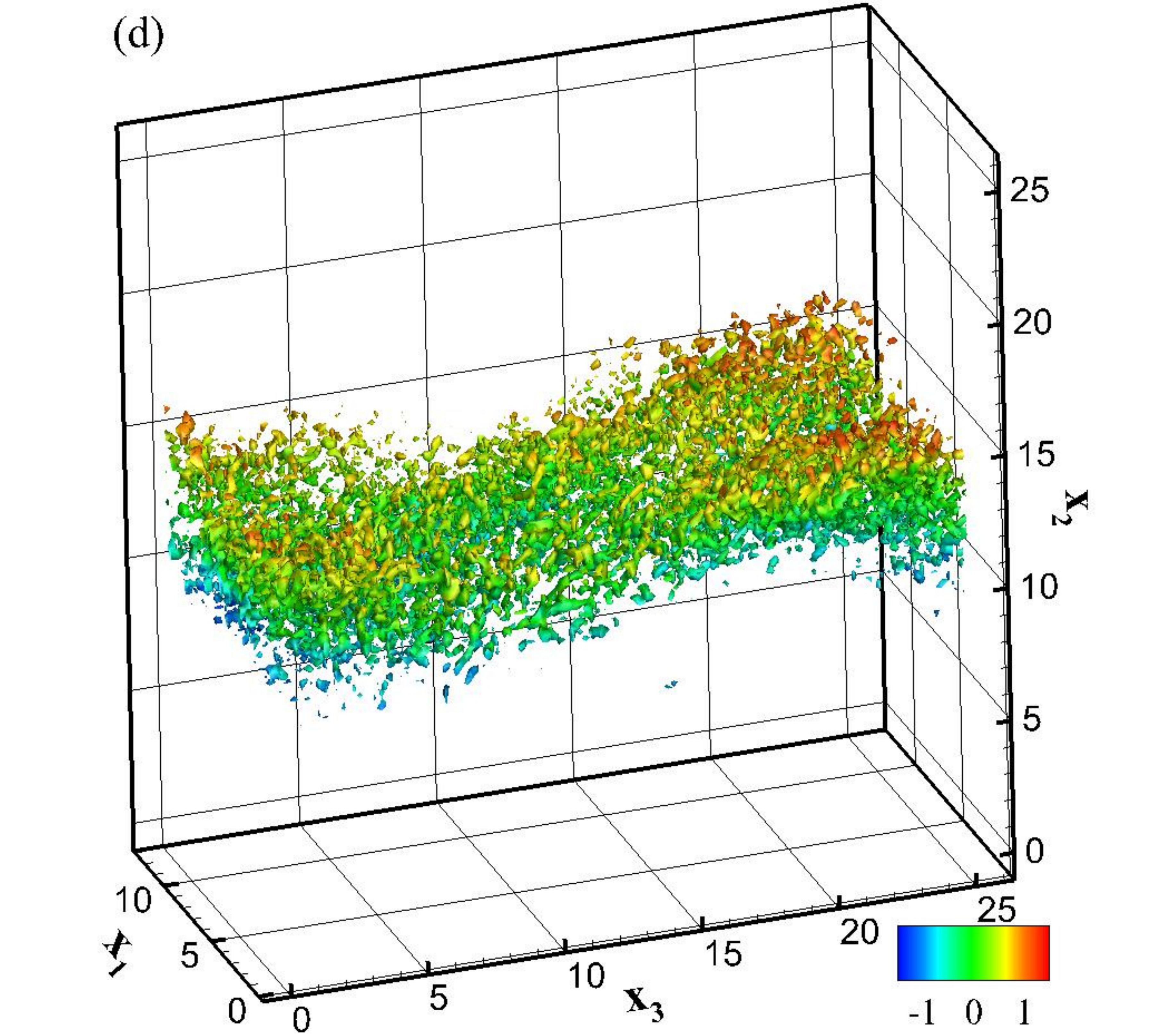}
\includegraphics[width=.45\textwidth]{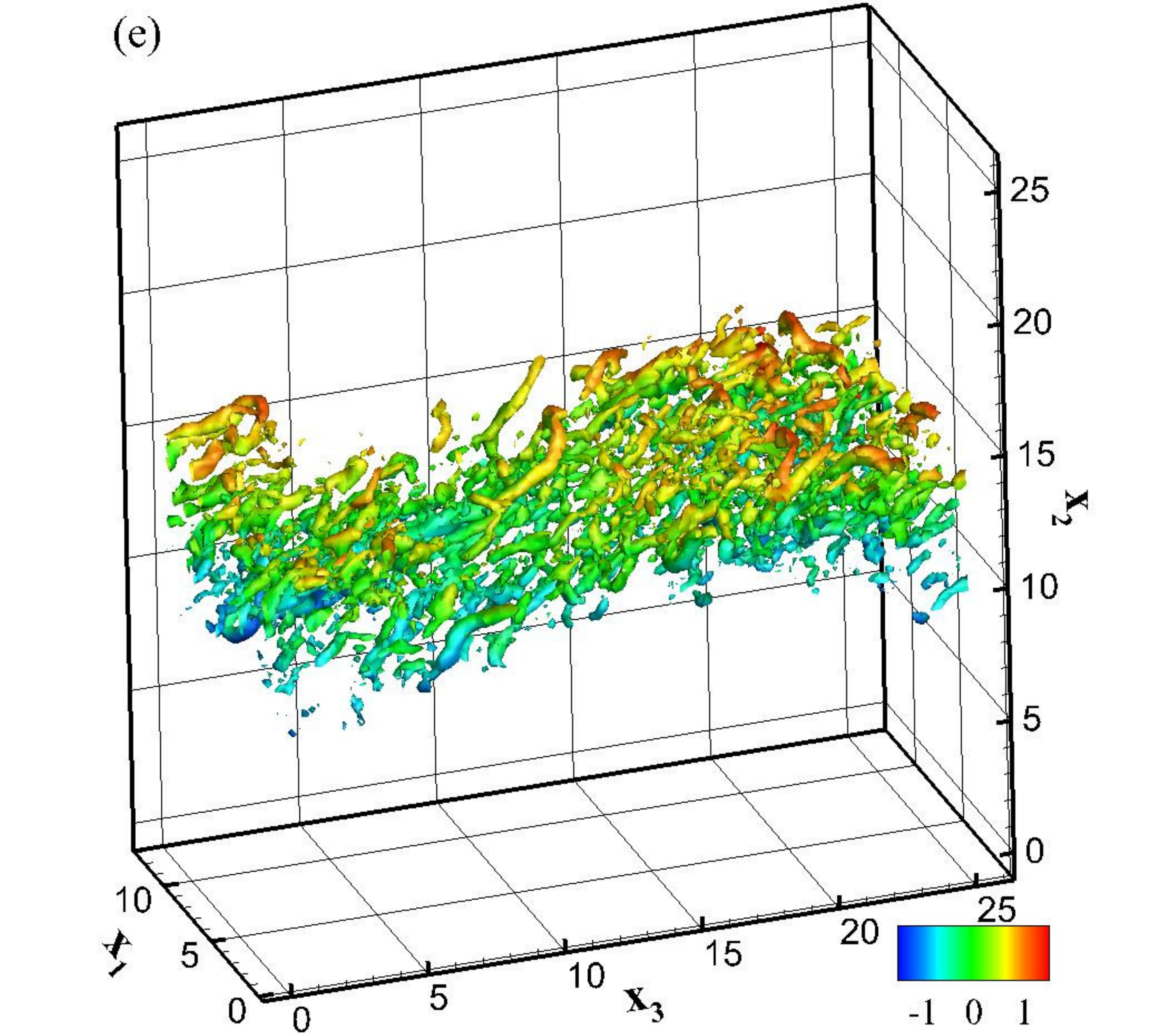}

 \caption{The iso-surface of the Q-criterion at $Q=1$ in the LES of TML using different SGS models. Here the iso-surface is colored by the streamwise velocity. (a) fDNS, (b) DSM, (c) DMM, (d) VGM, (e) EnKF-MM.}\label{fig_sl_Q}
\end{figure}

Finally, to examine the vortex structure in the TML field, we calculate the Q-criterion defined by\cite{Dubief2000,Zhao2021}

\begin{equation}
Q=\frac{1}{2}[\overline{\Omega}_{ij} \overline{\Omega}_{ij}-(\overline{S}_{ij}-\frac{1}{3}\delta_{ij}\overline{S}_{kk})(\overline{S}_{ij}-\frac{1}{3}\delta_{ij}\overline{S}_{ll})],
\end{equation}
where $\overline{\Omega}_{ij}=\frac{1}{2}(\partial{\overline{u}_{i}}/\partial{x_{j}}-\partial{\overline{u}_{j}}/\partial{x_{i}})$ is the rotation rate. The instantaneous iso-surface of $Q$ is displayed in Fig.~\ref{fig_sl_Q} at the end of the self-similar region $t/\tau_{\theta}=800$ for $Q=1$ . Here the iso-surface is colored by the streamwise velocity. As the figure depicts, the DSM and DMM models predict relatively larger vortex structures compared to the fDNS result since the small-scale structures are over-dissipated. On the other hand, the VGM model predicts much smaller structures compared to the fDNS result. As desired, the EnKF-MM model gives the best agreement with the fDNS result among all the tested models, demonstrating its advantage in predicting the spatial-temporal vortex structures of turbulence.

%%%%%%%%%%%%%%%%%%%%%%%%%%%%%%%%%%%%%%%%%%%%%%%%%%%%%%%%%%%%%%%%%%%%%%%%%%%%%%%%%%%%%%%%%%%%%%%%%%%%%%%%%%%%%%%%%%%%%%%%%%%%%%%%%%%%%%%%%%%%%%%%%%%%%
\section{Discussions}

As there are numerous data-driven SGS models in the literature, it is necessary to briefly discuss the merits of these methods. First of all, due to strong non-linear fitting ability of neural networks, most of the machine-learning-based SGS models have superb \emph{a priori} accuracy, i.e., the constitutive relation between the SGS stress and the filtered flow variables can be very accurately reconstructed using the fine-grid DNS data.\cite{Xie2020b,Xie2020d,Yuan2020} However, since the numerical errors in the coarse-grid LES are unknown and not considered in the \emph{a priori} machine-learning process, the obtained SGS models often under-perform and suffer from numerical instability in a practical LES (i.e. in the \emph{a posteriori} test). As a result, additional treatment must be adopted in the LES, such as adjusting a hyper-viscosity,\cite{Xie2020d,Yuan2020,Wang2021} regulating the negative SGS energy flux,\cite{Wang2022} or artificially configuring an additional filtering scheme.\cite{Yuan2022} Consequently, the efficiency of such SGS models are often lower than traditional models.

On the other hand, if we directly optimize the SGS models in the \emph{a posteriori} LES, data assimilation techniques are often necessary such as the nudging method,\cite{Buzzicotti2020} the adjoint-based variational method,\cite{He2018,Yuan2023} or the ensemble Kalman filter (EnKF)-based method.\cite{Kato2013,Deng2018} In the nudging method, the model coefficients must be evaluated manually based on the point-to-point error of the entire flow field. The adjoint-based variational methods can minimize the prediction error of a SGS model by evolving a set of adjoint equations in conjunction with a gradient decent scheme. However, the task of configuring and implementing the adjoint method for complex flow is numerically heavy. In comparison, the EnKF method is numerically more convenient to use, and its accuracy and efficiency have also be advocated.\cite{Zhang2022} 

In the present study, we have demonstrated the capacity of EnKF method in optimizing the mixed SGS model for LES, and the resulted EnKF-MM model performs reasonably well in the predictions of both the flow statistics and instantaneous flow structures. Nevertheless, we also emphasize that, for the current low-dimensional parameter estimation problem, other methods such as the finite difference method can also be applied.\cite{Plyasunov2007} However, for high-dimensional optimization problem, the EnKF method can be a more promising tool, which has already been demonstrated in RANS for optimizing multi-parameter models or neural networks. Recently, many multi-parameter models have been proposed in LES with proven accuracy,\cite{Xie2020d,Wang2022,Xu2023} and we shall apply EnKF to optimize such models in future works, so as to improve the performance of LES for more complex turbulent flow problems.

%%%%%%%%%%%%%%%%%%%%%%%%%%%%%%%%%%%%%%%%%%%%%%%%%%%%%%%%%%%%%%%%%%%%%%%%%%%%%%%%%%%%%%%%%%%%%%%%%%%%%%%%%%%%%%%%%%%%%%%%%%%%%%%%%%%%%%%%%%%%%%%%%%%%%
\section{Conclusions}

An ensemble Kalman filter (EnKF)-based mixed model (EnKF-MM) is proposed for the subgrid-scale (SGS) stress in the LES of turbulence. The VGM and Smagorinsky models are chosen as the structural and functional parts, respectively, in the mixed model. The model coefficients are determined through the EnKF-based data assimilation technique using fDNS results as the benchmark for LES.

The kinetic energy spectrum of the fDNS data has been adopted as the target for the EnKF to optimize the coefficient of the functional part in the mixed model. It has been observed that adding the functional part to the VGM model can significantly improve the prediction of LES on many important turbulent statistics and flow structures. 

The EnKF-MM framework is tested in the LES of both the incompressible homogeneous isotropic turbulence (HIT) and turbulent mixing layer (TML). In the LES, the prediction of the kinetic energy spectrum, the PDF of the SGS stress, the PDF of the strain rate and the PDF of the SGS flux are compared against the fDNS result among different SGS models. The structure functions, the evolution of turbulent kinetic energy, the mean flow and Reynolds stress profile, and the iso-surface of the Q-criterion are also examined to evaluate the spatial-temporal prediction ability of different SGS models. 

Overall, the LES results obtained using EnKF-MM model has a closer agreement to the benchmark fDNS result. The DSM and DMM models can over-dissipate the small-scale structures in the LES. On the other hand, the VGM model suffers from insufficient SGS dissipation. Compared to the traditional SGS models, the current EnKF-MM model shows more accuracy and consistency in predicting various flow statistics and spatial-temporal flow structures, demonstrating its great potential in improving the accuracy of LES of turbulence.

Finally, we would like to note that even though the EnKF has already been successfully applied to RANS model for wall-bounded turbulent flows, such as boundary-layer flow,\cite{Kato2013} and flow over periodic hills,\cite{Zhang2022} its application to LES is still in a preliminary state. To successfully apply EnKF in LES for more complex flows like wall-bounded turbulence, further adjustments of EnKF and improvements for computational efficiency are necessary.    

\begin{acknowledgments}
This work was supported by the National Natural Science Foundation of China (NSFC Grants No. 91952104, No. 92052301, No. 12172161, and No. 91752201), by the NSFC Basic Science Center Program (grant no. 11988102), by the Shenzhen Science and Technology Program (Grants No. KQTD20180411143441009), by Key Special Project for Introduced Talents Team of Southern Marine Science and Engineering Guangdong Laboratory (Guangzhou) (Grant No. GML2019ZD0103), and by Department of Science and Technology of Guangdong Province (No. 2020B1212030001). This work was also supported by Center for Computational Science and Engineering of Southern University of Science and Technology, and by National Center for Applied Mathematics Shenzhen (NCAMS).
\end{acknowledgments}

\section*{DATA AVAILABILITY}

The data that support the findings of this study are available from the corresponding author upon reasonable request.

% \appendix
\appendix

\section{The influence of filter type}

In this appendix, we explore the influence of the filter type. First of all, we note that the VGM part of the mixed model requires that the filter should have finite moments.\cite{Pope2000,Clark1979} The commonly adopted filters satisfying the `finite-moment' condition are the Gaussion filter given by Eq. (\ref{gauss}), and the top-hat (box) filter whose filter kernel in one dimension can be written as\cite{Pope2000}

\begin{equation}
  G(r)=\frac{1}{\Delta}H(\frac{1}{2}\Delta-|r|),
  \label{box}
\end{equation}
where $H$ is the Heaviside function. Since the second moments of both the Gaussion and top-hat filters are the same, and the transfer function for both filter are very similar at the dominant low wavenumbers,\cite{Pope2000} the LES and the corresponding SGS modeling using these two filters should not be very different. For this reason, the VGM formulation does not specify the exact type of the filter. To illustrate this, we re-run the LESs of forced HIT with initial conditions obtained using the top-hat filter, and the test filters in DSM and DMM models are also switched to top-hat filter.\cite{Xie2020d}

\begin{figure}\centering
\includegraphics[width=.45\textwidth]{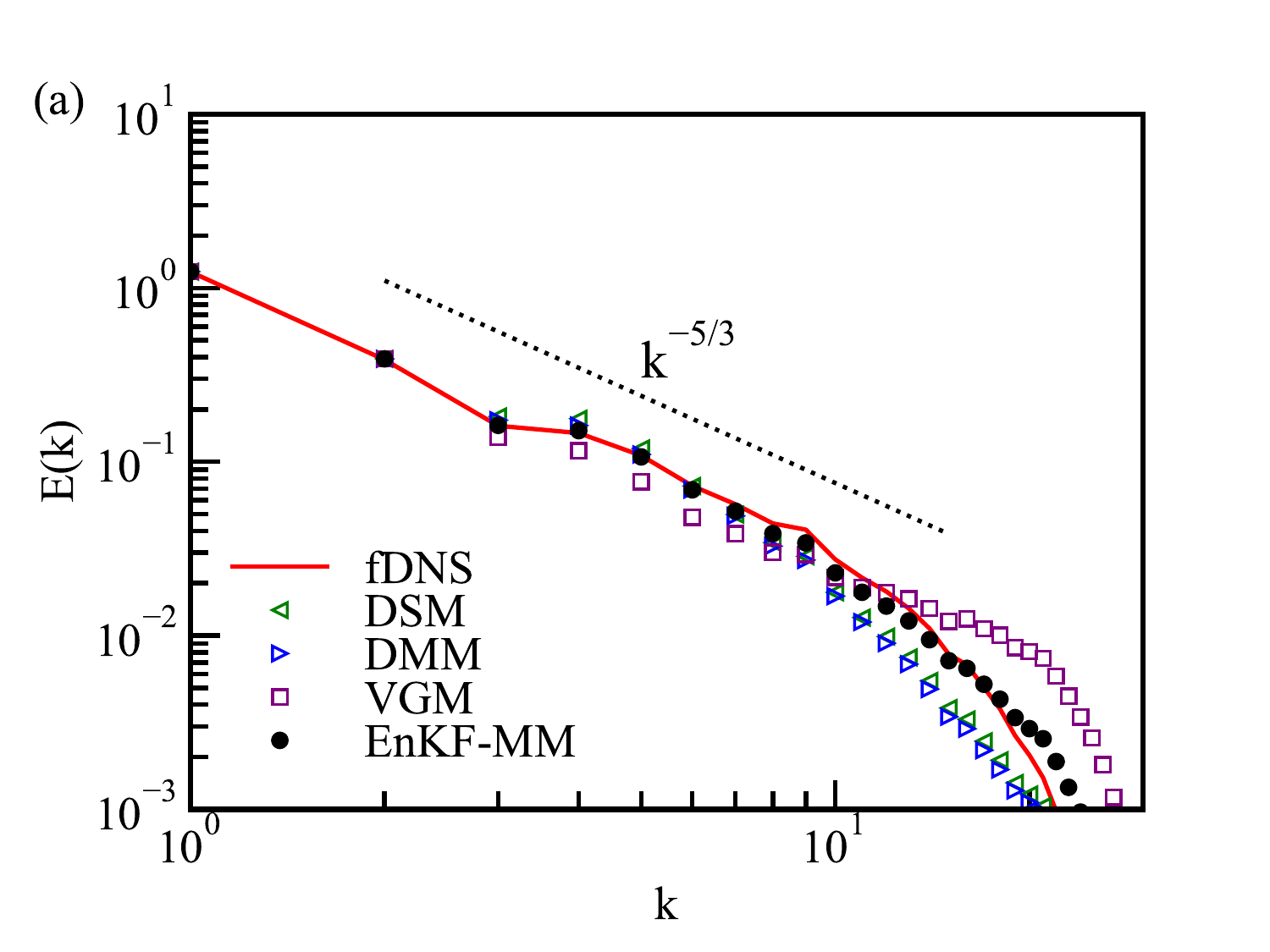}
\includegraphics[width=.45\textwidth]{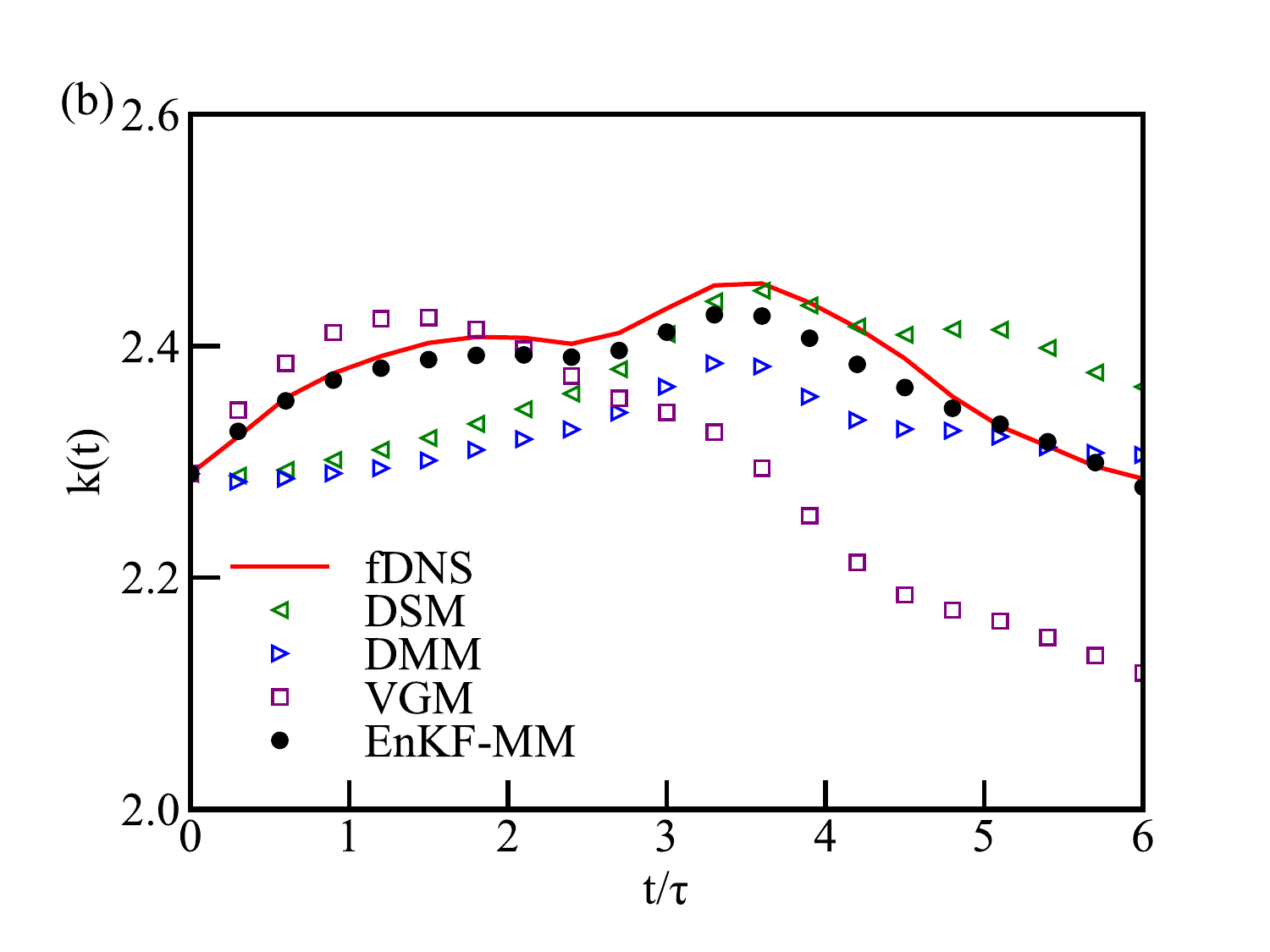}
 \caption{The kinetic energy spectrum and kinetic energy evolution in the LES of forced HIT using different SGS models with the top-hat filter: (a) kinetic energy spectrum, (b) evolution of kinetic energy.}\label{fig_hit_Ek_kt_box}
\end{figure}

In Fig. \ref{fig_hit_Ek_kt_box}, we compare the energy spectrum and the evolution of kinetic energy with the corresponding benchmark fDNS results obtained using the top-hat filter. As can be seen, both the predicted energy spectrum (Fig. \ref{fig_hit_Ek_kt_box}a) and evolution of kinetic energy (Fig. \ref{fig_hit_Ek_kt_box}b) by the EnKF-MM model are in better agreements with the fDNS results compared to those predicted by the other SGS models. We note that here the EnKF-MM model is not re-calibrated using the top-hat filter. Hence, the current results suggest that the EnKF-MM model obtained using the Gaussion filter-based fDNS data can also be applied in the case of the top-hat filter.

\section{The influence of the flow parameters on the performance of the EnKF-MM model}

\begin{figure}\centering
\includegraphics[width=.45\textwidth]{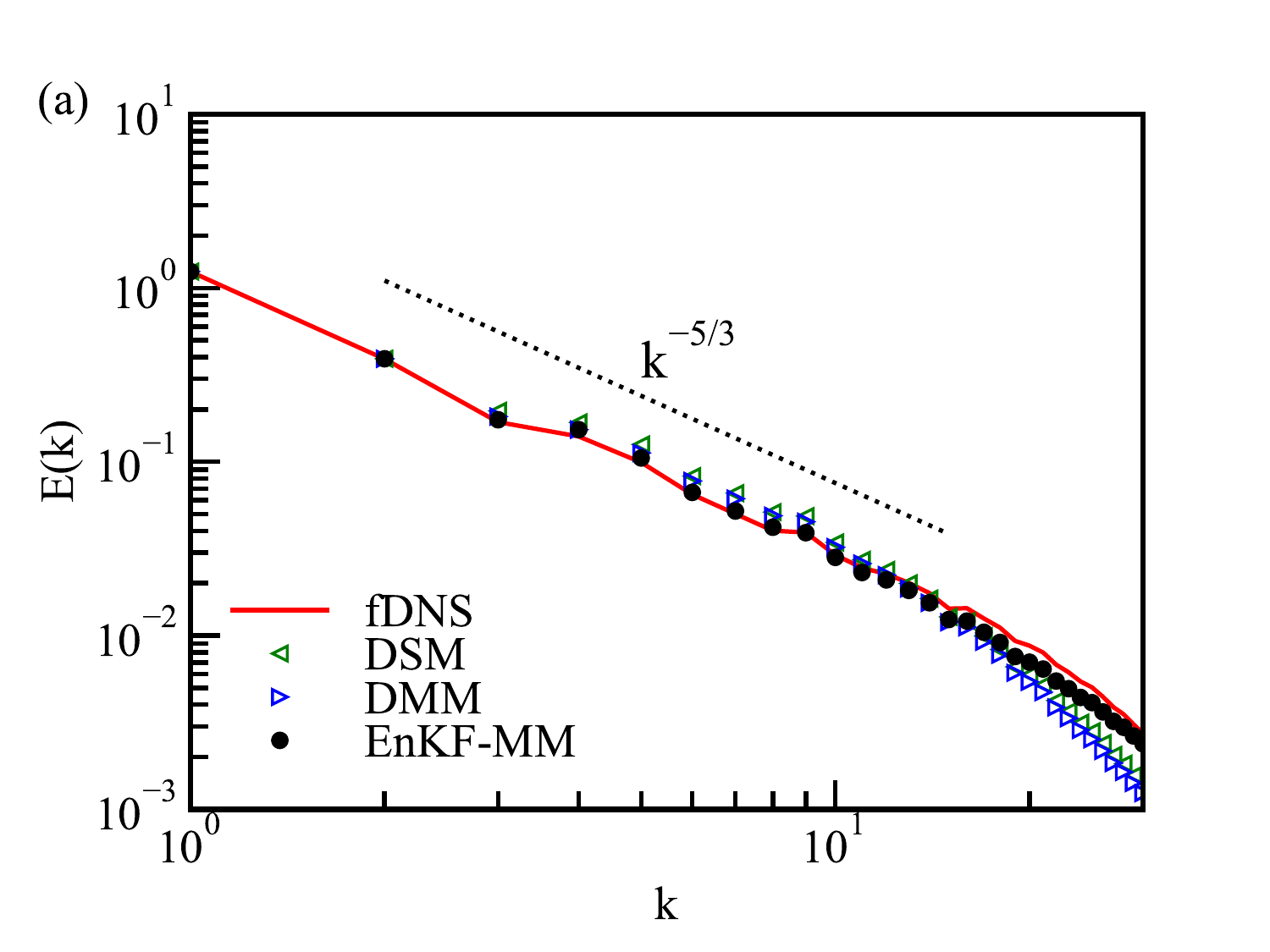}
\includegraphics[width=.45\textwidth]{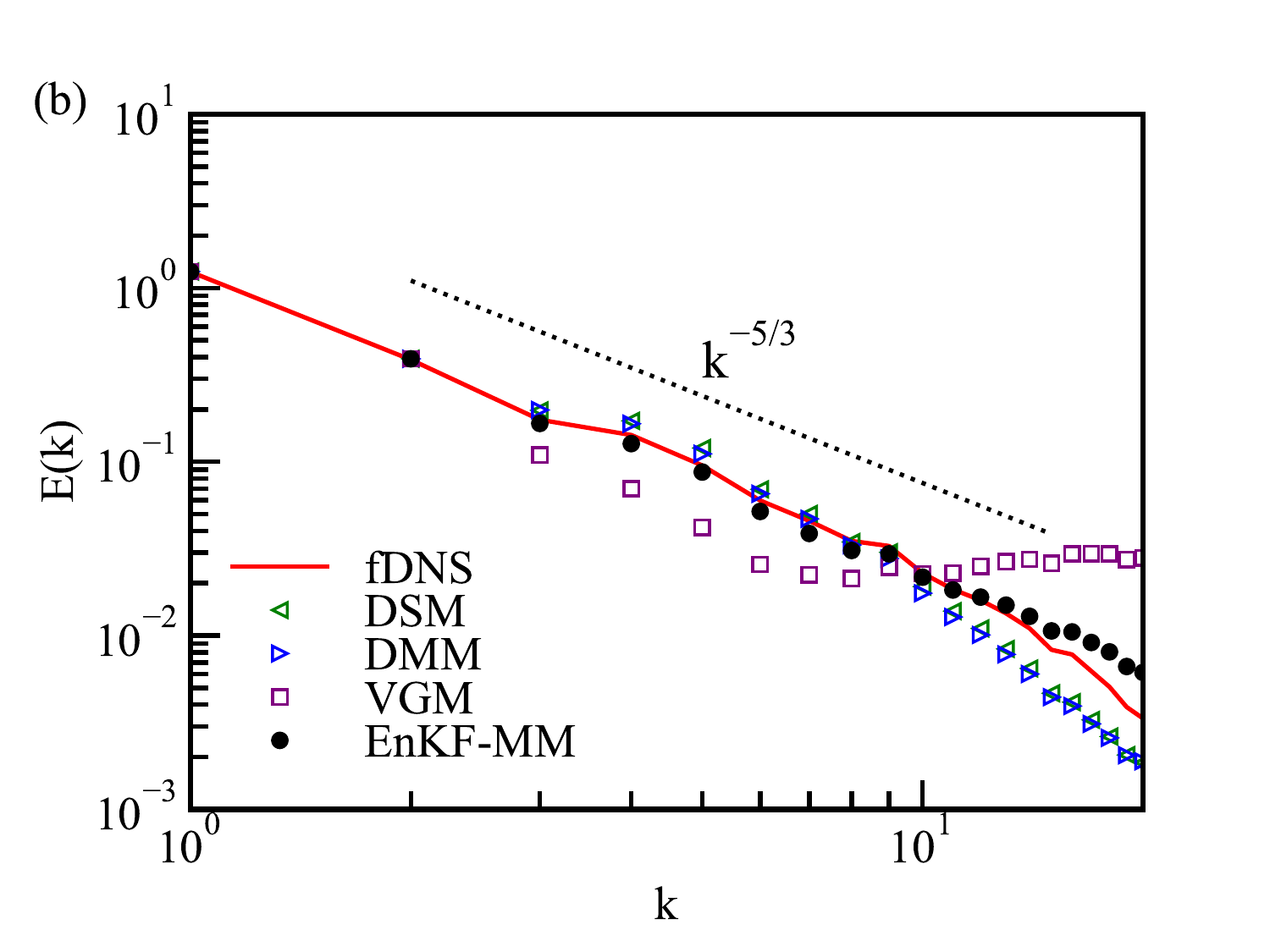}
 \caption{The kinetic energy spectrum in the LES of forced HIT using different SGS models at $Re_{\lambda}=259$: (a) $\Delta=16h_{DNS}$, (b) $\Delta=32h_{DNS}$.}\label{fig_hit_Ek_1632_259}
\end{figure}

In this appendix, we test the influence of the flow parameters on the performance of the obtained EnKF-MM model. More specifically, we shall examine the influences of the Reynolds number, the filter width and the large-scale forcing in the LES of HIT. First, we test the applicability of the EnKF-MM framework at a larger Reynolds number $Re_{\lambda}=259$ with two different filter widths $\Delta=16h_{DNS}$ and $\Delta=32h_{DNS}$. The details of the DNS is given in Table \ref{tab_hit_DNS259}. In the LES, we apply the EnKF-MM model which is optimized at $Re_{\lambda}=160$ and $\Delta=16h_{DNS}$. The kinetic energy spectra for $\Delta=16h_{DNS}$ and $\Delta=32h_{DNS}$ are shown in Figs.~\ref{fig_hit_Ek_1632_259}a and \ref{fig_hit_Ek_1632_259}b, respectively. We note that, at $\Delta=16h_{DNS}$, the results for VGM model diverges and thus not shown in Fig.~\ref{fig_hit_Ek_1632_259}a. As can be seen, at a larger Reynolds number and different filter widths, the EnKF-MM model invariably outperforms the traditional models, suggesting that the performance of the EnKF-MM framework is not much influenced by the Reynolds number and filter width.

\begin{table*}
\begin{center}
\small
\begin{tabular*}{0.85\textwidth}{@{\extracolsep{\fill}}cccccc}
\hline
Reso. &$Re_{\lambda}$   &$\nu$ &$\Delta t$ &$u^{rms}$ &$\epsilon$\\ \hline
$1024^{3}$ &259    &0.001 &0.0002 &2.29 &0.69\\ \hline
\end{tabular*}
\normalsize
\caption{Numerical simulation parameters and one-point statistical quantities for incompressible isotropic turbulence at $Re_{\lambda}=259$.}
\label{tab_hit_DNS259}
\end{center}
\end{table*}

\begin{figure}\centering
\includegraphics[width=.45\textwidth]{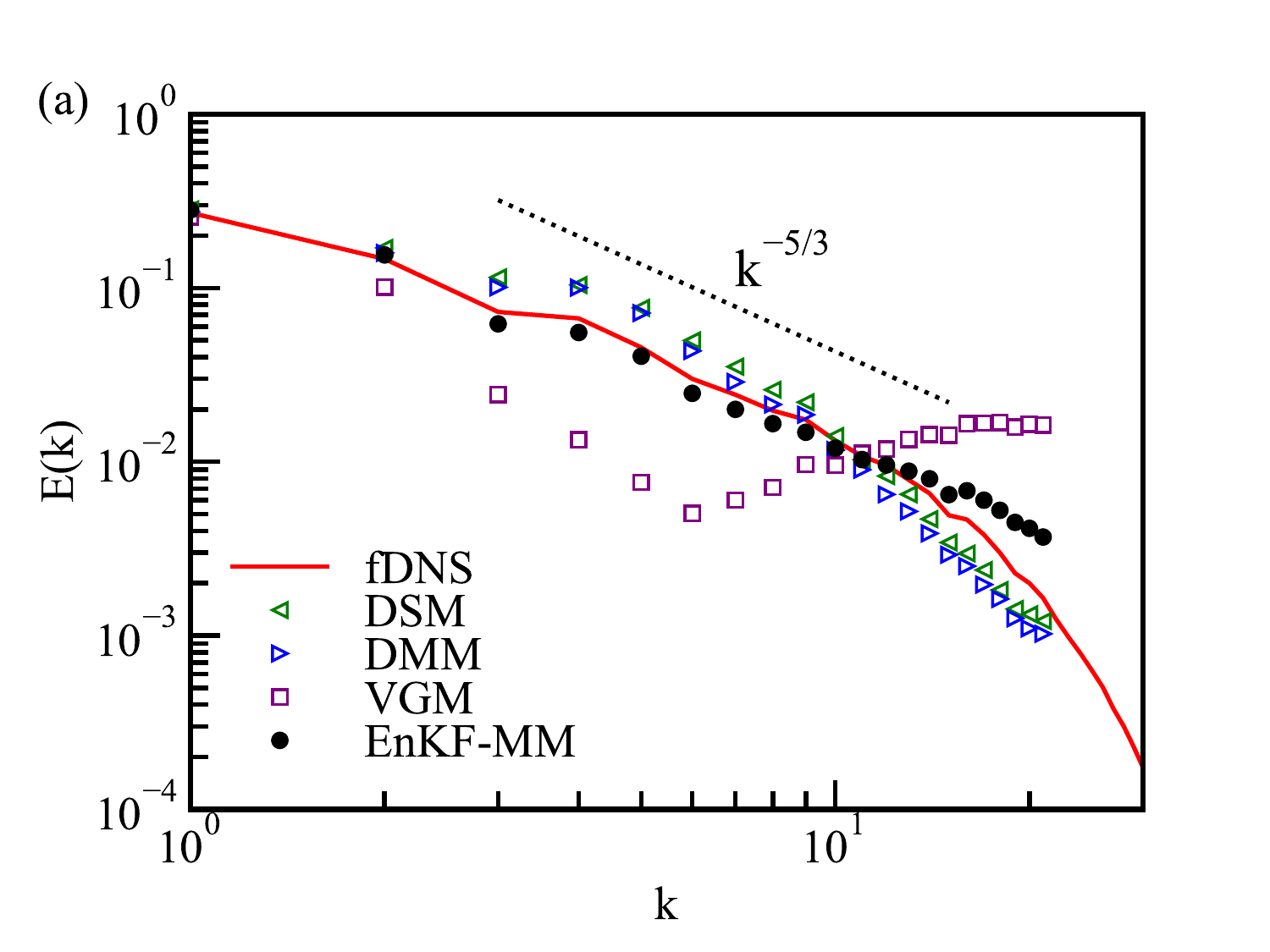}
\includegraphics[width=.45\textwidth]{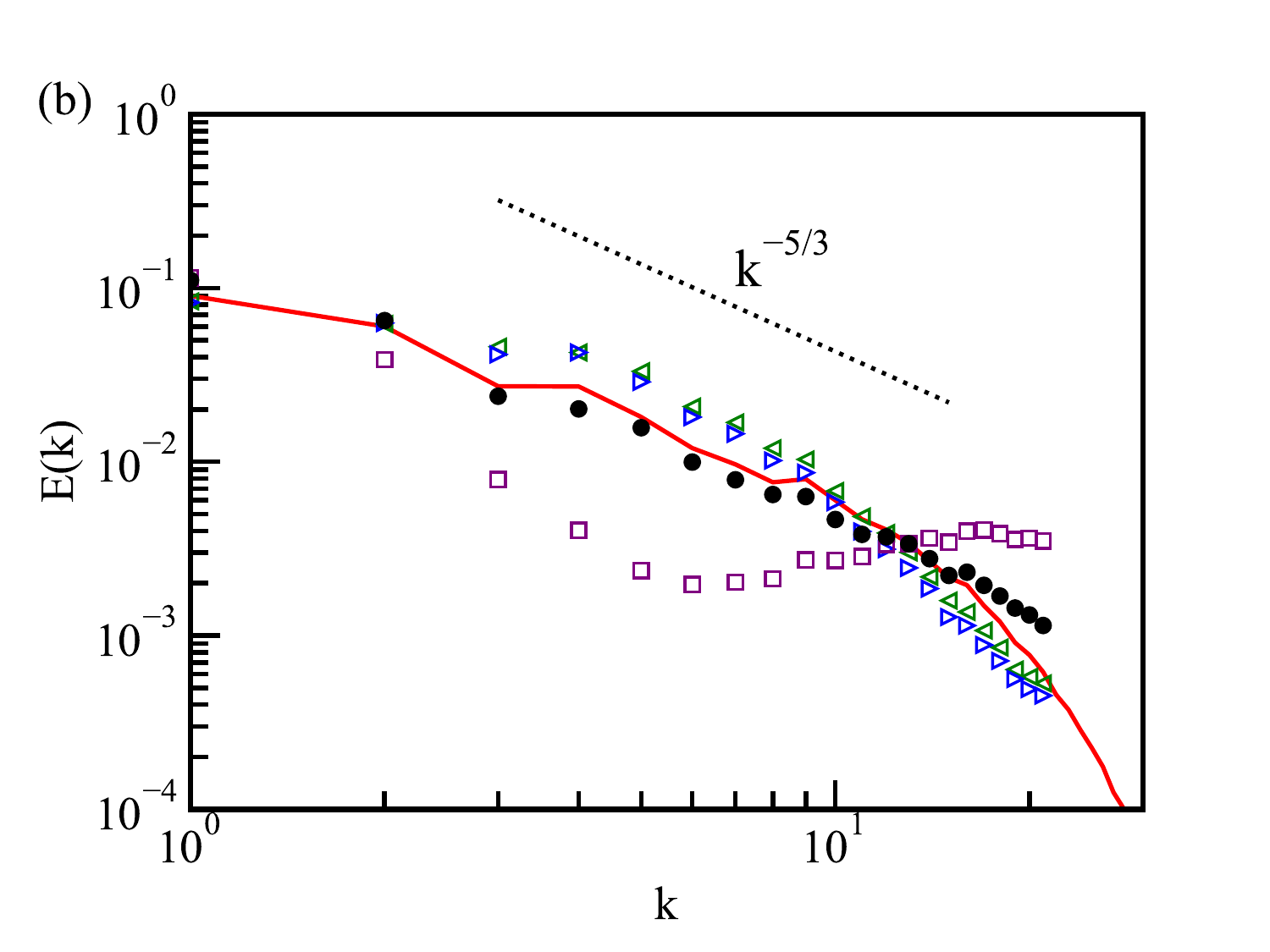}
 \caption{The kinetic energy spectrum in the LES of decaying HIT using different SGS models at initial Taylor Reynolds number $Re_{\lambda}=259$: (a) $t=4.8\tau$, (b) $t=9.6\tau$.}\label{fig_hit_Ek_decay_259}
\end{figure}

\begin{figure}\centering
\includegraphics[width=.5\textwidth]{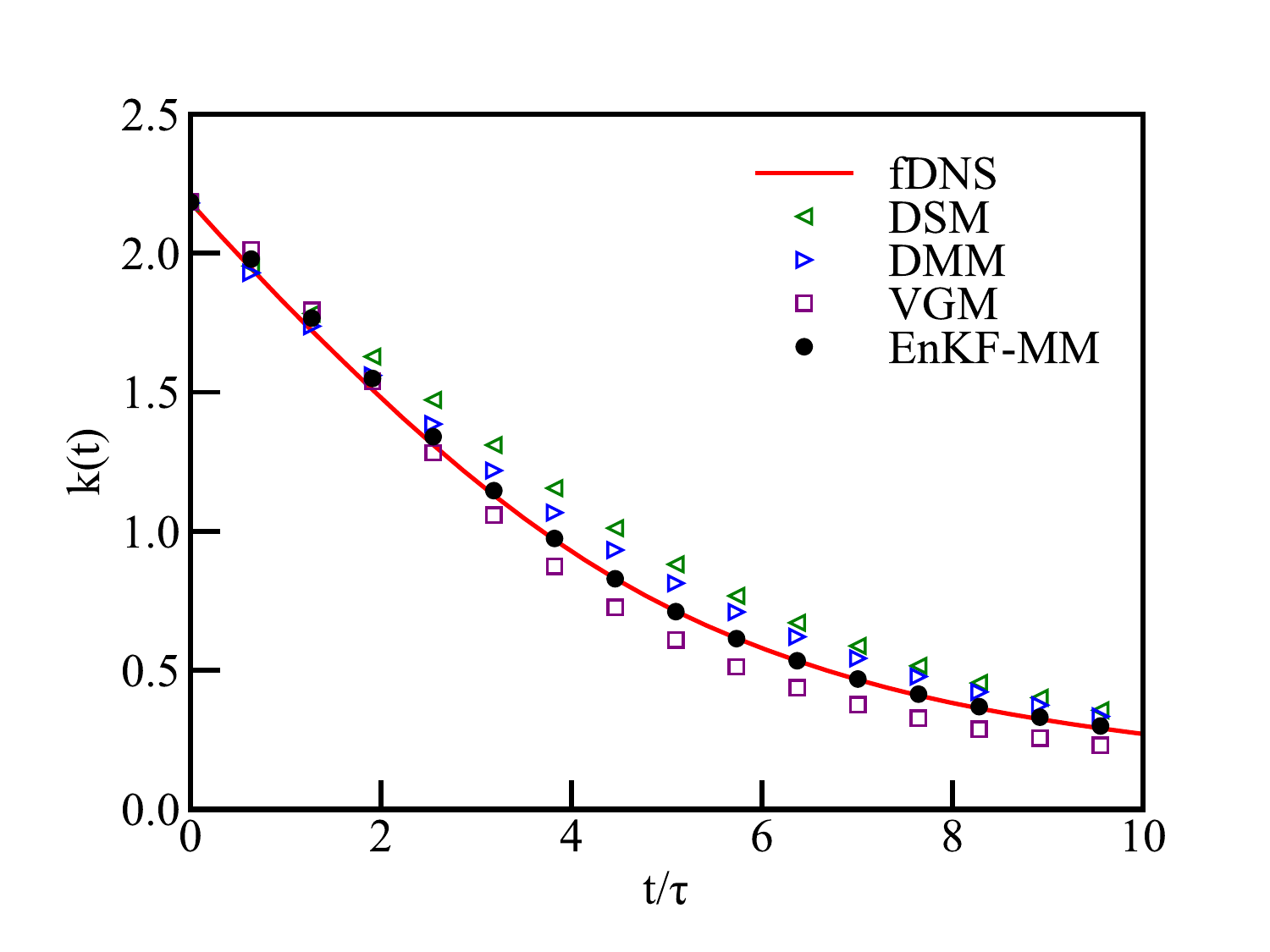}
 \caption{The evolution of turbulent kinetic energy in the LES of decaying HIT using different SGS models.}\label{fig_hit_kt_evo_decay}
\end{figure}

To examine the influence of the large-scale forcing on the current method, we apply the obtained EnKF-MM model in the decaying homogeneous isotropic turbulence, which is initiated at $Re_{\lambda}=259$. The parameter configuration and the initial statistics of the DNS is same as shown in Table \ref{tab_hit_DNS259}, while the large-scale forcing is set to zero during the simulation. The energy spectra are calculated at two time instants $t=4.8\tau$ and $t=9.6\tau$, and displayed in Figs. \ref{fig_hit_Ek_decay_259}a and \ref{fig_hit_Ek_decay_259}b, respectively. We note that, in the decaying HIT, the large-eddy turnover time $\tau$ is calculated based on the DNS field at $t=0$. As observed, the EnKF-MM model predicts the energy spectrum better compared to the other models at both time instants. Meanwhile, the energy spectra for both fDNS and LES at $t=9.6\tau$ are lower than those at $t=4.8\tau$ due to decaying of turbulence. The evolution of the total turbulent kinetic energy is shown in Fig. \ref{fig_hit_kt_evo_decay}, where the decreasing trend of kinetic energy can be clearly observed and the prediction by the EnKF-MM model follows most closely the fDNS result.

\section{Test on using the functional model alone with tuned coefficient}

In the current work, the mixed SGS model is constituted of the functional part and structural part. To take full advantage of the high accuracy of the structural model, its coefficient is kept at the original value.\cite{Clark1979,Pope2000} To enhance numerical stability, the coefficient of the functional part is calibrated by the EnKF framework. The tuned values for the functional coefficient in the forced HIT and TML cases are $C_D=0.00738$ and $0.01$, respectively. Meanwhile, the original value of $C_D$ can be calculated using the classical Smagorinsky coefficient $C_{Smag}=0.16$,\cite{Smagorinsky1963,Lilly1967,Pope2000} according to $C_D = 2C^{2}_{Smag} \approx 0.05$. Clearly, the tuned functional coefficient is no more than 20 percent of its original value. Hence, the contribution of the structural part is more dominant in the EnKF-MM model. In fact, our previous adjoint-based variational method has also shown that, the structural coefficient is very close to the original value while the functional coefficient is much less, even if both coefficients are re-calibrated in the mixed model.\cite{Yuan2023} Nevertheless, it is still interesting and meaningful to investigate the scenario where the functional model is used alone but with a tuned coefficient. 

For demonstration, we study the effect of the tuned functional model in the absence of the structural contribution for forced HIT. In the EnKF-MM framework, we set the coefficient of the VGM model to zero and re-calibrate the value of $C_D$. The resulted model shall be termed as the EnKF-based functional model (EnKF-FM). Using the proposed EnKF framework, we obtain $C_D =0.01$, which is about 36 percent larger than the corresponding value in the EnKF-MM model ($C_D=0.00738$). We note that this is not surprising since the structural part of the model also contributes to the SGS energy transfer,\cite{Garnier2009} while in EnKF-FM model the functional model is solely responsible for this task.

\begin{figure}\centering
\includegraphics[width=.5\textwidth]{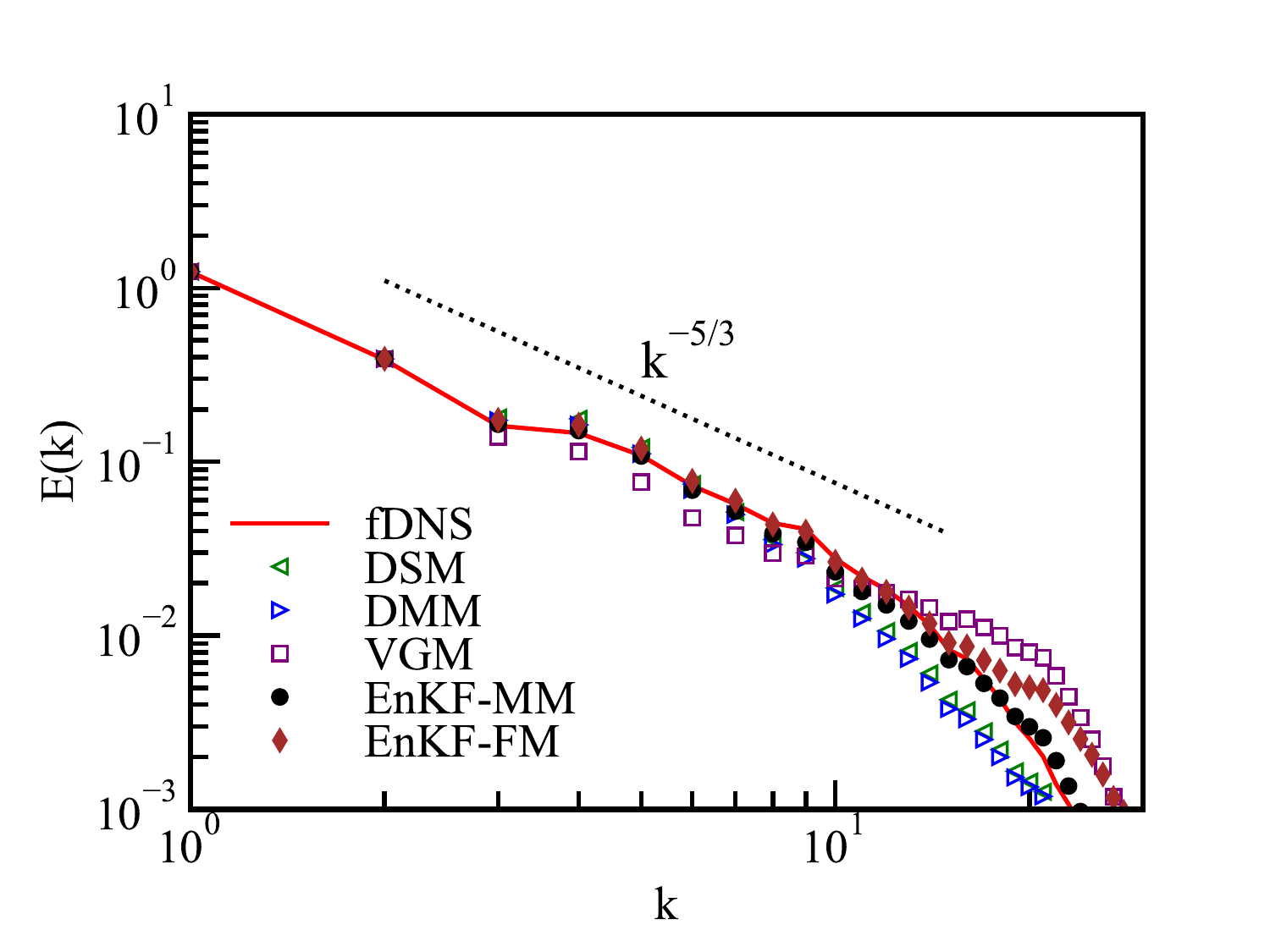}
 \caption{The kinetic energy spectrum in the LES of forced HIT using different SGS models.}\label{fig_hit_Ek_EnKF_FM}
\end{figure}

\begin{figure}\centering
\includegraphics[width=.45\textwidth]{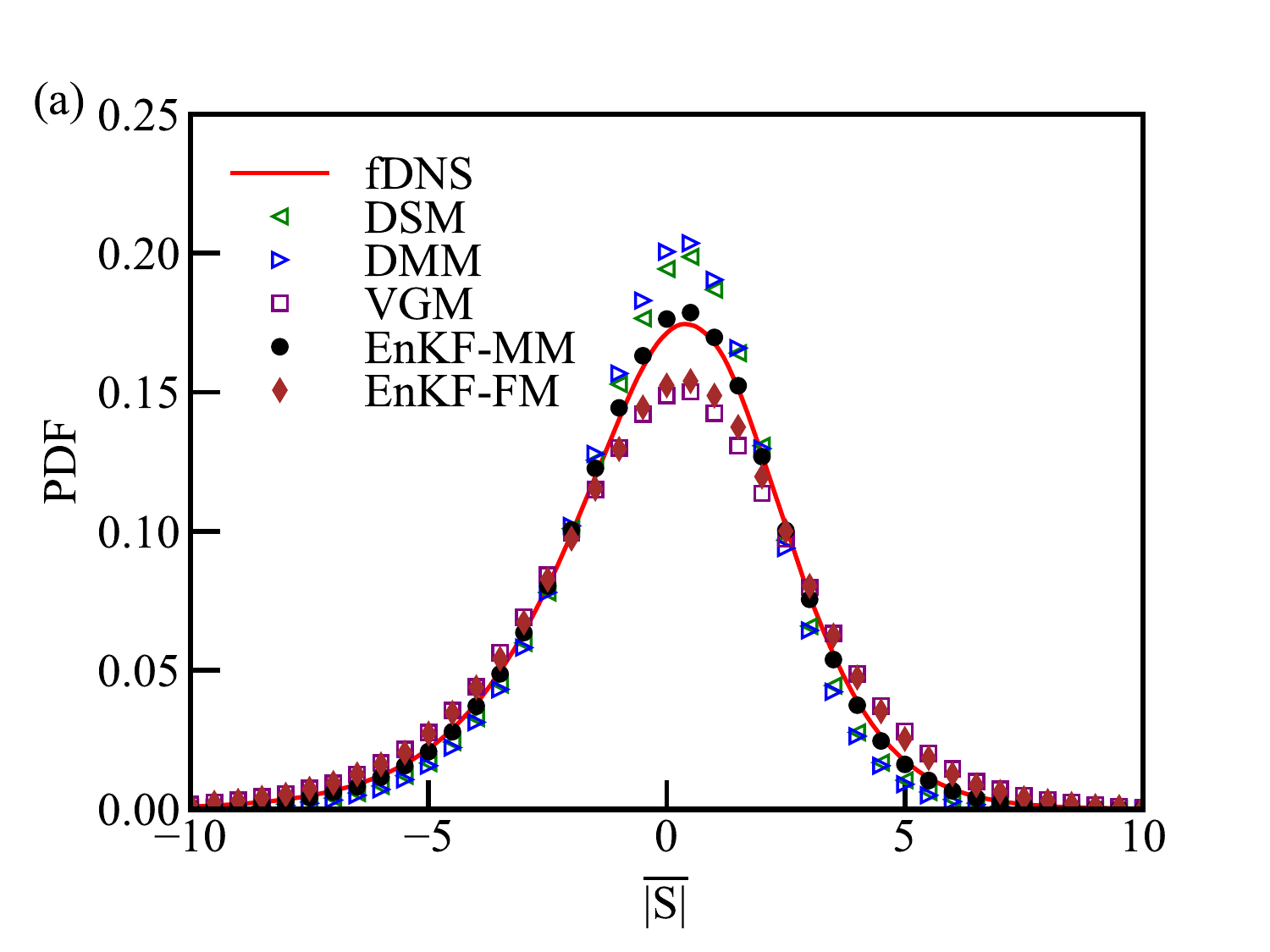}
\includegraphics[width=.45\textwidth]{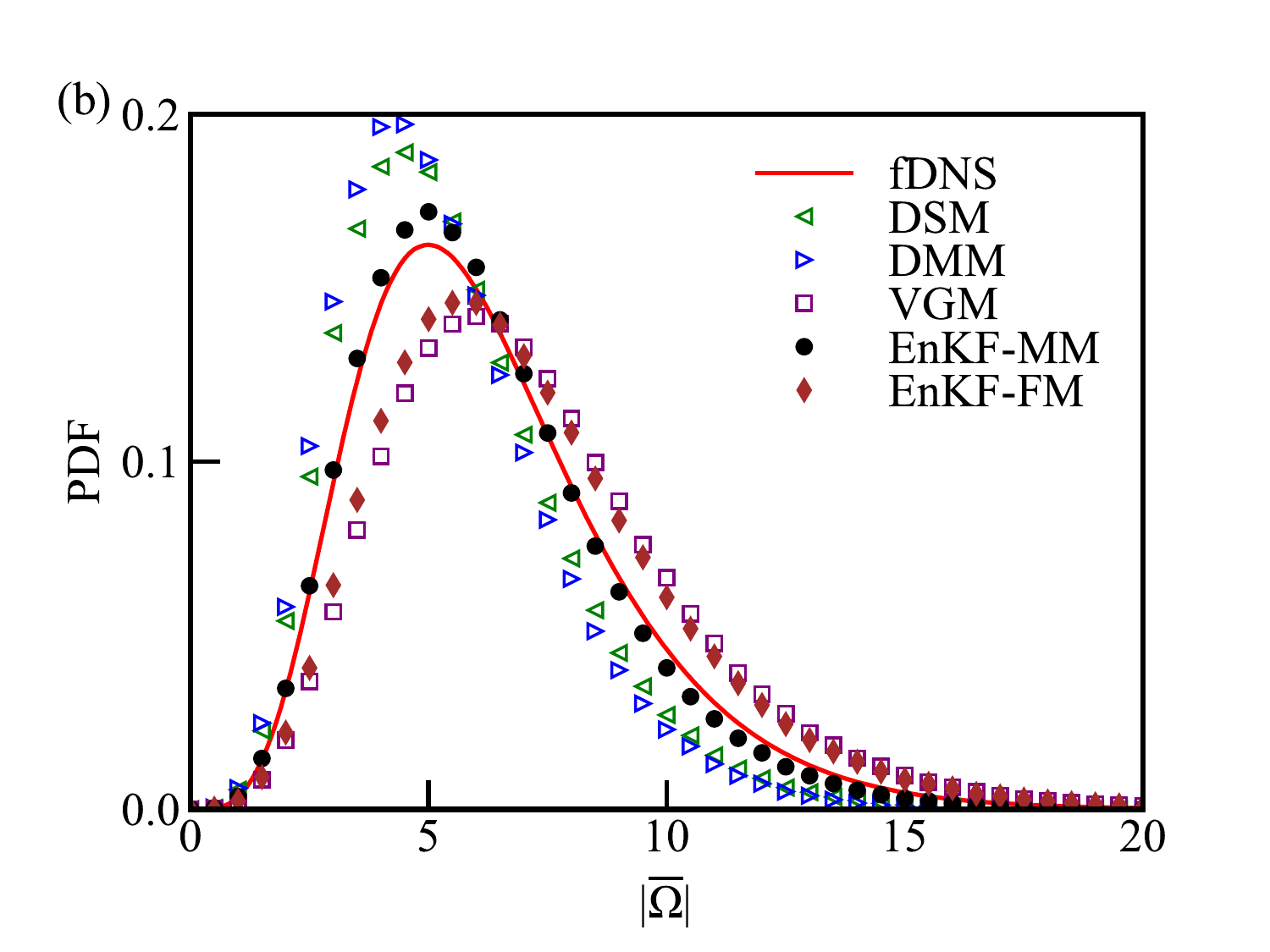}
 \caption{The PDFs of the characteristic filtered strain rate and characteristic filtered rotation rate in the LES of forced HIT using different SGS models: (a) PDFs of the characteristic filtered strain rate, (b) PDFs of the characteristic filtered rotation rate.}\label{fig_hit_shear_rotation_rate}
\end{figure}

\begin{figure}\centering
\includegraphics[width=.5\textwidth]{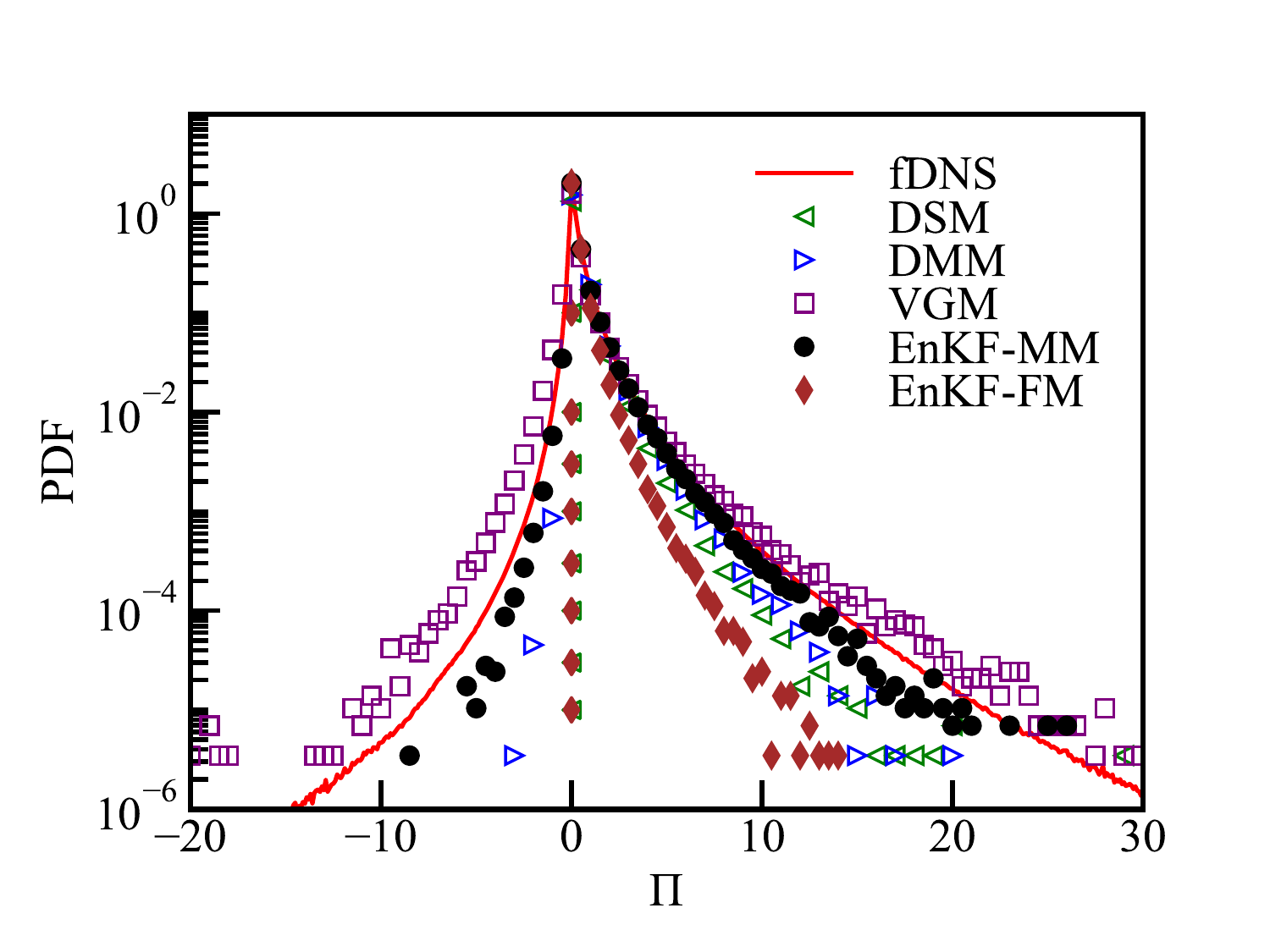}
 \caption{PDF of the SGS energy flux in the LES of forced HIT using different SGS models.}\label{fig_hit_pi_EnKF_FM}
\end{figure}

The results of LES using the EnKF-FM model are shown in Figs. \ref{fig_hit_Ek_EnKF_FM} to \ref{fig_hit_pi_EnKF_FM}, where the predictions by the EnKF-MM model and the traditional models are also displayed. The energy spectrum is shown in Fig. \ref{fig_hit_Ek_EnKF_FM}. As observed, with a tuned coefficient, the EnKF-FM model can also reasonably recover the energy spectrum except in the high-wavenumber range where the energy is overestimated. However, in the predictions for the PDFs of the characteristic strain rate $|\overline{S}|$ (Fig. \ref{fig_hit_shear_rotation_rate}a), characteristic rotation rate $|\overline{\Omega}|$ (Fig. \ref{fig_hit_shear_rotation_rate}b) and the SGS energy flux (Fig. \ref{fig_hit_pi_EnKF_FM}), the EnKF-FM model performs tangibly worse than the EnKF-MM model. In the predictions for $|\overline{S}|$ and $|\overline{\Omega}|$, the performance of the EnKF-FM model is similar to that of the VGM model. In the predictions for the SGS energy flux, the prediction by the EnKF-FM model is the worst among all the test models. These results suggest that the contribution of the structural model is indeed important in the mixed SGS model.

%\nocite{*}

\end{document}